\RequirePackage{fix-cm}
\documentclass{svjour3}                     
\smartqed  
\RequirePackage[T1]{fontenc}

\RequirePackage{graphicx}
\RequirePackage{mathptmx}      
\RequirePackage{flushend}
\RequirePackage[numbers,sort&compress]{natbib}
\RequirePackage[colorlinks,citecolor=blue,urlcolor=blue,linkcolor=blue]{hyperref}
\usepackage{pdflscape}
\usepackage{amsmath, amssymb, graphics}
\newcommand{\mathsym}[1]{{}}
\newcommand{\unicode}[1]{{}}
\usepackage{textcomp}
\usepackage{eurosym}
\usepackage{widetext}
\usepackage{amsfonts}
\usepackage{array}
\usepackage{bm}
\usepackage{helvet}

\usepackage{graphicx}
\usepackage{subfigure}
\usepackage{bm, array}
\usepackage[T1]{fontenc}
\def\be{\begin{equation}}
\def\ee{\end{equation}}
\def\beq{\begin{eqnarray}}
\def\eeq{\end{eqnarray}}

\newcommand{\ben}{\begin{eqnarray}}
\newcommand{\een}{\end{eqnarray}}

\usepackage{color}
\usepackage{enumerate}
\usepackage{soul}
\usepackage[normalem]{ulem}
\usepackage{color}
\usepackage{color}
\usepackage{mathtools, cuted}
\def\ex{e_1{}^1}
\begin{document}

\title{Static Spherically Symmetric Einstein-aether models I: Perfect fluids with a linear equation of state and scalar fields with an exponential self-interacting potential}

\titlerunning{Static Spherically Symmetric Einstein-aether models I} 

\author{Alan Coley     \and Genly Leon  
}

\institute{A. Coley \at
              Department of Mathematics and Statistics,
 Dalhousie University,  Halifax, Nova Scotia, Canada  B3H 3J5\\
\email{\href{mailto:aac@mathstat.dal.ca}{aac@mathstat.dal.ca} 
}  
           \and
           G. Leon \at
              Departamento  de  Matem\'aticas,  Universidad  Cat\'olica  del  Norte, Avda. Angamos  0610,  Casilla  1280  Antofagasta,  Chile. \\
\email{\href{mailto:genly.leon@ucn.cl}{genly.leon@ucn.cl}} }

\date{Received: date / Accepted: date}

\maketitle

\begin{abstract}
We investigate the field equations in the Einstein-aether theory for static spherically symmetric spacetimes and a perfect fluid source and subsequently with the addition of a scalar field (with an exponential self-interacting potential). We introduce more appropriate dynamical variables that facilitate the study of the equilibrium points of the resulting dynamical system and, in addition, we discuss the dynamics at infinity. We study the qualitative properties of the models with a particular interest in their asymptotic behaviour and whether they admit singularities. We also present a number of new solutions.

\end{abstract}

\tableofcontents
\section{Introduction}

Einstein-aether theory \cite{Jacobson:2000xp,Eling:2004dk,DJ,kann,
Zlosnik:2006zu,CarrJ,Jacobson,Carroll:2004ai,Garfinkle:2011iw}
is an effective field theory that consists
of General relativity (GR) coupled to a dynamical time-like unit vector field,
the aether. Since both the  dynamical  aether
vector field  and the geometric metric tensor  characterize the
spacetime structure  \cite{Jacobson}, the
Lorentz invariance is spontaneously broken  by the choice of a preferred frame at each spacetime point  (while
local rotational symmetry is maintained).
Such a
Lorentz violation has been proposed to model quantum gravity effects
at the microscopic level.
In addition, every hypersurface-orthogonal
Einstein-aether solution is a solution
of the IR limit of 
Horava-Lifshitz gravity \cite{Horava,TJab13}.

In a recent review some developments of Einstein-aether theory in general and Horava-Lifshitz  theory
in particular were discussed \cite{Wang:2017brl}.
This included a discussion of
universal horizons and black holes and their thermodynamics, non-relativistic gauge/gravity duality, and the quantization of the theory.
The well-posednessness of the Cauchy formulation of Einstein-aether theory
was recently studied in  \cite{Sarbach:2019yso} to ensure the stability of the numerical evolution of the initial value problem, and it was shown that, under suitable conditions on the parameters couplings, the governing equations can be cast into strongly hyperbolic form and even into symmetric hyperbolic form using a first-order formulation in the frame variables. 
Gravitational plane-waves in  Einstein-aether theory were also recently studied, and it was found 
that \cite{Oost:2018oww}
the vacuum Einstein-aether theory system of linearly polarized gravitational waves  is, in general, overdetermined,
and that there are further constraints  on the  coupling parameters $c_i$  in order to allow arbitrary gravitational plane waves.
In GR $c_i = 0$.

Cosmological scenarios in these theories were tested  against new observational constraints including updated Cosmic Microwave Background data from Planck and the expansion rates of elliptical and
lenticular galaxies, Joint Light-Curve Analysis data for
Type Ia supernovae and Baryon Acoustic Oscillations. Using priors
on the Hubble parameter and with an alternative parametrization of the equations in which
the curvature parameter is considered as a free parameter in the analysis, it was found  \cite{Nilsson:2018knn}
that the detailed-balance scenario exhibits
positive spatial curvature to more than $3\sigma$, whereas for further theory generalizations it was found that there is evidence for positive spatial curvature at $1\sigma$. In general, cosmologically viable extended Einstein-aether theories are known 
that are compatible with Planck Cosmic Microwave Background
temperature anisotropy, polarization, and lensing data \cite{Trinh:2018pcb}.

A number of exact solutions
and a qualitative analysis of Einstein-aether cosmological models have been presented \cite{Barrow:2012qy,Sandin:2012gq,Alhulaimi:2013sha,Coley:2015qqa,Latta:2016jix}.
An emphasis has been placed on whether Lorentz violation affects the inflationary scenario (in particular, in spatially anisotropic cosmological models) in Einstein-aether theory \cite{CarrJ,kann,Zlosnik:2006zu}. Einstein-aether cosmology has been studied for the FRW metric  \cite{Campista:2018gfi} (including contracting, expanding and bouncing solutions), for the Kantowski-Sachs metric \cite{Latta:2016jix,VanDenHoogen:2018anx}  and for spatially homogenous metrics \cite{Alhulaimi:2017ocb}. In all cases the matter source was assumed to be coupled to the expansion of the aether field through an exponential potential. The models have been generalized to include an additional scalar field source.

In a recent paper \cite{Coley:2015qqa} we studied spherically 
symmetric Einstein aether models 
with a  perfect fluid matter source. 
We begin by discussing the field equations.
In order to perform a dynamical systems analysis it is useful 
to introduce suitable normalized variables \cite{WE,Coley:2003mj,Copeland:1997et},
which also facilities their numerical study. 
We then derive the
equilibrium points of the algebraic-differential system in terms of
proper normalized variables \cite{Coley:2015qqa} and analyze their
stability. 
The Einstein-aether static model with a perfect
fluid was first introduced in Section 6.1 of \cite{Coley:2015qqa} 
utilizing the dynamical variables inspired by \cite{Nilsson:2000zf}. 
We attempt to find asymptotic
expansions for all of the solutions corresponding to the equilibrium points. 
In particular, 
explicit known exact spherically symmetric solutions are recovered \cite{Nilsson:2000zf,Tolman:1939jz,Oppenheimer:1939ne,Misner:1964zz,kasner} and a number of new solutions with
naked singularities or horizons are found and the line elements are
presented. We also investigate the
dynamics at infinity and we present some numerical results that support our
analytic results.

In addition to defining appropriate normalized variables, we also wish to utilize 
well defined coordinates and to exploit any  symmetries of the spacetimes. 
Since we use qualitative techniques  of dynamical systems theory that do not involve actually solving the field equations, some of the problems of coordinate choices and coordinate singularities are avoided. In particular, the local semi-tetrad splitting  \cite{Ganguly:2014qia} allows the field equations to be recast in the form of an autonomous system of covariantly defined quantities  \cite{Clarkson:2002jz, Clarkson:2007yp}.

\subsection{The models and spherical symmetry}
Spherically symmetric static and stationary  solutions are physically important.
The evolution equations
follow from the Einstein aether action \cite{Jacobson:2000xp,DJ}.
There are extra terms in the Einstein-aether field equations due to
the effects of the aether field on the spherically symmetric geometry, and
from an additional
stress tensor, $T_{ab}^{ae}$, which depends on a number of 
dimensionless parameters $c_i$. In the case of spherically symmetry the aether is hypersurface orthogonal 
\footnote{Aether fields are hypersurface orthogonal in the spherically symmetric case and, hence, all solutions
of Einstein aether theory will also be solutions of the IR limit of
Horava gravity. The converse is not true generally, but it is so for spherically symmetric solutions with a regular center \cite{Jacobson}.},
and so it has vanishing twist so that $c_4$ can be set to zero without loss of
generality \cite{Jacobson}, leaving a 3-dimensional parameter space. A renormalization of the  parameters in the
model can be then used to set $8 \pi G =1$, where  $G$ defines the effective Newtonian
gravitational constant, so that the model can consequently be
characterized by only two non-trivial constant parameters.
The remaining constraints on the $c_i$
have been summarized in \cite{Jacobson,Barausse:2011pu}.

Solutions which involve a static metric coupled to a stationary
aether are called ``stationary spherical symmetric'' models and, in
principle, must be treated separately.  If the spherically
symmetric aether is parallel to the Killing vector, the solutions
are referred to as ``static aether'' solutions (and an explicit
solution is known \cite{Eling:2006ec}).

Therefore, in Einstein-aether theory, and in contrast to GR, there
is an additional spherically symmetric mode corresponding to the
radial tilting of the aether. That is, the preferred aether  frame 
can be tilted relative to
the CMB rest frame in spherically symmetric models, which adds additional
terms to $T_{ab}^{ae}$, characterized by a so-called hyperbolic tilt angle, $\alpha$, which measures the boost of the aether relative to this rest frame. The tilt is anticipated to decay in spatially 
homogeneous models \cite{tilt}. For example,  it was shown that to linear order in the anisotropy
a Bianchi type I anisotropic  system (with a positive cosmological constant)
relaxes exponentially to the isotropic, de Sitter solution, and that the tilt decays  to the future \cite{kann}. 
The dynamics of a tilted aether in a Bianchi I cosmological model without the assumption of a small tilt was studied in \cite{CarrJ}, and it was found that when the  initial hyperbolic tilt angle $\alpha$ (and its time derivative) is sufficiently small, then $\alpha \rightarrow 0$ at late times (consistent with the linearized stability analysis in \cite{kann}).

 A number of time-independent
spherically symmetric solutions and, in particular, black hole
solutions, were studied in \cite{Eling:2006ec,Eling:2006df},
surveyed in \cite{Jacobson}, and recently revisited in
\cite{Barausse:2011pu}.  In general, the dynamics of the
perturbations in non-rotating neutron stars and black hole
solutions do not differ much from those in GR.  Although a fully
nonlinear positivity of the energy has been established for
spherically symmetric solutions at an instant of time symmetry
\cite{Garfinkle:2011iw}, a comprehensive investigation of the fully
nonlinear solutions has not yet been done.

In particular, in Einstein-aether theory there is a 3-parameter
family of spherically symmetric static vacuum solutions, since the
aether vector and its derivative add 2 extra degrees of freedom at
each spacetime point~\cite{Eling:2006df}.  In the case that we
assume asymptotic flatness, for a fixed mass there is then a single
parameter family of solutions \cite{Jacobson}, unlike the the
unique Schwarzschild solution in GR.  In addition, in GR asymptotic
flatness is a result of the vacuum field equations so that the 1-parameter
family of local (Schwarzschild) solutions is immediately
asymptotically flat.  Since the radial aether tilt constitutes an
additional local degree of freedom, spherical solutions in aether
theory are not necessarily time-independent (even in the stationary
case).  Spherically symmetric solutions are not generally static,
but even in the case of staticity they need not be asymptotically
flat.

The model is restricted to a single parameter (the total mass) when
the aether is aligned with the time-like Killing field
\cite{Eling:2006df}.  Therefore, in this case, for a given mass the
exterior solution for a static star is the unique ``static aether''
vacuum solution (depending on the $c_i$ parameters via only one
parameter) presented analytically in~\cite{Eling:2006df}; it has a
global time-like Killing vector, is asymptotically flat, and the
affine parameter distance to the singularity is finite along radial
null geodesics.  Although this static ``wormhole'' implies an
effective negative energy density in the field equations, all solutions in this
family actually have a positive total mass.  In addition, it was
found \cite{Seifert:2007fr} that such a static aether solution is,
in general, linearly stable under precisely the same conditions as
flat Minkowski spacetime.  In the pure GR limit, $c_1=0$, we just
have the Schwarzschild solution.  However, for small $r$ values the
static aether solutions can have quite different behaviour to that
of the Schwarzschild solution.  More recently, an analytic static
spherically symmetric vacuum Einstein-aether solution was obtained
numerically \cite{Eling:2006ec,Eling:2006df,Gao:2013im,Tamaki:2007kz}.

Unlike the case of a singular wormhole, the static solutions have
an origin that is regular \cite{Eling:2006df}.  It is also well
known that no asymptotically flat self-gravitating aether solutions
with a regular origin exist \cite{Eling:2006df}; i.e., there are no
pure aether stars.  In the presence of a perfect fluid, regular
asymptotically flat stellar solutions have been shown to exist
parameterized (for a given equation of state) by the central
pressure (in addition to the vacuum aether parameters).  If the
central pressure is fixed, then there is only a single parameter
that can be further tuned to obtain an asymptotically flat
solution.  Static aether star solutions with an interior with
constant energy density were obtained numerically in
\cite{Eling:2006df} by matching the interior solution to a specific
vacuum exterior.  The solution inside a fluid star has also been
found by numerical integration for more realistic neutron star
equations of state~\cite{Eling:2007xh}.  There are small
differences from GR in sufficiently compact stars.

Since the Killing vector cannot be time-like on or inside an
horizon, the aether cannot be aligned in the case of black holes.
Instead, at spatial infinity the aether is at rest but travels in
an inward direction at a finite radius.  A unique spherical
stationary solution from the 1-parameter family of solutions for a
given mass is selected if regularity is required at the so-called
spin-0 horizon~\cite{Eling:2006ec,Eling:2006df}.  This horizon
develops in a regular region of spacetime when a black hole forms
under graviational collapse.  Some particular examples of such a
collapse producing a nonsingular black hole horizon have been
confirmed in numerical simulations of scalar field
collapse~\cite{Garfinkle:2007bk}.  Black holes with a nonsingular
spin-0 horizon are, in general, very similar to the Schwarzschild
solution exterior to the horizon.  But in the region interior to
the horizon the solutions are typically different by a few percent.
However, they do contain a spacelike singularity like the
Schwarzschild spacetime.  Recently static spherically symmetric,
asymptotically flat, regular (non-rotating) black hole solutions in
Einstein-aether theory have been studied numerically
\cite{Barausse:2011pu}, generalizing the results of
\cite{Eling:2006ec,Eling:2006df} and \cite{Tamaki:2007kz}.
Quasi-normal modes of black holes in aether theory have also been
investigated in \cite{Konoplya:2006rv}.

This paper is the first of a series of papers devoted to the study of static and stationary Einstein-aether models, and it will be referred hereafter as Paper I. In Paper I here we will study, from the dynamical system point of view, the models and we shall classify the equilibrium points and comment on some particular interesting solutions. We note that the formalism employed facilitates a natural physical interpretation.  In some cases the matter configuration is enclosed in a finite radius and the models have an astrophysical application \cite{Nilsson:2000zf,Nilsson:2000zg,Carr:1999rv}. In the companion paper \cite{Leon:2019jnu}, referred as the Paper II of the series, we will apply the classical singularity analysis, which is summarized in the so-called ARS algorithm \cite{Abl}. Furthermore, the formulation of the modified Tolman-Oppenheimer-Volkoff (TOV) equations for perfect fluids with linear and polytropic equations of state (EoS) in the Einstein-aether theory is also of interest. The relativistic TOV equations are drastically modified in Einstein-aether theory \cite{Leon:2019jnu}. The addition of a scalar field, $\varphi$, with an  exponential or an harmonic potential is also of interest.

In future work in the series we will generalize the work to the conformally static (i.e., timelike selfsimilar) case and to scalar field models with generalized self-interacting potentials. In particular in \cite{paperIII}, referred to as paper III of the series, will be studied the general monomial potential \cite%
{Alho:2015cza}: 
\begin{equation}
W(\varphi)=\frac{1}{2n}(\mu \varphi)^{2n}, \mu >0, n=1,2,\ldots,
\end{equation}  which contains as a particular case the harmonic potential $W(\varphi)= \frac{1}{2} m^2 \varphi^2$.

The plan of the paper follows:
The basic definitions of the Einstein-aether gravity are given in Section %
\ref{einstein}. The stability
analysis for the static spherically
symmetric perfect fluid spacetime are presented in Section \ref{model2}, and the analysis with the additional scalar field model is discussed in Section \ref{sf}. 
Finally, in Section \ref{conclusions} we discuss our results
and draw our conclusions. 

\section{Einstein-aether Gravity}

\label{einstein}

In Einstein-aether theory the action is given by the
following expression \cite{Jacobson,Carroll:2004ai}:
\begin{equation}
S=S_{GR}+S_{u}+S_{m},  \label{action}
\end{equation}%
where~$S_{GR}=\int d^{4}x\sqrt{-g}\left( \frac{R}{2}\right) $ is the
Einstein-Hilbert term,~$S_{m}$ is the term which corresponds to the matter
source and
\begin{equation}
S_{u}=\int d^{4}x\sqrt{-g}\left( -K^{ab}{}_{cd}\nabla _{a}u^{c}\nabla
_{b}u^{d}+\lambda \left( u^{c}u_{c}+1\right) \right),
\end{equation}%
corresponds to the aether field. $\lambda $ is a Lagrange multiplier
enforcing the time-like constraint on the aether \cite{Garfinkle:2011iw},
for which we have introduced the coupling \cite{Jacobson}
\begin{equation}
K^{ab}{}_{cd}\equiv c_{1}g^{ab}g_{cd}+c_{2}\delta _{c}^{a}\delta
_{d}^{b}+c_{3}\delta _{d}^{a}\delta _{c}^{b}+c_{4}u^{a}u^{b}g_{cd},  \label{aeLagrangian}
\end{equation}%
which depends upon four dimensionless coefficients $c _{i}$. Finally $%
u^{a}$ is the normalized observer in which $u^{a}u_{a}=-1$.
For simplicity in the following we redefine the constants, $c_{i}$, as
follows:
\begin{equation*}
c_{\theta }=c_{2}+(c_{1}+c_{3})/3,\ c_{\sigma
}=c_{1}+c_{3},\ c_{\omega }=c_{1}-c_{3},\
c_{a}=c_{4}-c_{1}.
\end{equation*}

Variation with respect to the metric tensor in (\ref{action}) provides
the gravitational field equations%
\begin{equation}
{G_{ab}}=T_{ab}^{ae}+T_{ab}^{mat},  \label{EFE2}
\end{equation}%
in which $G_{ab}$ is the Einstein tensor, $T_{ab}^{m}$ corresponds to $S_{m}$
and $T_{ab}^{ae}$ is the aether tensor \cite{Garfinkle:2007bk}:
\begin{align}  
\label{aestress}
{T_{ab}^{ae}}& =2c_{1}(\nabla _{a}u^{c}\nabla _{b}u_{c}-\nabla
^{c}u_{a}\nabla _{c}u_{b})-2[\nabla _{c}(u_{(a}J^{c}{}_{b)})+\nabla
_{c}(u^{c}J_{(ab)})-\nabla _{c}(u_{(a}J_{b)}{}^{c})]  \notag \\
& -2c_{4}\dot{u}_{a}\dot{u}_{b}+2\lambda u_{a}u_{b}+g_{ab}\mathcal{L}%
_{u}, \quad 
{{J^{a}}_{m}}=-{{K^{ab}}_{mn}}{\nabla _{b}}{u^{n}}, \quad {\dot{u}_{a}}%
={u^{b}}{\nabla _{b}}{u_{a}.} 
\end{align}%
In addition, variation with respect to the vector field $u^{a}$ and the
Lagrange multiplier gives us
\begin{subequations}
\label{FE1}
\begin{eqnarray}
\lambda {u_{b}} &=&{\nabla _{a}}{{J^{a}}_{b}}+c _{4}\dot{u}_{a}\nabla
_{b}u^{a},  \label{evolveu} \\
{u^{a}}{u_{a}} &=&-1,  \label{unit}
\end{eqnarray}%
where from (\ref{evolveu}) we derive the Lagrange multiplier to be
\end{subequations}
\begin{equation}
\lambda =-u^{b}\nabla _{a}J_{b}^{a}-c_{4}\dot{u}_{a}\dot{u}^{a}.
\label{definition:lambda}
\end{equation}
Hence the compatibility conditions are
\begin{equation}
0=h^{bc}\nabla _{a}J_{b}^{a}+c_{4}h^{bc}\dot{u}_{a}\nabla _{b}u^{a}.
\label{restrictionaether}
\end{equation}

The energy momentum tensor of the matter source in the form of a
perfect fluid (with energy density $\mu $, and pressure $p$) in the 1+3
decomposition with respect to $u^{a}$ is given by:
\begin{equation}
{T_{ab}^{m}}\equiv \mu u_{a}u_{b}+p(g_{ab}+u_{a}u_{b}),
\end{equation}%
in which $h_{ab}=(g_{ab}+u_{a}u_{b})$ is the projective tensor where $%
h^{ab}u_{b}=0$. We shall use the equation \eqref{definition:lambda} as a
definition for the Lagrange multiplier, whereas the equation 
\eqref{restrictionaether} leads to a set of constraints that the aether
vector must satisfy.

The theory has additional degrees of freedom (model parameters) in flat space as compared with GR. The theory presents two spin-2 polarizations, as in GR, but also one spin-0 and two spin-1 polarizations. The squared propagation speeds on flat space are, respectively, given by  \cite{Jacobson:2004ts}:
\begin{align}
&s_2^2=\frac{1}{1-c_1-c_3}=-\frac{1}{c_\sigma-1},\\
&s_1^2=\frac{2 c_1-c_1^2+c_3^2}{2(c_1-c_4)(1-c_1-c_3)}=-\frac{c_\sigma c_\omega-c_\sigma-c_\omega}{2 (\beta -1) (c_\sigma-1)},\\
&s_0^2=\frac{(c_1+c_2+c_3)(2-c_1+c_4)}{(c_1-c_4)(2+c_1+c_3+3 c_2)}=-\frac{(\beta +1)
   (3 c_\theta+2 c_\sigma)}{3 (\beta -1) (3
   c_\theta+2)},
\end{align}
where we have introduced  the parameter redefinition $\beta=c_a+1$.

Stability at the classical and quantum levels requires  all of the $s_{i}^2$ ($i=0,1,2$) to be positive \cite{Garfinkle:2011iw,Jacobson:2004ts}. Ultra-high energy cosmic ray observations requires 
$s_i^2>1-\mathcal{O}(10^{-15})$ to prevent cosmic rays from losing energy into gravitational modes via Cherenkov-like cascade \cite{Elliott:2005va}. 
Additionally, from constraints on the PPN parameters it follows that $|\alpha_1|\lesssim 10^{-4}$ and $|\alpha_2|\lesssim 10^{-5}$,  
$\alpha_1=-\frac{8 \left({c_3}^2-{c_1}
  c_{4}\right)}{-c_{1}^2+2
  c_{1}+c_{3}^2}$,\\ $\alpha_2=-\frac{(-c_{1}+c_{4}+2)
   (c_1 +2 c-{3}+c_{4}) (2
   c-{1}+3 c-{2}+c_{3}-c_{4})}{c_{1}+c_{2}+c_{3}}+\frac{\alpha_1}{2}$ \cite{Foster:2005dk}. Combining all the above restrictions we find  
	$c_\omega\approx \mathcal{O}(10^{-15}), \beta \approx 1+\mathcal{O}(10^{-4}), c_\sigma \approx 3 c_\theta+\mathcal{O}(10^{-4})$. 
These bounds change if we consider static spherically symmetric curved space or if we change the matter content to 
include a perfect fluid or scalar field. Therefore, we assume no bounds on the model parameters.  
  
\section{Static spherically symmetric spacetime with a perfect fluid}

\label{model2}

In a static spherically symmetric spacetime with line element 
\begin{equation}
ds^{2}=-N^{2}(r)dt^{2}+\frac{dr^{2}}{r^2}+K^{-1}(r)(d\vartheta ^{2}+\sin
^{2}\vartheta d\varphi ^{2}),  \label{met2}
\end{equation}%
that is, we have fixed the spatial gauge to have $\ex(r)\equiv r$, 
the field equations are \cite%
{Coley:2015qqa}:
\begin{subequations}
\label{system_12}
\begin{align}
&r \frac{d x}{dr}=\frac{\mu +3p}{2\beta }+2(\beta-1)y^2+3xy+K
\label{eq.01},\\
&r \frac{dy}{dr}=\frac{\mu +3p}{2\beta }+2xy-y^{2}  \label{eq.02}
,\\
&r \frac{dp}{dr}=-y\left( \mu +p\right)   \label{eq.03}
,\\
&r \frac{dK}{dr}=2xK.  \label{eq.04}
\end{align}%
\end{subequations}
where $x\equiv \frac{1}{2} r \frac{d\ln(K)}{dr},~y \equiv r \frac{d\ln(N)}{dr}$, and $p$ is the pressure of the perfect fluid.\\ Furthermore there exists the constraint
equation%
\begin{equation}
x^{2}=(\beta-1)y^2+2xy+p+K,  \label{eq.05}
\end{equation}%

From (\ref{eq.03}) and (\ref{eq.04}) we have that $y=-\frac{r p^{\prime }}{\mu
+p}~\ ,~x=\frac{1}{2}\frac{r K^{\prime }}{K},~$and substituting into (\ref{eq.01}%
),~(\ref{eq.02}) we find a system of two second-order ordinary differential
equations,%
\begin{subequations}
\begin{align}
&r^2\left[\frac{p^{\prime \prime }}{\mu +p}-\frac{p^{\prime }\mu^{\prime }}{\left( \mu
+p\right)^2}-\frac{p^{\prime }K^{\prime }}{K\left( \mu
+p\right) }-\frac{2\left( p^{\prime }\right) ^{2}}{\left( \mu +p\right) ^{2}}\right]%
+\frac{r p^{\prime }}{\mu +p}+\frac{\mu +3p}{2\beta }=0,  \label{eq.07}\\
& r^2 \left[\frac{K^{\prime \prime }}{2K}-\frac{1}{2}\left( \frac{K^{\prime }}{K}\right)
^{2}+\frac{3}{2}\frac{p^{\prime }K^{\prime }}{K\left( \mu +p\right) }-\frac{%
2\left(\beta-1\right)\left( p^{\prime }\right) ^{2}}{\left( \mu +p\right) ^{2}}\right]+\frac{r K^{\prime }}{2 K}-K-\frac{\mu
+3p}{2\beta }=0,  \label{eq.08}
\end{align}
\end{subequations}
and (\ref{eq.05}) becomes
\begin{equation}
r^2\left[\left(\beta-1\right)\left( \frac{p^{\prime }}{\mu +p}\right) ^{2}+\frac{p^{\prime
}K^{\prime }}{K\left( \mu +p\right) }-\frac{1}{4}\left( \frac{K^{\prime }}{K}%
\right) ^{2}\right]+p+K=0,  \label{eq.10}
\end{equation}
\\ 
where prime means the derivative with respect $r$. 

\subsection{Phase-Space Evolution}
In this section we use the dynamical systems approach for investigating the
structure of the whole solution space of \eqref{system_12}. With this purpose, we introduce the quantities $\theta=y-x, \sigma=y$ as in \cite{Nilsson:2000zf}. {\footnote{Do not confuse these quantities with the usual expansion and shear scalars of homogeneous cosmologies.}}
The equations then read:
\begin{align}
&r \frac{d\theta}{dr}=-\beta  \sigma ^2-\theta ^2+\theta  \sigma +p,\\
&r \frac{d\sigma}{dr}=\frac{-4 \beta  \theta  \sigma +\beta  \sigma ^2+\theta ^2-K+\mu_0+(\eta+1)p}{2 \beta },\\
&r \frac{dp}{dr}=-\sigma  (\eta 
   p+\mu_{0}),\\
&r \frac{dK}{dr}=	2 K (\sigma -\theta ),
		\end{align}
where 
		\begin{align}
		& K+p= -\beta  \sigma ^2+\theta^2,
		\end{align}
\\and we have assumed a linear EoS
\begin{equation}
\mu =\mu _{0}+(\eta -1)p,  \label{ss1}
\end{equation}%
where the constants $\mu _{0}$ and $\eta $ satisfy $\mu _{0}\geq 0,\eta \geq
1.$ The case $\eta =1$ corresponds to an incompressible fluid with constant
energy density, while the case $\mu _{0}=0$ describes a scale-invariant
EoS.

Next, we introduce the scale invariant quantities:
\begin{equation}
\label{inv:qt}
Q=\frac{\theta}{\sqrt{\frac{\mu_0}{\eta }+\theta^2}}, \quad S=\frac{\sigma}{\sqrt{\frac{\mu_0}{\eta }+\theta^2}}, \quad C=\frac{\eta  K}{\mu_0+\eta  \theta^2},
\end{equation}
which are more appropriate for describing the dynamics than those used in \cite{Coley:2015qqa} as it covers new equilibrium points with $Q=S$ (i.e., $x=0$, where $x\equiv \frac{1}{2}  \partial_r \ln(K)$). Furthermore, to define the $x$-normalized dimensionless variables and the new independent coordinate $\tau$ given by $\partial_r(f) = -x\partial_\tau(f)$ in (6.7) of \cite{Coley:2015qqa}, it was presumably assumed that $x$ does not change sign during the whole evolution; but, when $x$ changes sign, the direction of the flow given by the independent variable $\tau$  in \cite{Coley:2015qqa} is lost. For this reason we use below the variables \eqref{inv:qt} and we introduce a new independent variable $\lambda$ given  by \cite{Nilsson:2000zf}:
\begin{equation}
\label{eqr}
\frac{dr}{r d\lambda}=\frac{1}{\sqrt{\frac{\mu_0}{\eta }+\theta^2}}
=
\left\{\begin{array}{cc}
\sqrt{\frac{\eta(1-Q^2)}{\mu_0}}, & \mu_0\neq 0\\
\sqrt{\frac{C}{K}}, & \mu_0=0, C\neq 0\\
\frac{S}{\sigma}, & \mu_0=0, C=0,
\end{array}
\right.
\end{equation}
which defines unequivocally the flow direction.  The ``past attractors'' ($\lambda\rightarrow -\infty$) corresponds to  
$r\rightarrow 0$ and the ``future attractors'' (  $\lambda\rightarrow \infty$) corresponds to
$r \rightarrow \infty$.

The relation between the variables $\{Q, S, C\}$ and   $\{U, V, Y\}$ to be used in the forthcoming paper \cite{Leon:2019jnu} is
\begin{subequations}
\label{TOV-vars}
\begin{align}
& U=\frac{C (\eta -1)-\eta +Q^2+\beta  (\eta -1) S^2}{C (\eta -2)-\eta +2 Q^2-2 Q S+S^2 (\beta  (\eta -1)+1)},\\
& V=\frac{\left(C-(Q-S)^2\right) \left(C (\eta -1)-\eta
   +Q^2+\beta  (\eta -1) S^2\right)}{\left(C+(Q-S)^2\right) \left(C-Q^2+\beta  S^2\right)+\eta  \left(C-(Q-S)^2\right) \left(C+\beta  S^2-1\right)},\\
& Y=\frac{C-Q^2+\beta  S^2}{\eta  \left(C+\beta  S^2-1\right)}.
\end{align}
\end{subequations}
We obtain then the evolution equations
\begin{subequations}
\label{inv:qt_evol}
\begin{align}
& \frac{d Q}{d\lambda}=\left(Q^2-1\right) (C+S (2 \beta S-Q)), \\
& \frac{d S}{d\lambda}=\frac{C (2 \beta Q S-\eta -2)+\left(\beta S^2-1\right) \left(4 \beta Q S-\eta -2 Q^2\right)}{2 \beta}, \\
& \frac{d C}{d\lambda}=2 C \left(Q \left(C+2 \beta S^2-Q S-1\right)+S\right).\end{align}
\end{subequations}
We have the useful relations
\begin{align}
\label{rel-30}
&\frac{\mu_0}{\eta}=\frac{(1-Q^2)K}{C}, \quad p= -\frac{K \left(C+\beta S^2-Q^2\right)}{C}, \quad \frac{C (\mu +p)}{\eta  K}=1-C-\beta S^2.
\end{align}

The equations \eqref{inv:qt_evol} reduce to the system (17) investigated in \cite{Nilsson:2000zf} for $\beta=1$.
Because $\mu_0\geq 0,\eta\geq 1$, it follows that $-1\leq Q\leq 1.$  Due to $K\geq 0$, it follows that $C\geq 0$. The condition $C+\beta S^2-Q^2=0$ defines the surface of zero-pressure. However, it is not an invariant set of \eqref{inv:qt_evol}, neither $C+\beta S^2-Q^2>0$. \footnote{A set of states $E\subset \mathbb{R}^n$ of a system of differential equations, say \eqref{inv:qt_evol}, is called an invariant set
of \eqref{inv:qt_evol} if for all $\mathbf{x}_0 \in E$ and for all $\lambda \geq  0$, $\mathbf{x}(\lambda; \mathbf{x}_0)\in E$, where by $\mathbf{x}(\lambda; \mathbf{x}_0)$ we understand the solution of \eqref{inv:qt_evol} satisfying the initial condition $\mathbf{x}(0; \mathbf{x}_0)=\mathbf{x}_0$, evaluated at $\lambda$.} If we assume that the weak energy condition $p+\mu\geq 0$ is satisfied, then we obtain the subset of the phase space $1-C-\beta S^2\geq 0$.
 Defining $\Omega=1-C-\beta S^2$ we obtain
\begin{equation}
\frac{d \Omega}{d \lambda}=\Omega\left[2 C Q+S \left(4 \beta Q S-\eta -2 Q^2\right)\right].
\end{equation}
Thus $1-C-\beta S^2\geq 0$ defines an invariant set.
In summary, the equations \eqref{inv:qt_evol} define a flow on the invariant set
\begin{equation}
\left\{\left(Q, S, C\right):-1\leq Q \leq 1, C\geq 0, C+\beta S^2\leq 1\right\}.
\end{equation}
This phase-space is compact for $\beta\geq 0$ and unbounded for $\beta< 0$.
The invariant sets $Q=\pm 1$ corresponds to $\mu_0=0$. The expression $\left(C+\beta S^2-Q^2\right)=0$ defines a surface on the phase space, which refers to the surface of vanishing pressure. This surface, however, is {\em{not}} an invariant surface for the flow. 

A monotonic function excludes equilibrium points, periodic orbits, recurrent orbits, and homoclinic orbits in is domain. 
As in \cite{Nilsson:2000zf} we introduce the function 
\begin{equation}
\label{eqZ}
Z=\frac{2 Q-S}{\sqrt{\left(2 Q-S\right)^2+3(1-Q^2)}},
\end{equation}
which satisfies
\begin{equation}
\frac{dZ}{d\lambda}=-\frac{3 \left(1-Q^2\right) \left(2 (2 \beta -1) C+2 (Q-2 \beta  S)^2+\eta  \Omega\right)}{2 \beta  \left(3 \left(1-Q^2\right)+(2 Q-S)^2\right)^{3/2}}.
\end{equation}

Since $C\geq 0, W\geq 0$ it is obvious that \eqref{eqZ} is a monotonic  decreasing function for $(2 \beta -1)\geq 0$. Furthermore, it is defined everywhere except on the scale-invariant boundaries $Q=\pm 1$. Hence, the ``past'' ($r\rightarrow 0$) and the ``future'' ($r \rightarrow \infty$) attractors lie on the $Q=\pm 1$ boundary sets. 
We also have the auxiliary equations
\begin{subequations}
\label{metric-eqs}
\begin{align}
\label{eqN}
&\frac{d\ln N}{d\lambda}=S,\\
\label{eqK}
&\frac{d \ln K}{d\lambda}=-2(Q-S),\\
\label{eqy}
& \frac{d \ln y}{d\lambda}= \frac{-C (\eta +2)+2 Q (Q-2 \beta S)-\beta \eta  S^2+\eta }{2 \beta S}.
\end{align}
\end{subequations}

The relation between the gravitational potential $\phi$ (related with the lapse function by $N=e^{\phi}$), and the matter field is given by 
\begin{align}
\frac{d \phi}{d p}=-\frac{1}{\mu + p}, \quad \mu=\mu_{0} +(\eta-1)p. 
\end{align}
Hence,
\begin{align}
\label{EQ-40}
e^\phi=e^{c_1} (\mu_{0}+\eta  p)^{-1/\eta }
=\alpha \left(\frac{1-Q^2}{1-C-\beta S^2}\right)^{\frac{1}{\eta}}. 
\end{align}
where $\alpha$ is a freely specifiable  constant corresponding to the freedom of scaling the time coordinate in the line element.

The line element expressed in the variables \eqref{inv:qt} and the dynamical system \eqref{inv:qt_evol} are invariant under the discrete symmetry
\begin{equation}
\label{discrete00}
(Q,S)\rightarrow  (-Q, -S).
\end{equation}
with a simultaneous reversal of the radial direction $\lambda\rightarrow -\lambda$.
With respect to the phase space dynamics  this implies that for two points related by this symmetry, say $A^+$ and $A^-$, one has the opposite dynamical behavior to the other; that is, if the equilibrium point $A^+$ is an attractor for a choice of parameters, then $A^-$ is a sink for the same choice of parameters.
On the other hand, as both the system and the line element are invariant under \eqref{discrete00}, a physical solution is represented by two orbits in the phase space. We can, however, without loss of generality, focus upon orbits entering the phase space from the ``upper'' boundary set $Q=+1$ \cite{Nilsson:2000zf}.

\subsubsection{Equilibrium points in the finite region of the phase space}\label{SECT_3.2.3}

The equilibrium points of the system \eqref{inv:qt_evol} are described in the Appendix \ref{phys-intepretation}. 
In table \ref{Tab1} we summarize the existence and stability conditions of the equilibrium points of physical interest of the system \eqref{inv:qt_evol}. At the relevant equilibrium points we discuss some regularity conditions of the corresponding physical solutions (see Appendix \ref{regular_pf}): 

\begin{enumerate}
\item $P_1^{+}$ is a source for $\beta=1, 1\leq\eta<2$. Since the conditions \eqref{Condition36} are fulfilled this solution has  a regular center as $\lambda\rightarrow -\infty$. Because of $C=0$ it belongs to the plane-symmetric boundary set. Furthermore, it belongs to the scale invariant boundary $Q=+1$.

\item $P_1^{-}$ is a sink for $\beta=1, 1\leq\eta<2$. Because of $C=0$ it belongs to the plane-symmetric boundary set. Furthermore, it belongs to the scale invariant boundary $Q=-1$. Since the conditions \eqref{asymp_flatness} are fulfilled this solution is asymptotically flat as $\lambda \rightarrow +\infty$. 

\item $P_2^{+}$ is a source for $\beta=1, \eta\geq 1$.  Since the first inequality of \eqref{Condition36} is not fulfilled, this solution does not have a regular center. Because of $C=0$ it belongs to the plane-symmetric boundary set. Furthermore, it belongs to the scale invariant boundary $Q=+1$. 

\item $P_2^{-}$ is a sink for $\beta=1, \eta\geq 1$. Because of $C=0$ it belongs to the plane-symmetric boundary set. Furthermore, it belongs to the scale invariant boundary $Q=-1$. This solution is not asymptotically flat  since the conditions \eqref{asymp_flatness} are not fulfilled as $\lambda \rightarrow +\infty$. 

\item $P_2^{+}$ is a sink for $\beta=-\frac{\eta+2}{4}\leq 1, \eta\geq 1$. Because of $C=0$ it belongs to the plane-symmetric boundary set. Furthermore, it belongs to the scale invariant boundary $Q=+1$. This solution is not asymptotically flat  since the conditions \eqref{asymp_flatness} are not fulfilled as $\lambda \rightarrow +\infty$. 

\item $P_2^{-}$ is a source for $\beta=-\frac{\eta+2}{4}\leq 1, \eta\geq 1$. Since the first inequality of \eqref{Condition36} is not fulfilled, this solution does not have a regular center as $\lambda\rightarrow -\infty$. Because of $C=0$ it belongs to the plane-symmetric boundary set. Furthermore, it belongs to the scale invariant boundary $Q=-1$. 

\item $P_5^{+}$ is a sink  for $\eta \geq 1, \beta<0$. Because of $C=0$ it belongs to the plane-symmetric boundary set. Furthermore, it belongs to the scale invariant boundary $Q=+1$.  This solution is not asymptotically flat  since the conditions \eqref{asymp_flatness} are not fulfilled as $\lambda \rightarrow +\infty$.

\item $P_5^{-}$ is a source for $\eta \geq 1, \beta<0$. Since the conditions \eqref{Condition36} are not fulfilled, this solution does not have a regular center as $\lambda\rightarrow -\infty$. Because of $C=0$ it belongs to the plane-symmetric boundary set. Furthermore, it belongs to the scale invariant boundary $Q=-1$. 

\item $P_6^{+}$ is a source for $\eta \geq 1, 16 \beta \geq (\eta +2)^2$. It has a regular center as $\lambda\rightarrow -\infty$ only when $\beta=1$ (i.e., when this point coincides with $P_1^{+}$). Otherwise the conditions \eqref{Condition36} are not fulfilled, and the solution does not have a regular center as $\lambda\rightarrow -\infty$. Because of $C=0$ it belongs to the plane-symmetric boundary set. Furthermore, it belongs to the scale invariant boundary $Q=+1$.
\item $P_6^{-}$ is a sink for $\eta \geq 1, 16 \beta \geq (\eta +2)^2$.  Because of $C=0$ it belongs to the plane-symmetric boundary set. Furthermore, it belongs to the scale invariant boundary $Q=-1$. It is not asymptotically flat as $\lambda \rightarrow +\infty$ unless $\beta=1$ (i.e., when $P_6^{-}$ merge with $P_1^{-}$ ).

\item $P_7^{+}$ is a source for $\eta \geq 1, \beta>0$. The conditions \eqref{Condition36} are not fulfilled, and the solution does not have a regular center  as $\lambda \rightarrow -\infty$. Because of $C=0$ it belongs to the plane-symmetric boundary set. Furthermore, it belongs to the scale invariant boundary $Q=+1$.

\item $P_7^{-}$ is a sink for  $\eta \geq 1, \beta>0$. Because of $C=0$ it belongs to the plane-symmetric boundary set. Furthermore, it belongs to the scale invariant boundary $Q=-1$.

\item $P_8^{+}$ is a sink for $\beta<0$. It is not asymptotically flat as $\lambda\rightarrow +\infty$. 

\item $P_8^{-}$ is a source for $\beta<0$. It has a regular center as $\lambda \rightarrow -\infty$ if $\eta >1, \beta <0, \mu_{0}>0,  p_c\geq  \frac{\mu_0\left(6 \beta-\beta  \eta  -3\right)}{\beta \eta(\eta-1)}.$
   
\end{enumerate}

There are relevant equilibrium points which are saddle points: 
\begin{enumerate}
\item The equilibrium point $P_3^{+}$ represents the Minkowski spacetime in spherical symmetric form. 
The idea now is to find approximated solutions for the regular orbit near $P_3^{+}$ as $\lambda\rightarrow -\infty$. In the limit $\lambda\rightarrow -\infty$ the unstable manifold of $P_3^{+}$ provides the necessary mathematical structure for constructing this approximated solution.  
For this reason we introduce the coordinate transformation 
\begin{equation}
Q=\frac{3 \beta  v_1}{2}-\frac{1}{4} (\eta+2) v_2+1, \quad S=u_1+v_1, \quad C=1-v_2. 
\end{equation}

Applying the Invariant Manifold theorem we find that the local unstable manifold of $P_3^{+}$,

$\left\{(u_1, v_2, v_3): u_1=h(v_1, v_2), h(0,0)=0, \frac{\partial h}{\partial v_1}(0,0)=0, \frac{\partial h}{\partial v_2}(0,0)=0, v_1^2+v_2^2\leq \delta\right\}$, $ \delta >0$, 
can be approximated up to third order by the graph 
\begin{equation}
u_1=h(v_1,v_2)\approx \frac{\eta  v_1 v_2 (-3 \beta +\eta +2)}{30 \beta }. 
\end{equation}
Now, substituting the approximated solutions (found by solving just the linear part of the differential equations along the unstable eigendirections):
\begin{equation}
v_1=\epsilon_1 e^{2\lambda}, v_2=\epsilon_2 e^{2\lambda},
\end{equation}
where $\epsilon_1, \epsilon_2$ are small positive constants, and keeping only the linear terms in $\epsilon$ we find the approximated solutions
\begin{equation}
Q=1+\frac{3\beta}{2} \epsilon_1 e^{2\lambda}-\frac{1}{4} (\eta+2) \epsilon_2 e^{2\lambda}, \quad S=\epsilon_1 e^{2\lambda}, \quad C=1-\epsilon_2 e^{2\lambda}. 
\end{equation}
Replacing
\begin{equation}
\epsilon_1 = -\frac{2}{3} (\varepsilon_1-\varepsilon_2), \quad \epsilon_2= \frac{4}{2+\eta}((1-\beta) \varepsilon_1+\varepsilon_2),
\end{equation}
where $\varepsilon_1$ and $\varepsilon_2$ are still small constants (we assume they are positive),
we find the more familiar equations
\begin{equation}
Q=1-\varepsilon_1 e^{2\lambda}, \quad S=\frac{2}{3} (\varepsilon_2-\varepsilon_1) e^{2\lambda}, \quad C=1-\frac{4}{2+\eta}((1-\beta) \varepsilon_1+\varepsilon_2) e^{2\lambda}, 
\end{equation}
that reproduce  equations (27a- 27c) of \cite{Nilsson:2000zf} for $\beta=1$. We see that $\varepsilon=\frac{\varepsilon_1}{\varepsilon_2}$  parametrize a 1-parameter family of regular solutions with an equation of state parameter at the center:
\begin{align}
\label{grav_strength}
&\frac{p_c}{\mu_c}=\lim_{\lambda \rightarrow -\infty }\frac{p}{\mu}=\lim_{\lambda \rightarrow -\infty }\frac{\mu-\mu_0}{(\eta-1)\mu}=\lim_{\lambda \rightarrow -\infty }\frac{C-Q^2+\beta  S^2}{C (\eta -1)-\eta +Q^2+\beta  (\eta -1) S^2}\nonumber\\
& =\frac{2-\varepsilon (2 \beta +\eta )}{\varepsilon  (2 \beta  (\eta -1)-3 \eta
   )+2 (\eta -1)}.
\end{align}

We see that there exists solutions with a regular center but negative pressure, so that we have to impose the condition \footnote{
This condition is reduced in GR, to $\eta >1, 0<\varepsilon <\frac{2}{\eta +2}$, when $\beta=1$.}
\begin{equation}
\label{cond_P3+a}
\frac{2-\varepsilon (2 \beta +\eta )}{\varepsilon  (2 \beta  (\eta -1)-3 \eta
   )+2 (\eta -1)}>0,
\end{equation}
that is:
\begin{enumerate}
\item $\eta >1, \beta \leq -\frac{\eta }{2}, \varepsilon >0$, or 
\item $\eta >1,  -\frac{\eta }{2}<\beta \leq \frac{3 \eta }{2 \eta -2}, 
   0<\varepsilon <\frac{2}{2 \beta +\eta }$, or 
\item $\eta >1, \beta >\frac{3 \eta }{2 \eta -2}, 0<\varepsilon <\frac{2}{2
   \beta +\eta }$, or 
\item $\eta >1, \beta >\frac{3 \eta }{2 \eta -2}, \varepsilon >\frac{2 \eta
   -2}{2 \beta  \eta -2 \beta -3 \eta }$.
\end{enumerate}
\begin{landscape}
\begin{table}[ph]
\caption{Eigenvalues,  stability and characterization of the equilibrium points of the dynamical system defined by equations \eqref{inv:qt_evol}. We use the coordinates $\left(t, \lambda, \vartheta, \varphi\right)$. $\textbf{diag}\left(\ldots\right)$ denotes the diagonal $4\times 4$ matrix. We assume $\eta\geq1$.  We have used the notation $\epsilon=\pm 1$ and $c_1, c_2$ are integration constants.}{
\centering
\scalebox{1}{
\begin{tabular}{|c|c|c|c|}
\hline
Labels & $(Q, S, C)$, $g_{\mu \nu }$ & Existence &   Stability  \\\hline
$P_1^{\pm}$ & $(\pm 1, \pm 1, 0)$ & & \\
            & $\left\{\begin{array}{c}
\textbf{diag}\left(-\bar{N}_0^2 e^{\pm 2\lambda}
, \frac{e^{\pm 2\lambda}}{c_1^2}, {\bar{K}_0}^{-1}, {\bar{K}_0}^{-1}\sin ^{2}\vartheta \right),\; \beta=1
\\
\textbf{diag}\left(-\bar{N}_0^2 e^{\pm 2\lambda}, \frac{e^{\pm \eta \lambda}}{c_1^2}, {\bar{K}_0}^{-1}, {\bar{K}_0}^{-1}\sin ^{2}\vartheta \right),\;  \beta=\frac{\eta +2}{4}\end{array}\right.$ & $\begin{array}{c}
\beta=1\; \text{or}\\
\beta=\frac{\eta +2}{4}\leq 0.
\end{array}$ & $\begin{array}{c}
P_1^{+} \;\text{source for}\; \beta=1, 1\leq\eta<2\\
P_1^{-} \;\text{sink for}\; \beta=1, 1\leq\eta<2\\
P_1^{\pm}\; \text{saddle for}\; \beta=1, \eta>2\\
P_1^{\pm} \;\text{saddle for}\; \beta=\frac{\eta +2}{4}\leq 0 
\end{array}$    \\\hline
$P_2^{\pm}$ & $(\pm 1, \mp 1, 0)$ & & \\
& $\left\{\begin{array}{c}
\textbf{diag}\left(-\bar{N}_0^2 e^{\mp 2\lambda}, \frac{e^{\pm 6\lambda}}{c_1^2}, {\bar{K}_0}^{-1} e^{\pm 4\lambda}, {\bar{K}_0}^{-1} e^{\pm 4\lambda}\sin ^{2}\vartheta\right),\;\beta=1 \\
\textbf{diag}\left(-\bar{N}_0^2 e^{\mp 2\lambda}, \frac{e^{\mp \eta \lambda}}{c_1^2}, {\bar{K}_0}^{-1} e^{\pm 4\lambda}, {\bar{K}_0}^{-1} e^{\pm 4\lambda}\sin ^{2}\vartheta\right),\; \beta=-\frac{\eta+2}{4}
\end{array}\right.$ & $\begin{array}{c}
\beta=1\; \text{or}\\
4 \beta+\eta +2=0,  \beta \leq 1\end{array}$. & $\begin{array}{c}
P_2^{+} \; \text{source for}\;  \beta=1, \eta\geq 1\\
P_2^{-} \; \text{sink for} \; \beta=1, \eta\geq 1 \\
P_2^{+}  \; \text{sink for}\; \beta=-\frac{\eta+2}{4}\leq 1, \eta\geq 1\\
P_2^{-}  \; \text{source for}\; \beta=-\frac{\eta+2}{4}\leq 1, \eta\geq 1
\end{array}$ \\\hline
$P_3^{\pm}$ & $(\pm 1, 0, 1)$ && \\
 &  $
\textbf{diag}\left(
	-\bar{N}_0^{2}, \bar{K}_0^{-1} e^{\pm 2\lambda}, \bar{K}_0^{-1} e^{\pm 2\lambda},  \bar{K}_0^{-1} e^{\pm 2\lambda} \sin ^{2}\vartheta \right)$ & always& saddle \\
\hline 
$P_4^{\pm}$ &$\left(\pm 1, \pm\frac{2}{\eta +2},1-\frac{8 \beta}{(\eta +2)^2}\right)$ && \\
& $
\textbf{diag}\left(-\bar{N}_0^{2}e^{\pm \frac{4\lambda}{2+\eta}}, \frac{1}{\bar{K}_0}\left(1-\frac{8\beta}{(\eta +2)^2}\right)e^{\pm \frac{2\eta\lambda}{2+\eta}}, \frac{1}{\bar{K}_0} e^{\pm \frac{2\eta\lambda}{2+\eta}}, \frac{1}{\bar{K}_0} e^{\pm \frac{2\eta\lambda}{2+\eta}} \sin ^{2}\vartheta\right)$ &$0\leq \beta \leq \frac{1}{8} (\eta +2)^2$ & saddle
\\\hline
 $P_5^{\pm}$ & $\left(\pm 1,\pm\frac{\eta +2}{4 \beta},0\right)$ && \\
 & $
\textbf{diag}\left(-\bar{N}_0^2 e^{\pm\frac{(\eta+2)\lambda}{2\beta}}, \frac{(\eta +2)^2 e^{\pm\frac{\eta  (\eta +2) \lambda}{4 \beta}}}{16 c_1^2 \beta^2} , {\bar{K}_0}^{-1} e^{\pm 2 \left(1-\frac{\eta+2}{4\beta}\right)\lambda},  {\bar{K}_0}^{-1} e^{\pm 2 \left(1-\frac{\eta+2}{4\beta}\right)\lambda} \sin ^{2}\vartheta\right)$ & $\begin{array}{c}\eta\geq 1, \beta<0\;\text{or} \\
\eta >1, 16 \beta \geq (\eta +2)^2
\end{array}$ & $\begin{array}{c}P_5^{+}\; \text{sink  for}\;  \eta \geq 1, \beta<0\\
 P_5^{-} \;\text{source for}\;  \eta \geq 1, \beta<0 \\
\text{saddle otherwise}\end{array}$   
\\\hline
$P_6^{\pm}$ & $\left(\pm 1, \pm \frac{1}{\sqrt{\beta}}, 0\right)$ && \\
& $\textbf{diag}\left(-\bar{N}_0^2 e^{\pm \frac{2\lambda}{\sqrt{\beta}}},  \frac{ e^{\pm\frac{2 \left(2 \sqrt{\beta}+1\right) \lambda}{\sqrt{\beta}}}}{c_1^2 \beta}, {\bar{K}_0}^{-1} e^{\pm \frac{2 \left(\sqrt{\beta}-1\right)\lambda}{\sqrt{\beta}}}, {\bar{K}_0}^{-1} e^{\pm \frac{2 \left(\sqrt{\beta}-1\right)\lambda}{\sqrt{\beta}}} \sin ^{2}\vartheta\right)$ & $\eta\geq 1, \beta>0$ & $\begin{array}{c}
P_6^{+} \;\text{source for}\; \eta \geq 1, 
16 \beta \geq (\eta +2)^2 \\
P_6^{-} \;\text{sink for}\; \eta \geq 1, 
16 \beta \geq (\eta +2)^2\\
\text{saddle for}\;  \eta \geq 1, 0<\beta< \frac{1}{4} \; \\
\text{or}\; \eta \geq 1, \frac{1}{4}<\beta<\frac{1}{16} (\eta +2)^2\\  \text{non-hyperbolic otherwise}\end{array}$ 
 \\\hline
$P_7^{\pm}$ & $\left(\pm 1, \mp \frac{1}{\sqrt{\beta}}, 0\right)$ & & \\
& $
\textbf{diag}\left(
	-\bar{N}_0^2e^{\mp \frac{2\lambda}{\sqrt{\beta}}}, \frac{e^{\pm \frac{2 \left(1+2\sqrt{\beta}\right)\lambda}{\sqrt{\beta}}}}{c_1^2 \beta}, {\bar{K}_0}^{-1}  e^{\pm \frac{2 \left(1+\sqrt{\beta}\right)\lambda}{\sqrt{\beta}}}, {\bar{K}_0}^{-1}  e^{\pm \frac{2 \left(1+\sqrt{\beta}\right)\lambda}{\sqrt{\beta}}} \sin ^{2}\vartheta 
\right)$&$\eta\geq 1, \beta>0$ & $\begin{array}{c} P_7^{+} \; \text{source for}\; \eta \geq 1,
\beta>0\\
P_7^{-} \text{sink for}\; \eta \geq 1,
\beta>0
\end{array}$ \\
\hline
 $P_8^{\pm}$ &$\left(\pm\frac{1}{\sqrt{1-\beta}}, \pm\frac{1}{\sqrt{1-\beta}},\frac{2 \beta -1}{\beta -1}\right)$ & & \\
&  $\textbf{diag}\left(-\bar{N}_0^2 e^{\pm \frac{2\lambda}{\sqrt{1-\beta}}}, \frac{\eta}{\mu_0}\left(\frac{\beta }{\beta -1}\right), {\bar{K}_0}^{-1}, {\bar{K}_0}^{-1}\sin
^{2}\vartheta\right)$ & $\beta\leq 0$ &
$\begin{array}{c}
P_8^{+} \;\text{sink for}\; \beta<0\\
P_8^{-} \;\text{source for}\; \beta<0
\end{array}$\\\hline
$P_9^{\pm}$ & $\left(\pm 2 \sqrt{\beta}, \pm \frac{1}{\sqrt{\beta}}, 0\right)$ & &\\
& $\textbf{diag}\left(-\bar{N}_0^2e^{\pm \frac{2\lambda}{\sqrt{\beta}}}, \frac{\eta}{\mu_0}(1-4\beta), {\bar{K}_0}^{-1} e^{\pm \frac{2 \left(2\beta-1\right)\lambda}{\sqrt{\beta}}}, {\bar{K}_0}^{-1} e^{\pm \frac{2 \left(2\beta-1\right)\lambda}{\sqrt{\beta}}} \sin ^{2}\vartheta\right)$ &  $\eta \geq 1, 0<\beta \leq \frac{1}{4}$ & saddle 
\\\hline
\end{tabular}}\label{Tab1}}
\end{table}
\end{landscape}
For  $C-(Q-S)^2>0$, the first and second the Buchdahl conditions are satisfied at the solution as $\lambda\rightarrow -\infty$, if
\begin{align}
\label{cond_P3+b}
&\frac{(\beta -1) (\eta +2) \varepsilon }{\varepsilon  (6 \beta +\eta
   -4)+2 (\eta -1)}\geq 0,\\
   \label{cond_P3+c}
&   \frac{\eta  (\eta +2) \varepsilon  (\mu_{0}+(\eta -1) p_c)}{\mu_{0} (\varepsilon  (6 \beta +\eta
   -4)+2 (\eta -1))}\leq 1.
\end{align} 
Additionally, taking the limit $\lambda\rightarrow -\infty$ we have
\begin{equation}
\frac{1}{9} \left(7 C-3 Q^2+3 \beta  S^2\right)+2 \sqrt{C \left(C+3 Q^2-3
   \beta  S^2\right)}-C+(Q-S)^2\rightarrow \frac{40}{9}>0,
\end{equation}
such that the third Buchdahl condition is also satisfied.
Thus, combining the conditions \eqref{cond_P3+a}, \eqref{cond_P3+b}, and \eqref{cond_P3+c}, we have the conditions for the existence of regular solution at the center associated to $P_3^{+}$.

The quotient, $\frac{p_c}{\mu_c}$ in \eqref{grav_strength}  is a gravitational strength parameter. In GR where the parameter $\beta=1$, the maximal value of the gravitational strength, $\frac{1}{\eta+1}$, is obtained when $\varepsilon_1=0$, which corresponds to the subset $Q=1$. However, in the Einstein-aether theory the parameter $\beta$ is a freely specifiable parameter, and for $\eta >1, \beta >\frac{3 \eta }{2 \eta -2}, \frac{\varepsilon_1}{\varepsilon_2} >\frac{2 \eta
   -2}{2 \beta  \eta -2 \beta -3 \eta }$, the maximal strength is not $\frac{p_c}{\mu_c}=\frac{1}{\eta+1}$ anymore as it is in GR.  

\item The equilibrium point $P_4^{+}$ generalizes the so called Tolman point (which corresponds to $\beta=1$), which now is promoted to a 1-parameter solution. This solution exists for $0\leq \beta \leq \frac{1}{8} (\eta +2)^2$. The eigenvalues are 
\begin{small}
\begin{equation}
\lambda_1=\frac{2 \eta }{\eta +2},   \lambda_2=-\frac{\eta +2+\sqrt{64 \beta -7 (\eta +2)^2}}{2 (\eta +2)},   \lambda_3= -\frac{\eta+2-\sqrt{64 \beta -7 (\eta +2)^2}}{2 (\eta +2)}. 
\end{equation}
\end{small}
Notice that $\lambda_{2}+\lambda _{3}+1=0$.

The eigenvalue $\lambda_1$ is always real and positive. 

 The eigenvalues  $\lambda_2, \lambda_3$ are both real and negative for
 \begin{enumerate}
 \item $\frac{63}{64}<\beta \leq \frac{9}{8}, 1<\eta \leq -2+\frac{8 \sqrt{\beta }}{\sqrt{7}}$, or 
 \item $\beta >\frac{9}{8}, 2 \sqrt{2} \sqrt{\beta }-2<-2+\eta \leq \frac{8 \sqrt{\beta }}{\sqrt{7}}$.
 \end{enumerate}
The eigenvalues $\lambda_2, \lambda_3$ are complex conjugated with negative real part for
 \begin{enumerate}
 \item $0<\beta \leq \frac{63}{64}, \eta >1$, or 
 \item $\beta >\frac{63}{64}, \eta >-2 +\frac{8 \sqrt{\beta }}{\sqrt{7}}$.
  \end{enumerate}

Following the same method as for the analysis of $P_3^{+}$ we can explore approximated solutions related to $P_4^{+}$ by constructing the unstable manifold of this equilibrium point. 

{\bf{Case 1:}}\\
When $\lambda_{2}$, $\lambda_{3}$ are both reals and negative, that is whenever $\frac{63}{64}<\beta \leq \frac{9}{8}, 1<\eta \leq -2+\frac{8 \sqrt{\beta }}{\sqrt{7}}$, or 
 $\beta >\frac{9}{8}, 2 \sqrt{2} \sqrt{\beta }-2<-2+\eta \leq \frac{8 \sqrt{\beta }}{\sqrt{7}}$, we can define the real quantities 
\begin{small}
\begin{subequations}
\begin{align}
& u=\frac{\lambda_{2} (\lambda_{2}+1) (1-Q) \left((\eta +2)^2 \lambda_{2}^2+(\eta +2)^2 \lambda_{2}+2 \eta  (\eta +6)\right)}{(\eta  (\lambda_{2}-2)+2 \lambda_{2}) ((\eta +2)
   \lambda_{2}+3 \eta +2)},\\
  & v_1=  \frac{\lambda_{2}(\lambda_{2}+2) \left(2 C+\lambda_{2}^2+\lambda_{2}\right)}{4 \lambda_{2}+2}\nonumber\\
   & +\frac{\lambda_{2}(\lambda_{2}+1) (Q-1) \left(\eta 
   \left(\lambda_{2}^2+\lambda_{2}+2\right)+2 (\lambda_{2}-2) (\lambda_{2}+1)\right)}{(2 \lambda_{2}+1) ((\eta +2) \lambda_{2}+3 \eta +2)}\nonumber\\
   & +\frac{\lambda_{2}(\lambda_{2}+1)
   \left(\lambda_{2}^2+\lambda_{2}+2\right) ((\eta +2) S-2)}{4 \lambda_{2}+2},\\
 &  v_2=-\frac{\left(\lambda_{2}^2-1\right) \left(2 C+\lambda_{2}^2+\lambda_{2}\right)}{4 \lambda_{2}+2}\nonumber\\
   & +\frac{\lambda_{2} (\lambda_{2}+1) (Q-1) \left(\eta  \left(\lambda_{2}^2+\lambda_{2}+2\right)+2 \lambda_{2} (\lambda_{2}+3)\right)}{(2 \lambda_{2}+1) (\eta 
   (\lambda_{2}-2)+2 \lambda_{2})} \nonumber \\
   & -\frac{\lambda_{2} (\lambda_{2}+1) \left(\lambda_{2}^2+\lambda_{2}+2\right) ((\eta +2) S-2)}{4 \lambda_{2}+2}
\end{align}
\end{subequations}
\end{small}
where we have used the relation 
\begin{equation}
\beta = \frac{1}{16} (\eta +2)^2 \left(\lambda_{2}^2+\lambda_{2}+2\right).
\end{equation}

Now, applying the Invariant Manifold theorem we find that the local unstable manifold of $P_4^{+}$ is 
\begin{equation*}
\left\{(u, v_2, v_3): v_1=h_1(u),v_2=h_2(u), h_1(0)=0, h_1'(u)=0, h_2(0)=0, h_2'(u)=0, u\leq \delta\right\}, 
\end{equation*}
with $\delta >0$.
Calculating the unstable manifold up to second order in powers of $u$, neglecting the higher order terms and substituting back to the equations of $Q, S, C$, in terms of u, through $v_1=h_1(u), v_2=h_2(u)$, we obtain that any solution near the unstable manifold of $P_4^{+}$, satisfies 
\begin{small}
\begin{subequations}
\begin{align}
&Q=1-\frac{(\eta +2)^2 (-4 \beta +\eta  (2 \eta +3)+2)}{2 \left((4 \beta +\eta -2) \left((\eta +2)^2-8 \beta \right)\right)} \varepsilon e^{\frac{2 \eta  \lambda }{\eta +2}} ,\\
&S=\frac{2}{\eta +2}-\frac{ (\eta +2) \left(-8 \beta ^2-2 (\eta -2) \beta +\eta  (\eta +2)^2\right)}{2 \left(\beta  (4 \beta +\eta -2) \left((\eta +2)^2-8
   \beta \right)\right)} \varepsilon e^{\frac{2 \eta  \lambda }{\eta +2}} \nonumber \\
	& +\frac{\eta  (\eta +2)^4 (-4 \beta +\eta +2) }{8 \beta ^2 (4 \beta +\eta -2)^2 \left((\eta +2)^2-8 \beta \right)^2 (-8 \beta +\eta  (11 \eta +8)+4)} \times \\
	& \Big[-32 \beta ^3+4 \beta ^2 (\eta  (7 \eta -2)+8)
	\nonumber \\
	& -\beta  (\eta  (\eta  (\eta  (2 \eta +33)+102)+28)+8)+\eta  (\eta +2)^2 (\eta  (\eta +12)+4)\Big]\varepsilon ^2 e^{\frac{4 \eta  \lambda }{\eta +2}},\end{align}
	\begin{align}
   & C=1-\frac{8 \beta }{(\eta +2)^2}+\varepsilon e^{\frac{2 \eta}{\eta +2}\lambda} -\Big[128 \beta ^4-48 \beta ^3 (\eta  (5 \eta +4)+4) \nonumber\\
	& +8 \beta ^2 \left(\eta  \left(\eta  \left(7 \eta ^2+\eta +19\right)+24\right)+12\right) \nonumber \\
	& -\beta  (\eta  (\eta  (\eta  (\eta  (42 \eta +251)+300)+80)+48)+16)+2 \eta ^2 (\eta +2)^3  (9 \eta +2)\Big] \times \nonumber\\
	& \frac{(\eta +2)^2}{4 \beta  (4 \beta +\eta -2)^2 \left((\eta +2)^2-8 \beta \right) (-8 \beta +\eta  (11 \eta +8)+4)}\varepsilon ^2 e^{\frac{4 \eta  \lambda }{\eta +2}},
\end{align}
\end{subequations} 
\end{small}
where we have used the original parameters $\beta$ and $\eta$ and we have substituted the approximated solution $u_1= \varepsilon e^{\frac{2 \eta}{\eta +2}\lambda}$, that is obtained by integrating the linearized equation along the unstable direction. This expansion is accurate as long as $\lambda\rightarrow -\infty$. 

Using this solution, we find 
\begin{small}
\begin{align}
& \frac{p_c}{\mu_c}=\lim_{\lambda \rightarrow -\infty }\frac{C-Q^2+\beta  S^2}{C (\eta -1)-\eta +Q^2+\beta  (\eta -1) S^2}\nonumber\\
   & = \lim_{\lambda \rightarrow -\infty } \frac{1}{\eta-1}-\frac{\varepsilon  \left(\eta  (\eta +2)^4 (-4 \beta +\eta  (2 \eta +3)+2) e^{\frac{2 \eta  \lambda }{\eta +2}}\right)}{4 \left(\beta  (\eta -1)^2 (4 \beta +\eta -2) \left((\eta +2)^2-8 \beta  \right)\right)}+O\left(\varepsilon ^2\right)\nonumber\\
   &    = \frac{1}{\eta-1} >0.
\end{align}
\end{small}
Furthermore, the Buchdahl conditions can be expressed as 
\begin{align}
& 1\geq \frac{-(\eta -1) \left(C+\beta  S^2\right)+\eta -Q^2}{3 \left(C-(Q-S)^2\right)},\\
& 1\leq \frac{\eta  \left(1-Q^2\right) (\mu_{0}+(\eta -1) p_c)}{3 \mu_{0} \left(C-(Q-S)^2\right)},\\
& \frac{1}{9} \left(7 C-3 Q^2+3 \beta  S^2\right)+2 \sqrt{C \left(C+3 Q^2-3
   \beta  S^2\right)}-C+(Q-S)^2\geq 0.
\end{align}
As $\lambda\rightarrow -\infty$, applying the above conditions we have that the second one is satisfied; and the first and third imply
\begin{align}
& \frac{\beta -\beta  \eta }{6 \beta -3 (\eta +1)}\geq 1,\\
& \frac{4 (7 \beta +9)}{9 (\eta +2)^2}+2 \sqrt{\left(4-\frac{20 \beta }{(\eta +2)^2}\right) \left(1-\frac{8 \beta }{(\eta +2)^2}\right)}-\frac{4}{\eta
   +2}+\frac{4}{9}\geq 0.
\end{align}
These conditions are not satisfied for $\beta=1$ (that is, for GR). But in AE-theory $\beta$ is a free  parameter, such that the above inequalities can be satisfied for $\eta >1, \frac{3 \eta +3}{\eta +5}\leq \beta <\frac{\eta +1}{2}, 64 \beta -7 (\eta +2)^2\geq 0$.

{\bf{Case 2:}}\\
For the choice  $0<\beta \leq \frac{63}{64}, \eta >1$, or 
 $\beta >\frac{63}{64}, \eta >-2 +\frac{8 \sqrt{\beta }}{\sqrt{7}}$, the eigenvalues $\lambda_2, \lambda_3$ are complex conjugates with negative real part. 
Indeed, 
\begin{equation}
\Re(\lambda_2)=\Re(\lambda_3)=-\frac{1}{2}, -\Im(\lambda_2)=\Im(\lambda_3)=\frac{\sqrt{7 (\eta +2)^2-64 \beta }}{2 (\eta +2)}.
\end{equation}
For the analysis we introduce the parametrization
\begin{small}
\begin{subequations}
\begin{align}
& Q= 1-\frac{(\eta +2)^2 u (-4 \beta +\eta  (2 \eta +3)+2)}{2 (4 \beta +\eta -2) \left(8 \beta -(\eta +2)^2\right)},\\
& S=\frac{2}{\eta +2}+\frac{(\eta +2) u \left(-8 \beta ^2-2 \beta  (\eta
   -2)+\eta  (\eta +2)^2\right)}{2 \beta  (4 \beta +\eta -2) \left(8 \beta -(\eta +2)^2\right)} \nonumber \\
	& +\frac{(\eta +2) v_1 \left(5 (\eta +2)^2-32 \beta \right)}{16 \beta  \left(8 \beta -(\eta +2)^2\right)} -\frac{(\eta +2)^2
   v_2 \sqrt{7 (\eta +2)^2-64 \beta }}{16 \beta  \left(8 \beta -(\eta +2)^2\right)},\\
	& C= 1-\frac{8 \beta }{(\eta +2)^2}+u+2 v_1, 
\end{align}
\end{subequations}
\end{small}
where $u, v_1, v_2$ are reals.

Calculating the unstable manifold up to second order in powers of $u$, neglecting the higher order terms and substituting back to the equations of $Q, S, C$, in terms of $u$, through $v_1=h_1(u), v_2=h_2(u)$, we obtain that any solution near the unstable manifold of $P_4^{+}$, satisfies 
\begin{small}
\begin{subequations}
\begin{align}
& Q= 1-\frac{(\eta +2)^2  (-4 \beta +\eta  (2 \eta +3)+2) }{2 (4 \beta +\eta -2) \left(8 \beta -(\eta +2)^2\right)}\varepsilon e^{\frac{2 \eta  \lambda }{\eta +2}},\\
& S=\frac{2}{\eta +2}+\frac{(\eta +2) \varepsilon  \left(-8 \beta ^2-2 \beta  (\eta
   -2)+\eta  (\eta +2)^2\right) e^{\frac{2 \eta  \lambda }{\eta +2}}}{2 \beta  (4 \beta +\eta -2) \left(8 \beta -(\eta +2)^2\right)} \nonumber \\
	& -\frac{(\eta +2) }{9216 \beta ^2 (7 \eta +2) \left((\eta +2)^2-8 \beta \right)^2} \Big[2048 \beta ^3 (347 \eta +612)\nonumber \\
	& -32 \beta ^2 (\eta  (\eta  (8025 \eta +18818)+7740)-840) \nonumber \\
	& -4 \beta  (\eta  (\eta  (15457 \eta +67932)+96052)+49440) (\eta +2)^2 \nonumber \\
	& +9 (\eta  (\eta  (826 \eta +3043)+3628)+1612) (\eta +2)^4\Big]\varepsilon ^2 e^{\frac{4 \eta  \lambda }{\eta +2}},\\
& C=1-\frac{8 \beta }{(\eta +2)^2}+\varepsilon  e^{\frac{2 \eta  \lambda }{\eta +2}} \nonumber \\
	& -\frac{\left(-32 \beta ^2 (6 \eta +53)+4 \beta  (\eta  (\eta  (6 \eta +77)-126)-184)+9 (\eta +2)^2 (5 \eta +14)\right) }{72 \beta  (7 \eta +2) \left((\eta +2)^2-8 \beta \right)}\varepsilon ^2 e^{\frac{4 \eta  \lambda
   }{\eta +2}},
	\end{align}
	\end{subequations}
\end{small}

where we have substituted the approximated solution $u_1= \varepsilon e^{\frac{2 \eta}{\eta +2}\lambda}$, that is obtained by integrating the linearized equation along the unstable direction. This expansion is accurate as long as $\lambda\rightarrow -\infty$. At the stable manifold the orbits spiral in and tends asymptotically to the origin with modes $\cos(\frac{\sqrt{7 (\eta +2)^2-64 \beta }}{2 (\eta +2)} \lambda ) e^{-\frac{\lambda}{2}}$, $\sin(\frac{\sqrt{7 (\eta +2)^2-64 \beta }}{2 (\eta +2)} \lambda ) e^{-\frac{\lambda}{2}}$. 

We have the estimates
\begin{small}
\begin{align*}
& \frac{p_c}{\mu_c}=\lim_{\lambda \rightarrow -\infty }\frac{C-Q^2+\beta  S^2}{C (\eta -1)-\eta +Q^2+\beta  (\eta -1) S^2}\nonumber\\
   & = \lim_{\lambda \rightarrow -\infty } \frac{1}{\eta -1}+\frac{\eta  (\eta +2)^4 \varepsilon  \left(-4 \beta +2 \eta ^2+3 \eta +2\right) e^{\frac{2 \eta  \lambda }{\eta +2}}}{4 \beta  (\eta -1)^2 (4 \beta +\eta -2) \left(-8 \beta +\eta ^2+4 \eta
   +4\right)}+O\left(\varepsilon ^2\right)
\nonumber\\
   &    = \frac{1}{\eta - 1} >0.
\end{align*}
\end{small}
 Furthermore, for  $C-(Q-S)^2>0$, the Buchdahl conditions reduce to 
\begin{small}
\begin{equation*}
\frac{2 (\beta -1) \beta }{-2 \beta +\eta +1}+\beta \leq 3, 2 \beta \mu_0\geq (\eta +1) \mu_0,
\end{equation*} 
\begin{equation*} \frac{9 \sqrt{40 \beta ^2-13 \beta  (\eta +2)^2+(\eta +2)^4}+7 \beta +\eta ^2-5 \eta -5}{2 \beta -\eta
   -1}\leq 0,
	\end{equation*}
	\end{small}
as $\lambda\rightarrow -\infty$, respectively. 
	That is, when $1<\eta \leq 1.04725, \beta <\frac{1}{64} \left(7 \eta ^2+28 \eta +28\right), \mu_0\leq 0$ or $\eta >1.04725, \beta \leq \frac{3 \eta +3}{\eta +5}, \mu_0\leq 0$. We are assuming $\mu_0\geq 0$, therefore the conditions are fulfilled if $\mu_0=0$.

 \end{enumerate}

\begin{figure*}[t!]
	\centering
	\subfigure[\label{fig:Fig1a}]{\includegraphics[width=0.3\textwidth]{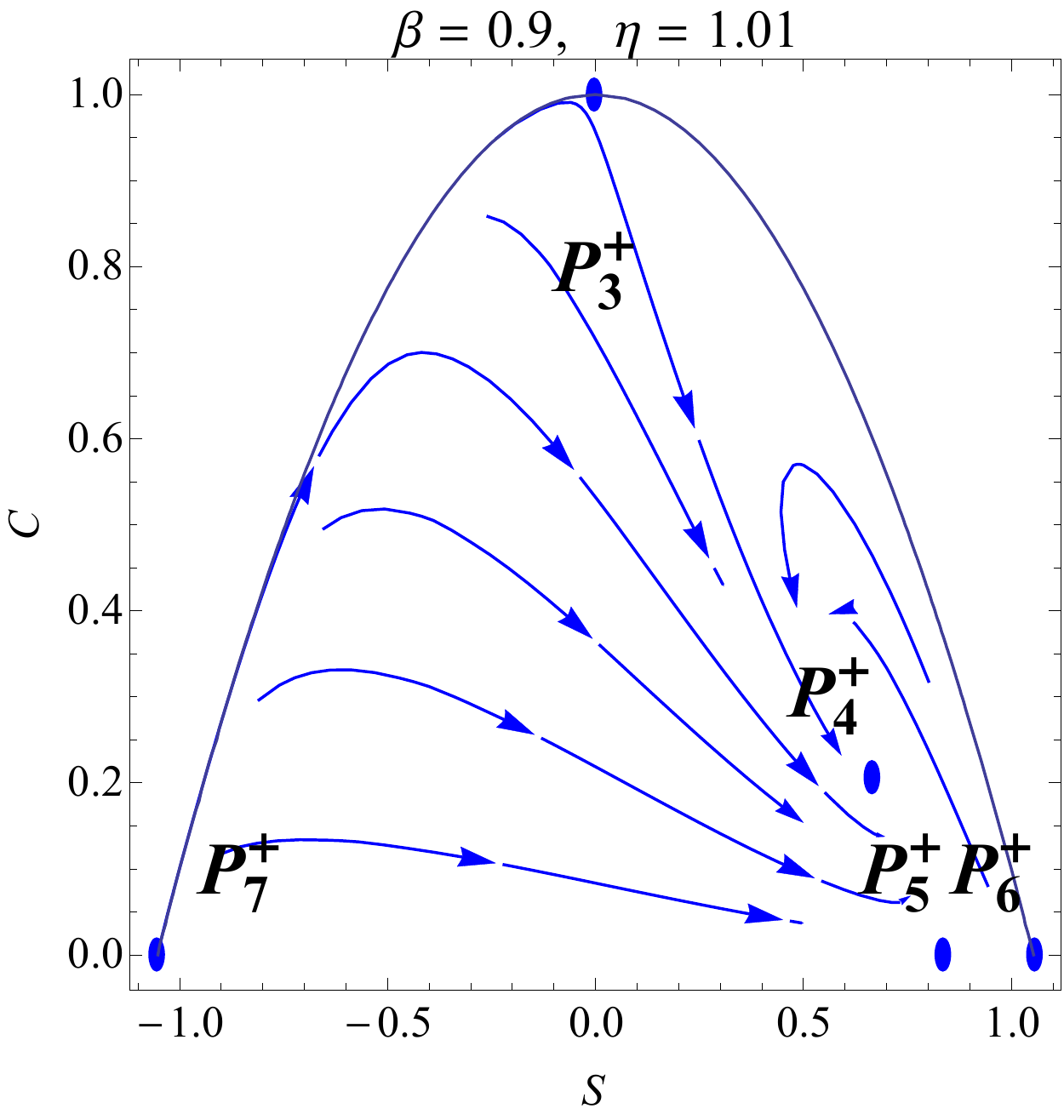}} \hspace{1cm}
	\subfigure[\label{fig:Fig1b}]{\includegraphics[width=0.3\textwidth]{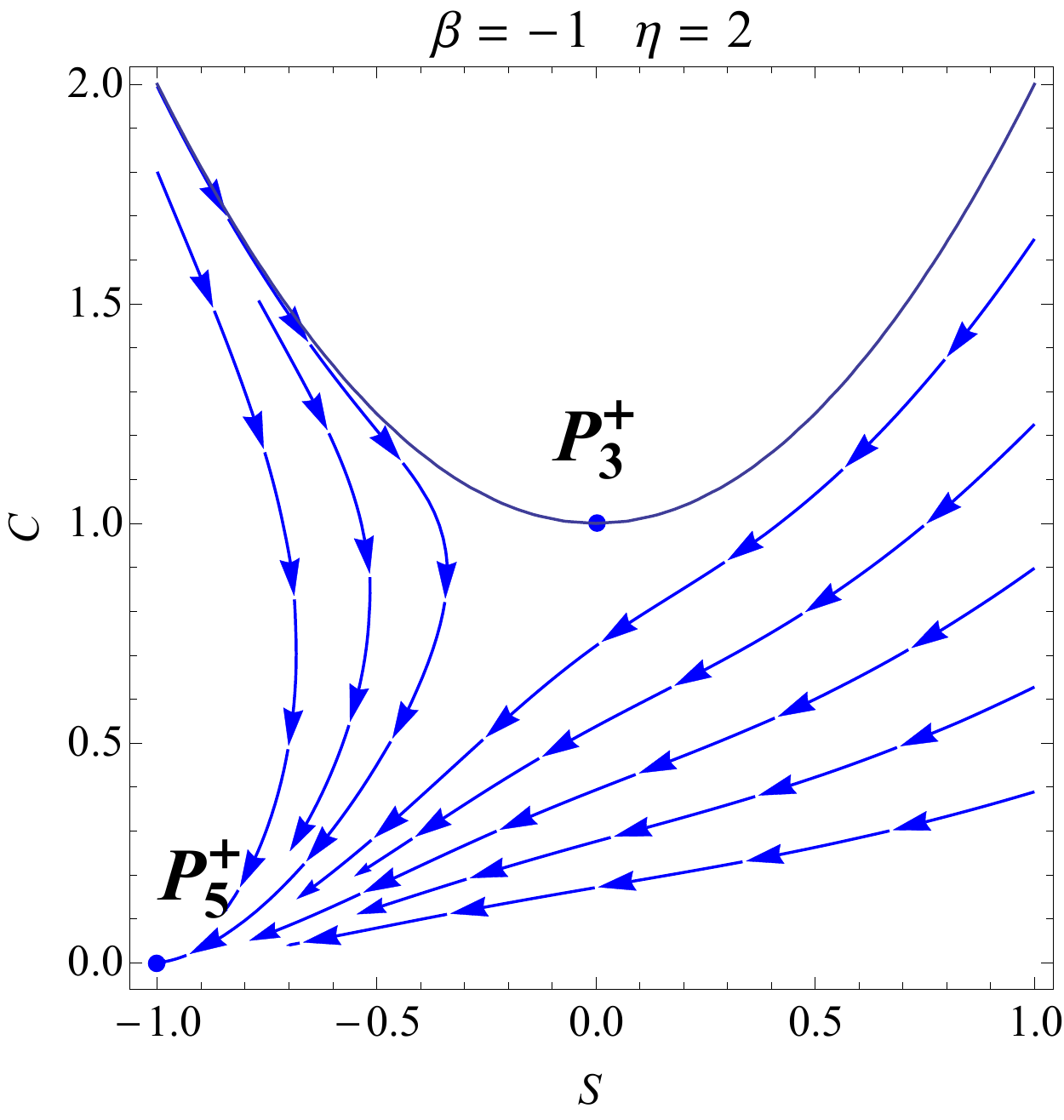}}
	\\
	\subfigure[\label{fig:Fig1c}]{\includegraphics[width=0.3\textwidth]{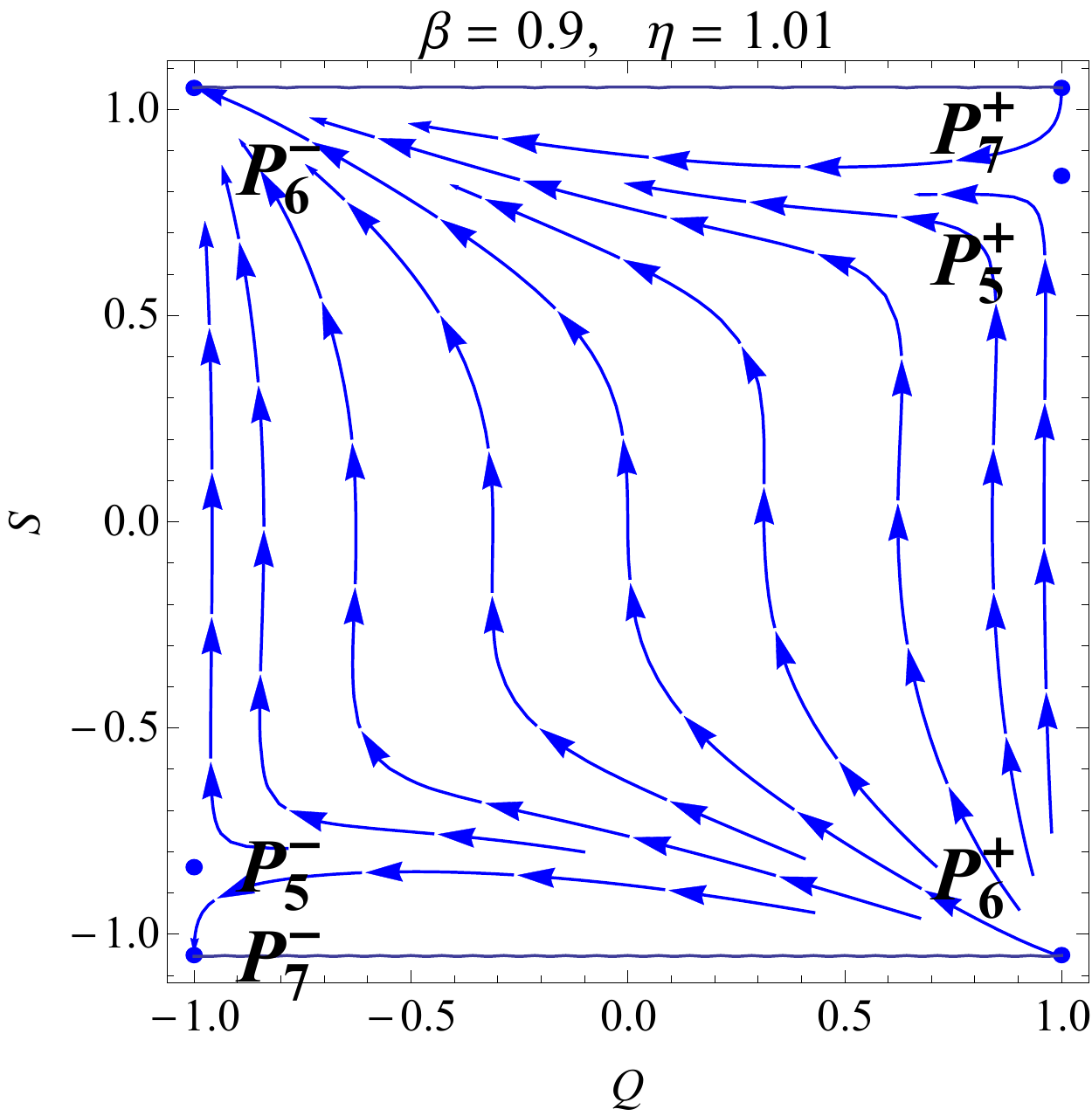}} \hspace{1cm}
	\subfigure[\label{fig:Fig1d}]{\includegraphics[width=0.3\textwidth]{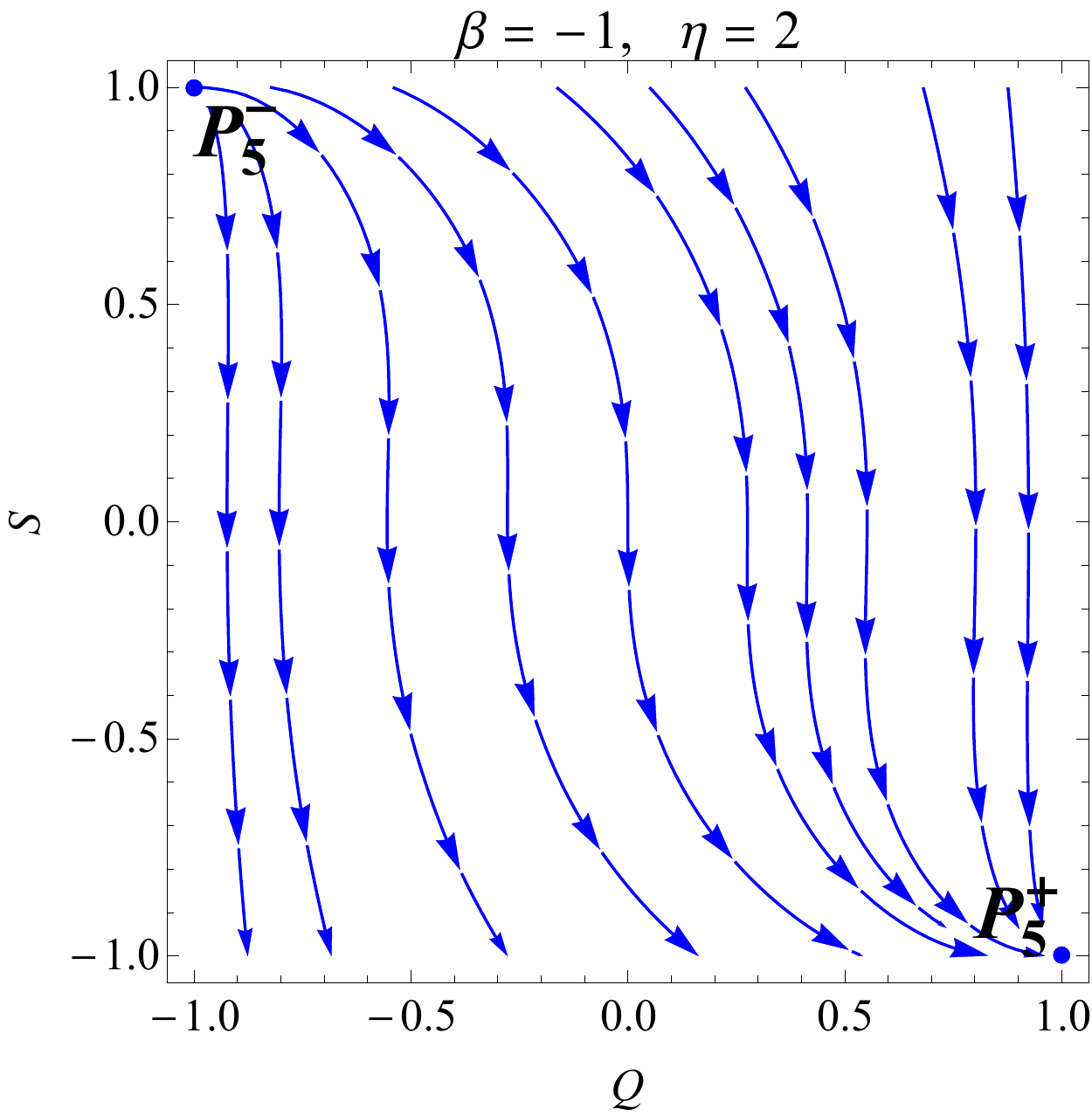}}
	\\
	\subfigure[\label{fig:Fig1e}]{\includegraphics[width=0.3\textwidth]{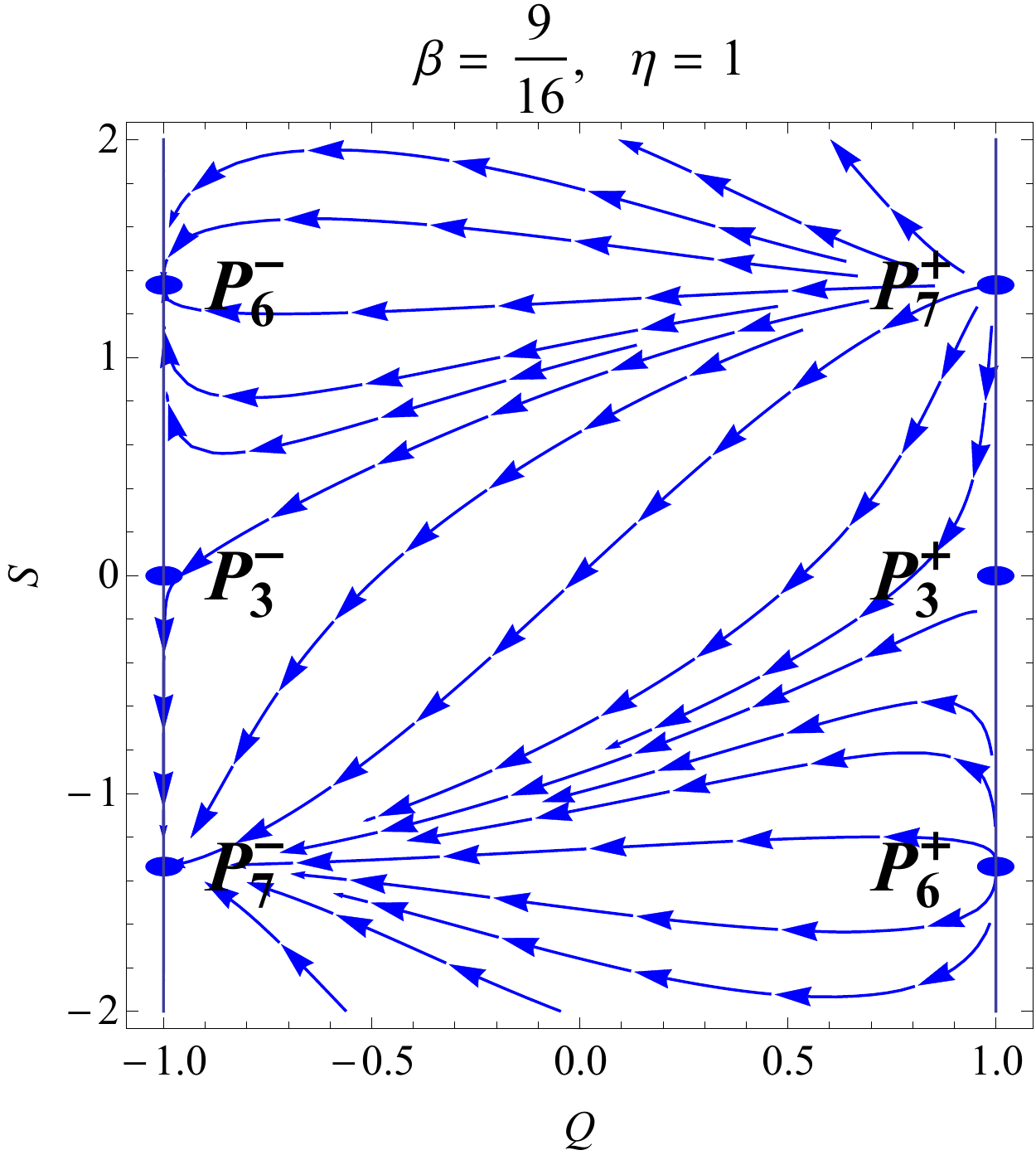}} \hspace{1cm}
	\subfigure[\label{fig:Fig1f}]{\includegraphics[width=0.3\textwidth]{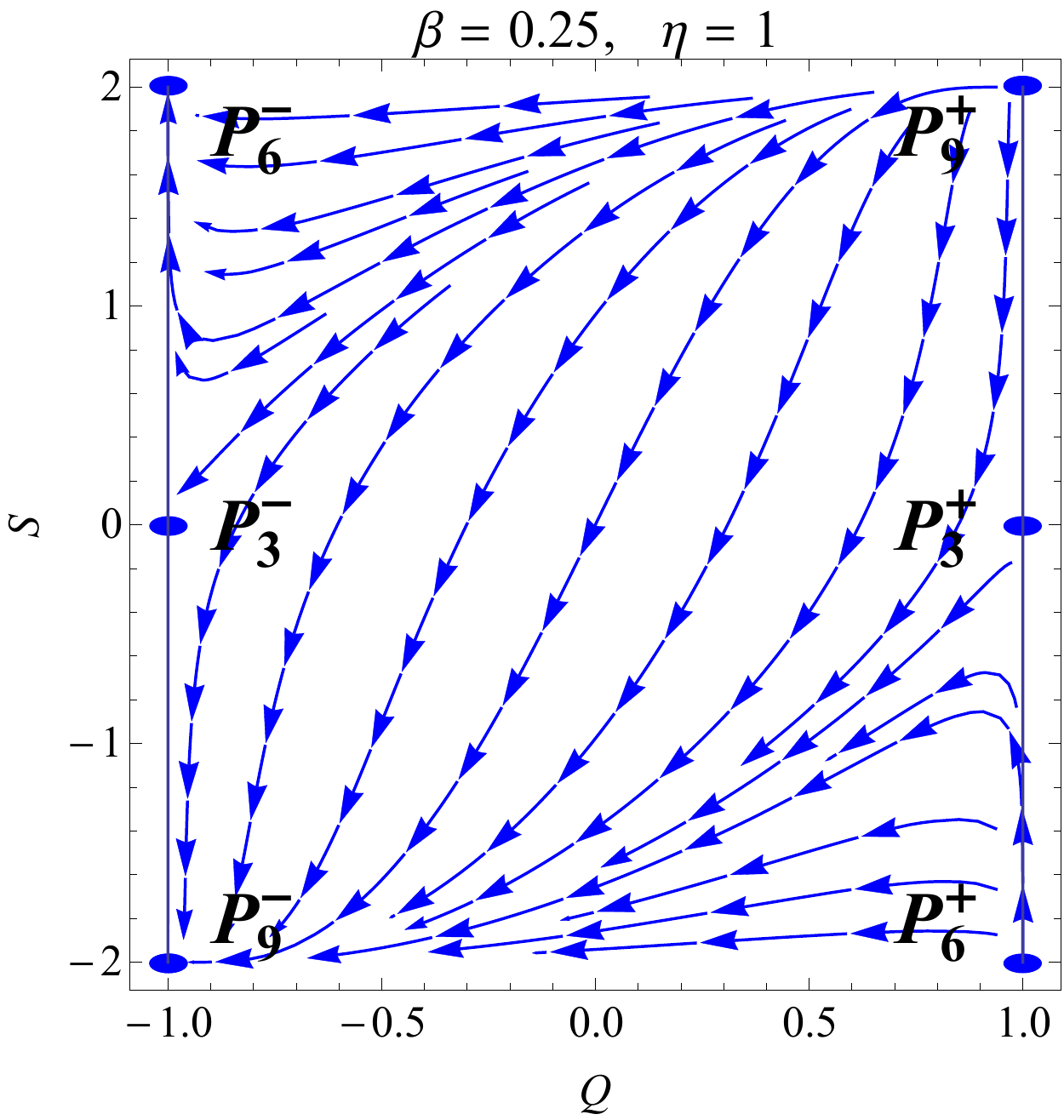}}
	\\
	\subfigure[\label{fig:Fig1g}]{\includegraphics[width=0.3\textwidth]{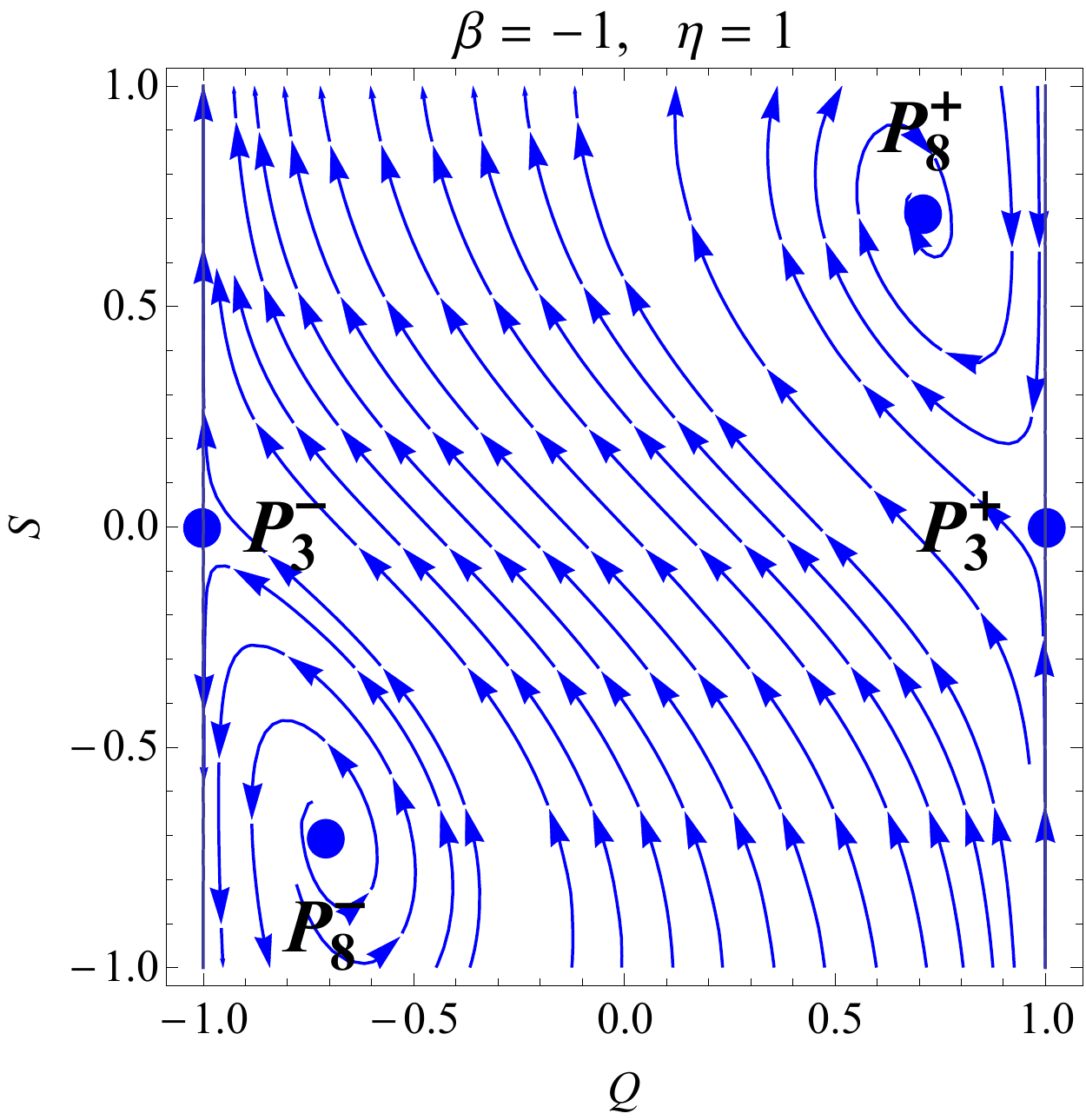}} \hspace{1cm}
	\subfigure[\label{fig:Fig1h}]{\includegraphics[width=0.3\textwidth]{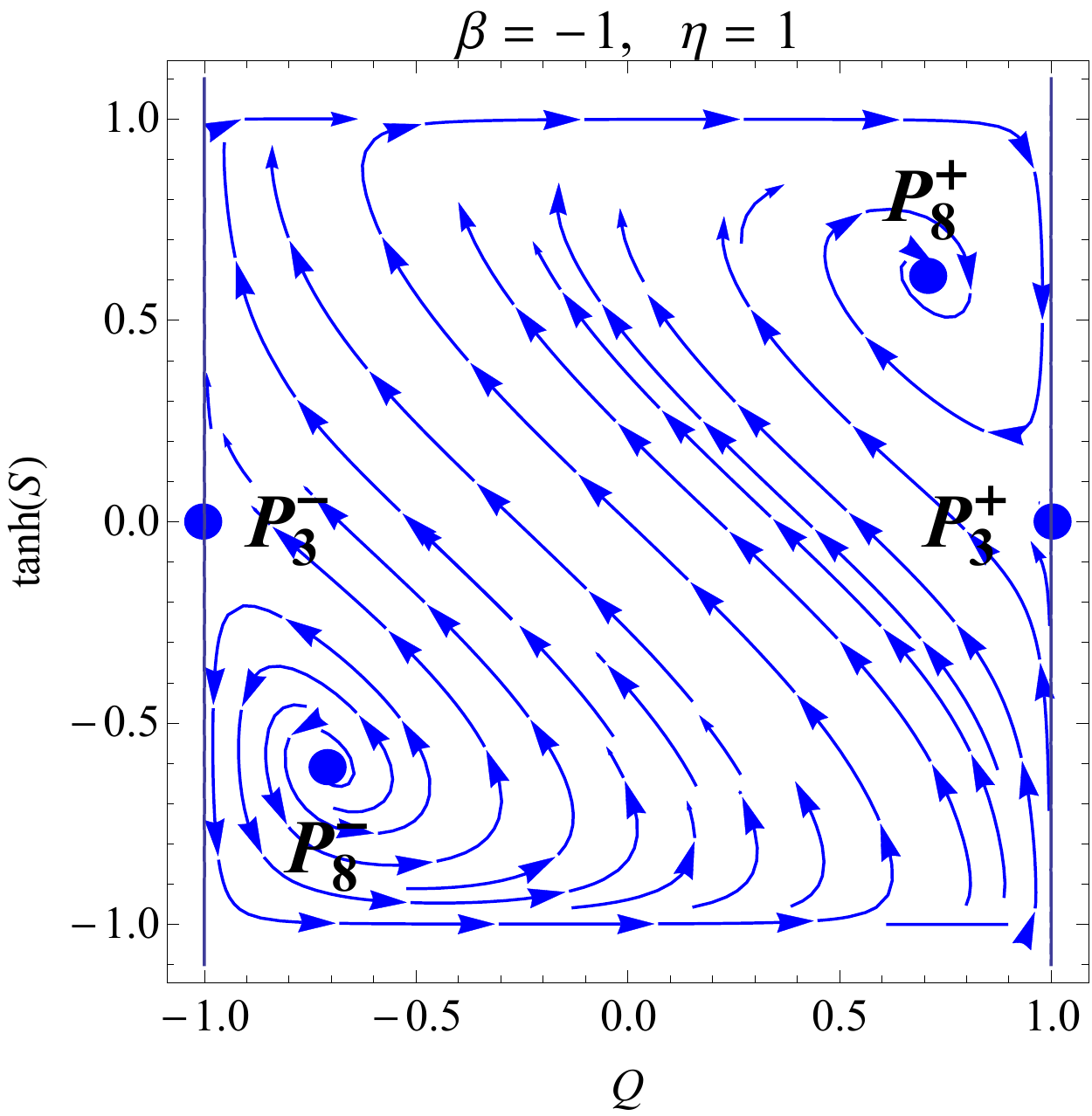}}
		\caption{\label{Fig1} {\footnotesize{Streamlines of the system \eqref{inv:qt_evol} for different choices of the parameters $\beta, \eta$.}}}
\end{figure*}


Fig. \ref{Fig1} shows the flow of the system \eqref{inv:qt_evol} for different choices of the parameters $\beta, \eta$. Figs. \ref{fig:Fig1a}, \ref{fig:Fig1b} represents the scale-invariant ($\mu_0=0$) boundary $Q=+1$. 
In Fig.  \ref{fig:Fig1a} the local attractor is $P_4^{+}$ (but it is a saddle point for the 3D dynamical system). At Fig. \ref{fig:Fig1b} the attractor is $P_5^{+}$. Figs. \ref{fig:Fig1c}, \ref{fig:Fig1d},  show the behavior on the plane-symmetric boundary $C=0$. In \ref{fig:Fig1c} the stable (respectively, unstable) points are $P_7^{-}$ and $P_6^{-}$ (respectively, $P_7^{+}$ and $P_6^{+}$).  The $P_5^{\pm}$ are saddles. In Fig. \ref{fig:Fig1d} the attractor is $P_5^{+}$.  Figs. \ref{fig:Fig1e}, \ref{fig:Fig1f}, \ref{fig:Fig1g}, \ref{fig:Fig1h}, show the dynamics on the invariant set $C=1-\beta S^2$ corresponding to fluids satisfying $\mu+p=0$. In Fig. \ref{fig:Fig1f} $P_3^{\pm}$ are saddles. The stable (respectively, unstable) points in the physical regions are $P_7^{-}$ and $P_6^{-}$ (respectively, $P_7^{+}$ and $P_6^{+}$). 
In Fig. \ref{fig:Fig1f} is presented the dynamics for $\beta=\frac{1}{4}$ and $\eta=1$. For these values $P_3^{\pm}$ are saddles.  $P_7^{-}$ (respectively $P_7^{+}$) is a stable (respectively, unstable) node on the physical region. At the bifurcation value $\beta=\frac{1}{4}$, $P_6^{\pm}$ merges with $P_9^{\pm}$ and becomes saddles. 
In the figure \ref{fig:Fig1g} show the dynamics on the invariant set $C=1-\beta S^2$ for $\beta=-1,\eta=1$. The sink is $P_8^{+}$ and the source is $P_8^{-}$. $P_3^{\pm}$ are saddles.  In figure \ref{fig:Fig1h} it is shown that the points at infinity $(Q=\pm1, S=\mp \infty)$ are saddle. Thus, $P_8^{+}$ is a global attractor for this choice of parameters.

\subsubsection{Compactification procedure}

The variables $C, S$ satisfy $C+\beta S^2\leq 1$ and $Q\in [-1,1]$. Therefore, in case of $\beta>0$ the phase space is compact. However, $S, C$ can be infinite values for $\beta<0$. 
Assuming $\beta<0$, we have $C \leq 1+|\beta| S^2$, such that we can use the dominant quatity, $ 1+|\beta| S^2$, to normalize.    
By introducing the compact variables, where we have assumed $ 1+|\beta| S^2\neq 0$ 
\begin{equation}
u=\frac{\sqrt{|\beta|}S}{\sqrt{1+|\beta| S^2}}, \quad v=\frac{C}{1+|\beta| S^2},
\end{equation}
with $$(Q,u,v)\in [-1,1]\times[-1,1]\times [0,1],$$
and the radial derivative 
\begin{equation}
\frac{d f}{d \xi}=\frac{|\beta|}{\sqrt{1+|\beta| S^2}}\frac{d f}{d\lambda},
\end{equation}
we obtain the dynamical system 
\begin{subequations}
\label{infinity_static}
\begin{align}
&\frac{d Q}{d \xi}=\left(Q^2-1\right) \left[\beta  \left(2 u^2-v\right)-Q  u \sqrt{1-u^2} \sqrt{\left| \beta \right| }\right],\\
&\frac{d u}{d \xi}=\left(u^2-1\right) \left[\beta  Q u (v-2)+\frac{1}{2} \left(\eta +2 Q^2-(\eta +2)
   v\right)\sqrt{1-u^2} \sqrt{\left| \beta \right| } \right],\\
&\frac{d v}{d \xi}=(v-1) v\left[2 \beta  Q \left(u^2-1\right) -(\eta +2) u \sqrt{1-u^2} \sqrt{\left| \beta \right| }\right].
\end{align}
\end{subequations}

The system \eqref{infinity_static} covers all the equilibrium points in the finite region, with the exception of $P_4^{\pm}$, $P_6^{\pm}$, $P_7^{\pm}$, and $P_9^{\pm}$ (which do not exist for $\beta<0$), and incorporates the equilibrium points at infinity. That is, in the new coordinates, the equilibrium points in the finite region are: $P_1^{\pm}: (Q, u,v)=\left(\pm 1 , \pm\frac{\sqrt{\left| \beta \right| }}{\sqrt{1+\left| \beta \right|}},0\right)$, $P_2^{\pm}: (Q, u,v)=\left(\pm 1 ,\mp\frac{\sqrt{\left| \beta \right| }}{\sqrt{1+\left| \beta \right|}},0\right)$, $P_3^{\pm}: (Q, u,v)=(\pm 1, 0, 1)$, $P_5^{\pm}: (Q, u,v)=\left(\pm 1 ,\mp\frac{(\eta +2)}{\sqrt{(\eta +2)^2+16 \left| \beta \right|}},0\right)$, $P_8^{\pm}: (Q, u,v)=\left(\frac{\epsilon }{\sqrt{1+\left| \beta \right|}},\sqrt{\frac{\left| \beta \right|}{1+2 \left| \beta \right|}} \epsilon ,1\right)$. The lines of equilibrium points at infinity are $I_1^\pm: (Q,u,v)=(\pm 1, \pm 1, v_c)$, and 
$I_2^\pm: (Q,u,v)=(\pm 1, \mp 1, v_c)$; that is, $|S|\rightarrow \infty$, and $C$, arbitrary, such that $\frac{C}{1+|\beta| S^2}\rightarrow v_c$, constant. The eigenvalues are $\pm 2(2-v_c)\beta, \mp 2(2-v_c)\beta,0$. Therefore, they behave as saddles.

\section{Stationary comoving aether with perfect fluid and scalar field in
static metric}
\label{sf}

In this section we investigate a stationary comoving aether with perfect fluid and scalar field in
static metric
\begin{equation}
ds^{2}=-N^{2}(r)dt^{2}+e_{1}{}^{1}(r)^{-2}dr^{2}+e_{2}{}^{2}(r)^{-2}(d%
\vartheta ^{2}+\sin ^{2}\vartheta d\varphi ^{2}),  \label{met2-2}
\end{equation}%
and we define $x=\mathbf{e}_{1}\ln e_{2}{}^{2},y=\mathbf{e}_{1}\ln N$, and the
differential operator $\mathbf{e}_{1}=e_{1}{}^{1}\partial _{r}$. 

The
equations for the variables $x,y,p,\varphi ,K$ are:
\begin{subequations}
\label{static}
\begin{align}
& \mathbf{e}_{1}\left( x\right) =\frac{\mu +3p}{2\beta }+%
\mathbf{e}_{1}(\varphi ){}^{2}-\frac{W(\varphi)}{\beta}+2(\beta-1)y^2+3xy+K,
\label{statica} \\
& \mathbf{e}_{1}\left( y\right) =\frac{\mu +3p}{2\beta }-%
\frac{W(\varphi)}{\beta}+2xy-y^{2},  \label{staticb} \\
& {\mathbf{e}_{1}}\left( p\right) =-y(\mu +p),  \label{staticc1} \\
& \mathbf{e}_{1}(\mathbf{e}_{1}(\varphi ))=-\left( y-2x\right) \mathbf{e}%
_{1}(\varphi )+W'(\varphi),  \label{staticd1} \\
& \mathbf{e}_{1}(K)=2xK,  \label{eqKstatic}
\end{align}%
where $W(\varphi)$ is the scalar field self-interacting potential. The system satisfies the
restriction
\end{subequations}
\begin{equation}
\label{static_sf}
-(x-y)^{2} +\beta y^2+p+\frac{1}{2}\mathbf{e}_{1}(\varphi ){}^{2}-W(\varphi)+K=0.
\end{equation}%
Equation \eqref{static_sf} is called Gauss
constraint, and it corresponds to a first integral of the system \eqref{static}. This can be proven by applying the differential operator $\mathbf{e}_{1}(...)$ to both sides of %
\eqref{static_sf}, and then using the equations \eqref{static} to eliminate the
spatial derivatives. Therefore, by using again the restriction \eqref{static_sf}
solved for $K,$ we obtain an identity. On the other hand, the aether constraint \eqref{restrictionaether} is
identically zero. 

As before, we assume an EoS parametrized by \eqref{ss1}.
 We next consider the case of an exponential self-interacting potential; we shall study the case of a harmonic
potential $W(\varphi)= \frac{1}{2} m^2 \varphi^2$ elsewhere, e.g., as particular example in paper III \cite{paperIII}.

\subsection{Phase-Space Evolution: Exponential potential $W(\varphi)=W_0 e^{-k \varphi}$.}

For the analysis of the system of equations \eqref{static} one can use methods to obtain exact solutions. Additionally one can use the dynamical systems approach for investigating the
structure of the whole solution space. Using the quantities $\theta=y-x, \sigma=y$ as in \cite{Nilsson:2000zf}, the equations read
\begin{subequations}
\begin{align}
& e_{1}{}^{1}\frac{d \theta}{d r}= -\beta  \sigma ^2-\theta ^2+\theta  \sigma -W_0
   e^{-k \varphi }+p-\frac{\Phi ^2}{2},\\
& e_{1}{}^{1}\frac{d \sigma}{d r}= \frac{2 \beta  \sigma 
   (\sigma -2 \theta )-2 W_0 e^{-k \varphi }+\mu_0+(\eta +2)
   p}{2 \beta },\\
& e_{1}{}^{1}\frac{d K}{d r}	= -2 (\sigma -\theta ) \left(\beta  \sigma ^2-\theta
   ^2-W_0 e^{-k \varphi }+p+\frac{\Phi ^2}{2}\right),\\
& e_{1}{}^{1}\frac{d \varphi}{d r}= \Phi,\\
&  e_{1}{}^{1}\frac{d \Phi}{d r}= \Phi (\sigma -2 \theta )-k W_0 e^{-k \varphi },\\
&  e_{1}{}^{1}\frac{d p}{d r}=-\sigma  (\eta p+ \mu_0),
\end{align}
\end{subequations}
where
\begin{subequations}
\begin{align}
& K+p= -\beta  \sigma ^2+\theta^2+W_0 e^{-k \varphi }-\frac{\Phi ^2}{2},\\
& \mu = \mu_0+(\eta -1) p. 
\end{align}
\end{subequations}
Now we define the radial variable $\lambda$
\begin{equation}
 e_{1}{}^{1}\frac{d \lambda}{d r}=\sqrt{\frac{\mu_0}{\eta }+\theta^2}. 
\end{equation}
For convenience, we set one of the metric components as $e_{1}{}^{1}=r$. This is equivalent to set $\mathbf{e}%
_{1}=\partial _{{\ell}}$, where 
$r=e^{\ell}$, such that $\ell\rightarrow -\infty$ as $r\rightarrow 0$ and $\ell\rightarrow \infty$ as $r\rightarrow \infty$.
In other words, $\lambda$ unequivocally defines the flow direction. That is, the ``past attractors'' ($r\rightarrow 0$) correspond to $\lambda\rightarrow -\infty$ and the ``future attractors'' ($r \rightarrow \infty$) correspond to  $\lambda\rightarrow \infty$. 
Defining the scale invariant quantities:
\begin{equation}
\label{inv:qt2}
Q=\frac{\theta}{\sqrt{\frac{\mu_0}{\eta }+\theta^2}},   S=\frac{\sigma}{\sqrt{\frac{\mu_0}{\eta }+\theta^2}},  C=\frac{\eta  K}{\mu_0+\eta \theta^2},
A_{\varphi}=\frac{\Phi}{\sqrt{2}\sqrt{\frac{\mu_0}{\eta }+\theta^2}},  A_{W}=\frac{\sqrt{W_0} e^{-\frac{k}{2} \varphi }}{\sqrt{\frac{\mu_0}{\eta }+\theta^2}},
\end{equation}
we obtain the evolution equations
\begin{subequations}
\label{inv:qt_evol2}
\begin{align}
  & \frac{d Q}{d \lambda}=\left(Q^2-1\right) \left(2 A_\varphi^2+C+S (2 \beta
   S-Q)\right),\\
	&
	\frac{d S}{d\lambda}=\frac{C (2 \beta Q S-\eta -2)-2 Q \left(\beta
   S^2-1\right) (Q-2 \beta S)-\beta \eta  S^2+\eta }{2 \beta}\nonumber \\
	& \;\;\;\;\;\;\;\;\;\; \frac{A_{W}^2 \eta +A_\varphi^2 (4 \beta Q S-\eta -2)}{2 \beta},\\
	& \frac{d C}{d\lambda}= 2 C \left(Q \left(2 A_\varphi^2+C+2 \beta S^2-Q S-1\right)+S\right),\\
	& \frac{d A_\varphi}{d\lambda}=A_\varphi \left(Q \left(2 A_\varphi^2+C+2 \beta S^2-Q S-2\right)+S\right)-\frac{\sqrt{2}}{2}{A_{W}^2 k
   },\\
	&\frac{dA_{W}}{d\lambda}=\frac{1}{2} A_{W} \left(4 A_\varphi^2 Q-\sqrt{2} A_\varphi
   k +2 Q (C+S (2 \beta S-Q))\right).
\end{align}
\end{subequations}

We have the useful relations
\begin{subequations}
\begin{align}
& \frac{\mu_0}{\eta}=\frac{(1-Q^2)K}{C}, \\
&  p=  -\frac{K \left(-A_{W}^2+A_\varphi^2+C+\beta S^2-Q^2\right)}{C}, \\
& \quad \frac{C (\mu +p)}{\eta  K}=A_{W}^2-A_\varphi^2-C-\beta S^2+1.
\end{align}
\end{subequations}
Since $\mu_0\geq 0,\eta\geq 1$ it follows $-1\leq Q\leq 1.$ Since $K\geq 0$ it follows $C\geq 0$. The condition $$\left(-A_{W}^2+A_\varphi^2+C+\beta S^2-Q^2\right)=0$$ defines the surface of zero-pressure. However, it is not an invariant set of \eqref{inv:qt_evol2}, neither is $$\left(-A_{W}^2+A_\varphi^2+C+\beta S^2-Q^2\right)>0.$$ If we assume that the weak energy condition $p+\mu\geq 0$ is satisfied, then we obtain the subset of the phase space
$$1+A_{W}^2-A_\varphi^2-C-\beta S^2\geq 0.$$ Defining $\Omega=1+A_{W}^2-A_\varphi^2-C-\beta S^2$ we obtain
\begin{equation}
\frac{d \Omega}{d \lambda}=\Omega\left[4 A_\varphi^2 Q+2 Q (C+S (2 \beta S-Q))-\eta  S\right].
\end{equation}
Thus, $1+A_{W}^2-A_\varphi^2-C-\beta S^2\geq 0$ defines an invariant set.
Summarizing, the equations \eqref{inv:qt_evol2} defines a flow on the invariant set
\begin{equation}
\left\{(Q, S, C,  A_\varphi, A_{W}):-1\leq Q \leq 1, C\geq 0, -A_{W}^2+A_\varphi^2+C+\beta S^2\leq 1, A_{W}\geq 0\right\}.
\end{equation}
This phase-space is unbounded.
The invariant sets $Q=\pm 1$ corresponds to $\mu_0=0$.

We have the auxiliary equations
\begin{subequations}
\label{metric-eqs-exp}
\begin{align}
\label{eqN-exp}
&\frac{d\ln N}{d\lambda}=S,\\
\label{eqK-exp}
&\frac{d \ln K}{d\lambda}=-2(Q-S),\\
\label{eqy-exp}
& \frac{d \ln y}{d\lambda}= -\frac{-\left(A_{W}^2+1\right) \eta +A_\varphi^2 (\eta +2)+C (\eta +2)+4
   \beta Q S+\beta \eta  S^2-2 Q^2}{2 \beta S}.
\end{align}
\end{subequations}

Both the line element expressed in the variables \eqref{inv:qt} and the dynamical system \eqref{inv:qt_evol2} are invariant under the discrete symmetry
\begin{equation}
\label{discrete}
(Q, S,  A_\varphi, \lambda)\rightarrow  (-Q, -S, -A_\varphi, -\lambda).
\end{equation}

The  equilibrium points are discussed in Appendix \ref{phys-intepretation2}.

\subsubsection{Equilibrium points in the finite region of the phase space}\label{SECT_5.2.2}

We have recovered the previous results for the points $P_1^{\pm}$-$P_9^{\pm}$ (when no scalar field is present). 
For further details about the derivation of the physical interpretation of the equilibrium points $P_1^{\pm}$-$P_9^{\pm}$ we submit the reader to Appendix
\ref{phys-intepretation}. The equilibrium points with non-trivial scalar field  are discussed in Appendix \ref{phys-intepretation2}.

Now, we discuss the more interesting points (in the sense that they have the highest dimensional stable/unstable manifold).

\begin{enumerate}
\item $P_5^{+}$ is a sink for $\beta<0, \eta \geq 1$. 
\item $P_5^{-}$ is source for $\beta<0, \eta \geq 1$. 
\item $P_8^{+}$ is non-hyperbolic with a 4D stable manifold. 
\item $P_8^{-}$ is non-hyperbolic with a 4D unstable manifold.

\item The following subsets (arcs, or specific equilibrium points) of the line $P_{10}^{+}(S_c)$ are unstable for the given conditions: 
 \begin{enumerate}
   \item $S_c<0, \beta <\frac{1}{{S_c}^2}, \eta >1, k<\sqrt{\frac{-2 {S_c}^2+8 S_c-8}{\beta {S_c}^2-1}}$, or 
  \item $S_c<0, \beta =\frac{1}{{S_c}^2}, \eta >1, k\in \mathbb{R}$, or         
  \item $S_c=0, \beta \in \mathbb{R}, \eta>1, k<2 \sqrt{2}$, or 
  \item $0<S_c<\frac{4}{3}, \beta <\frac{1}{{S_c}^2}, 1< \eta <\frac{4-2 S_c}{S_c},  k<\sqrt{\frac{-2 {S_c}^2+8 S_c-8}{\beta  {S_c}^2-1}}$, or  
  \item $0<S_c<\frac{4}{3}, \beta =\frac{1}{{S_c}^2}, 1< \eta <\frac{4-2
  S_c}{S_c}, k\in \mathbb{R}$.
 \end{enumerate}

\item The following subsets (arcs, or specific equilibrium points) of the line $P_{10}^{-}(S_c)$ are stable for the given conditions: 
 \begin{enumerate}
   \item $S_c<0, \beta <\frac{1}{{S_c}^2}, \eta >1, k<\sqrt{\frac{-2 {S_c}^2+8 S_c-8}{\beta {S_c}^2-1}}$, or 
  \item $S_c<0, \beta =\frac{1}{{S_c}^2}, \eta >1, k\in \mathbb{R}$, or         
  \item $S_c=0, \beta \in \mathbb{R}, \eta>1, k<2 \sqrt{2}$, or 
  \item $0<S_c<\frac{4}{3}, \beta <\frac{1}{{S_c}^2}, 1< \eta <\frac{4-2 S_c}{S_c},  k<\sqrt{\frac{-2 {S_c}^2+8 S_c-8}{\beta  {S_c}^2-1}}$, or  
  \item $0<S_c<\frac{4}{3}, \beta =\frac{1}{{S_c}^2}, 1< \eta <\frac{4-2
  S_c}{S_c}, k\in \mathbb{R}$.
 \end{enumerate}

\item The following subsets (arcs, or specific equilibrium points) of the line $P_{11}^{+}(S_c)$ are unstable for the given conditions: 
\begin{enumerate}
   \item ${S_c}<0, \beta <\frac{1}{{S_c}^2}, \eta
   \geq 1, k>-\sqrt{2} \sqrt{-\frac{{S_c}^2-4 {S_c}+4}{\beta  {S_c}^2-1}}$, or 
   \item ${S_c}<0, \beta =\frac{1}{{S_c}^2}, \eta \geq 1, k\in \mathbb{R}$, or 
   \item ${S_c}=0, \beta \in \mathbb{R}, \eta \geq 1, k>-2 \sqrt{2}$, or 
  \item $0<{S_c}<\frac{4}{3}, \beta <\frac{1}{{S_c}^2}, 1\leq \eta <\frac{4-2{S_c}}{{S_c}}, k>-\sqrt{2} \sqrt{-\frac{{S_c}^2-4 {S_c}+4}{\beta  {S_c}^2-1}}$, or
  \item $0<{S_c}<\frac{4}{3}, \beta =\frac{1}{{S_c}^2}, 1\leq \eta
   <\frac{4-2 {S_c}}{{S_c}}, k\in \mathbb{R}$.
\end{enumerate}

\item The following subsets (arcs, or specific equilibrium points) of the line  $P_{11}^{-}(S_c)$ are stable for the given conditions: 
\begin{enumerate}
   \item ${S_c}<0, \beta <\frac{1}{{S_c}^2}, \eta
   \geq 1, k>-\sqrt{2} \sqrt{-\frac{{S_c}^2-4 {S_c}+4}{\beta  {S_c}^2-1}}$, or 
   \item ${S_c}<0, \beta =\frac{1}{{S_c}^2}, \eta \geq 1, k\in \mathbb{R}$, or 
   \item ${S_c}=0, \beta \in \mathbb{R}, \eta \geq 1, k>-2 \sqrt{2}$, or 
  \item $0<{S_c}<\frac{4}{3}, \beta <\frac{1}{{S_c}^2}, 1\leq \eta <\frac{4-2{S_c}}{{S_c}}, k>-\sqrt{2} \sqrt{-\frac{{S_c}^2-4 {S_c}+4}{\beta  {S_c}^2-1}}$, or
  \item $0<{S_c}<\frac{4}{3}, \beta =\frac{1}{{S_c}^2}, 1\leq \eta
   <\frac{4-2 {S_c}}{{S_c}}, k\in \mathbb{R}$.
\end{enumerate}

\item $P_{12}^{+}$ is non-hyperbolic with a 4D unstable for 
   \begin{enumerate}
   \item $\beta <0, \eta \geq 1, k>-\sqrt{2} \sqrt{\frac{(8 \beta -\eta -2)^2}{\beta  \left(16 \beta -\eta ^2-4 \eta -4\right)}}$, or 
    \item $\beta >\frac{9}{16}, 1\leq \eta <2 \left(2
   \sqrt{\beta }-1\right), k<\sqrt{2} \sqrt{\frac{(8 \beta -\eta -2)^2}{\beta  \left(16 \beta -\eta ^2-4 \eta -4\right)}}$. 
    \end{enumerate}

\item $P_{12}^{-}$ is non-hyperbolic with a 4D stable manifold for 
   \begin{enumerate}
   \item $\beta <0, \eta \geq 1, k>-\sqrt{2} \sqrt{\frac{(8 \beta -\eta -2)^2}{\beta  \left(16 \beta -\eta ^2-4 \eta -4\right)}}$, or 
    \item $\beta >\frac{9}{16}, 1\leq \eta <2 \left(2
   \sqrt{\beta }-1\right), k<\sqrt{2} \sqrt{\frac{(8 \beta -\eta -2)^2}{\beta  \left(16 \beta -\eta ^2-4 \eta -4\right)}}$. 
    \end{enumerate}
		
		\item  $P_{13}^{+}$ is non-hyperbolic with a 4D unstable manifold for
  \begin{enumerate}
  \item $\beta <0, \eta \geq 1, k<\sqrt{2} \sqrt{\frac{(8 \beta -\eta -2)^2}{\beta  \left(16 \beta -\eta ^2-4 \eta -4\right)}}$, or 
  \item $\beta >\frac{9}{16}, 1\leq \eta <2 \left(2
   \sqrt{\beta }-1\right), k>-\sqrt{2} \sqrt{\frac{(8 \beta -\eta -2)^2}{\beta  \left(16 \beta -\eta ^2-4 \eta -4\right)}}$.
  \end{enumerate}

\item $P_{13}^{-}$ is non-hyperbolic with a  4D  stable manifold for 
  \begin{enumerate}
  \item $\beta <0, \eta \geq 1, k<\sqrt{2} \sqrt{\frac{(8 \beta -\eta -2)^2}{\beta  \left(16 \beta -\eta ^2-4 \eta -4\right)}}$, or 
  \item $\beta >\frac{9}{16}, 1\leq \eta <2 \left(2
   \sqrt{\beta }-1\right), k>-\sqrt{2} \sqrt{\frac{(8 \beta -\eta -2)^2}{\beta  \left(16 \beta -\eta ^2-4 \eta -4\right)}}$.
  \end{enumerate}

\item $P_{16}^{+}$ is a source for 
\begin{enumerate}
\item $\eta \geq 1, 0<\beta <\frac{\eta +2}{8}, -\sqrt{\frac{\eta }{\beta }}<k<-2 \sqrt{2} \sqrt{\frac{\eta ^2}{-16 \beta +\eta ^2+4 \eta +4}}$, or
\item $\eta \geq 1, 0<\beta <\frac{\eta +2}{8}, 2 \sqrt{2} \sqrt{\frac{\eta ^2}{-16 \beta +\eta ^2+4 \eta +4}}<k<\sqrt{\frac{\eta }{\beta }}$.
\end{enumerate}

\item $P_{16}^{-}$ is a sink for 
\begin{enumerate}
\item $\eta \geq 1, 0<\beta <\frac{\eta +2}{8}, -\sqrt{\frac{\eta }{\beta }}<k<-2 \sqrt{2} \sqrt{\frac{\eta ^2}{-16 \beta +\eta ^2+4 \eta +4}}$, or
\item $\eta \geq 1, 0<\beta <\frac{\eta +2}{8}, 2 \sqrt{2} \sqrt{\frac{\eta ^2}{-16 \beta +\eta ^2+4 \eta +4}}<k<\sqrt{\frac{\eta }{\beta }}$.
\end{enumerate}

\item $P_{18}^{+}$ is a source  for 
\begin{enumerate}
\item $\eta \geq 1, \beta <0, k<-\sqrt{2} \sqrt{\frac{4 \beta -1}{\beta }}$, or
\item $\eta \geq 1, \beta <0, k>\sqrt{2} \sqrt{\frac{4 \beta -1}{\beta }}$, or
\item $\eta \geq 1, 0<\beta \leq \frac{\eta +2}{8}, k<-\sqrt{\frac{\eta }{\beta }}$, or
\item $\eta \geq 1, 0<\beta \leq \frac{\eta +2}{8}, k>\sqrt{\frac{\eta }{\beta }}$, or
\item $\eta \geq 1, \beta >\frac{\eta +2}{8}, k<-\sqrt{2} \sqrt{\frac{4 \beta -1}{\beta }}$, or 
\item $\eta \geq 1, \beta >\frac{\eta +2}{8}, k>\sqrt{2} \sqrt{\frac{4 \beta -1}{\beta }}$. 
\end{enumerate}

\item $P_{18}^{-}$ is a sink for 
\begin{enumerate}
\item $\eta \geq 1, \beta <0, k<-\sqrt{2} \sqrt{\frac{4 \beta -1}{\beta }}$, or
\item $\eta \geq 1, \beta <0, k>\sqrt{2} \sqrt{\frac{4 \beta -1}{\beta }}$, or
\item $\eta \geq 1, 0<\beta \leq \frac{\eta +2}{8}, k<-\sqrt{\frac{\eta }{\beta }}$, or
\item $\eta \geq 1, 0<\beta \leq \frac{\eta +2}{8}, k>\sqrt{\frac{\eta }{\beta }}$, or
\item $\eta \geq 1, \beta >\frac{\eta +2}{8}, k<-\sqrt{2} \sqrt{\frac{4 \beta -1}{\beta }}$, or 
\item $\eta \geq 1, \beta >\frac{\eta +2}{8}, k>\sqrt{2} \sqrt{\frac{4 \beta -1}{\beta }}$. 
\end{enumerate}

\end{enumerate}

As can be seen in the Appendix \ref{phys-intepretation2}, all the possible metrics (with nontrivial scalar field)
can be written in a compact form as 
\begin{equation}
ds^2=-e^{2a \lambda} dt^2+e^{2b \lambda }d\lambda^2+e^{2c \lambda} \left(d\vartheta ^{2}+\sin ^{2}\vartheta d\varphi ^{2}\right);
\end{equation}
Hence for $c \neq 0$ always there is a singularity at $l=\infty$. You can make that easily $r=0$, through the transformation $\exp(2c l)=r^2$.
The same for $a=0$, or  $b=0$ just $c \neq 0$.
Now on the other hand the other possible case is for $c=0$. In that case:
\begin{enumerate}
 \item  We do not have a singularity when: $b<0$ or when $b>0$ and $a^2=a b$. 
 \item When $b=0$ and $a^2 \neq a b$ we have a singularity. 
\end{enumerate}

\section{Discussion}

\label{conclusions}

In this paper we have investigated the field equations in the Einstein-aether
model in a static spherically symmetric spacetime. The static model with perfect
fluid, first introduced in Section 6.1 of \cite{Coley:2015qqa}, has been investigated 
using more appropriate dynamical variables inspired by \cite{Nilsson:2000zf} with a direct physical interpretation which lead to the system \eqref{inv:qt_evol}, for which we have presented further results.  The results of \cite{Nilsson:2000zf} for GR have been extended to the Einstein-aether setup. In particular, we are interested in models which are asymptotically vacuum and asymptotically flat, and which admit singularities.
We  have found 
asymptotic expansions for all of the equilibrium points in the finite region. 
We have shown that the Minkowski spacetime can be given in explicit spherically symmetric form \cite{Nilsson:2000zf} irrespectively on the   aether parameter. We have shown that we can have nonregular self-similar perfect fluid solutions like those in \cite{Tolman:1939jz,Oppenheimer:1939ne,Misner:1964zz},
self-similar plane-symmetric perfect fluid models and Kasner
plane-symmetric vacuum solutions \cite{kasner}. We have discussed the existence of new solutions related with
naked singularities or with horizons. The line elements have been presented in explicit form. In addition, we have discussed the
dynamics at infinity and presented some numerical results supporting our
analytical results. In the next subsection we will summarize all of the sources and sinks in the perfect fluid model. We have also 
investigated Einstein-aether perfect fluid cosmological
models and a scalar field with an exponential self-interaction potential (we shall study the case of a harmonic potential in the Paper III \cite{paperIII}).

In a subsequent paper \cite{Leon:2019jnu}, referred as Paper II, we present a singularity analysis for these models. That is, we study if the gravitational field equations possesses the Painleve property; consequently one can find if an  analytic explicit integration can be performed for the field equations. Then,  we can apply the classical treatment for the singularity analysis which is summarized in the ARS algorithm. Furthermore, it is of interest the formulation of the modified Tolman-Oppenheimer-Volkoff equations for perfect fluids with linear and polytropic equations of state in the Einstein-aether theory, and the addition of scalar field with exponential or an harmonic potentials. 
One special application which we are interested in is to use dynamical system tools to  determine conditions under which stable stars can form. By using the Tolman-Oppenheimer-Volkoff (TOV) approach \cite{Tolman:1939jz,Oppenheimer:1939ne,Misner:1964zz}, the relativistic TOV equations are drastically modified in Einstein-aether theory, and we can explore the physical implications of that modification. Then we can construct a 3D dynamical system in compact variables and obtain a picture of the entire solution space for a linear EoS, that can visualized in a geometrical way. This study can be extended to a wide class of EoS, for example polytropic EoS. For higher dimensional systems we still can find information by numerical integrations and the use of projections. The results obtained can be inserted coherently into the physical models, obtaining an appropriate description of the universe both in local and larger scales. More of this analysis can be found in the paper \cite{Leon:2019jnu}.

\subsection{Summary of relevant saddles}

There are relevant equilibrium points which are saddle points: 
\begin{enumerate}
\item The equilibrium point $P_3^{+}$ represents the Minkowski spacetime in spherical symmetric form. For which 
we find the more familiar equations
\begin{equation}
Q=1-\varepsilon_1 e^{2\lambda}, \quad S=\frac{2}{3} (\varepsilon_2-\varepsilon_1) e^{2\lambda}, \quad C=1-\frac{4}{2+\eta}((1-\beta) \varepsilon_1+\varepsilon_2) e^{2\lambda}, 
\end{equation} where $\varepsilon_1$ and $\varepsilon_2$ are still small constants (we assume they are positive),
that reproduce  equations (27a- 27c) of \cite{Nilsson:2000zf} for $\beta=1$. We see that $\varepsilon=\frac{\varepsilon_1}{\varepsilon_2}$  parametrize a 1-parameter family of regular solutions with an equation of state parameter at the center:
\begin{align*}
&\frac{p_c}{\mu_c}=\lim_{\lambda \rightarrow -\infty }\frac{p}{\mu}=\lim_{\lambda \rightarrow -\infty }\frac{\mu-\mu_0}{(\eta-1)\mu}=\lim_{\lambda \rightarrow -\infty }\frac{C-Q^2+\beta  S^2}{C (\eta -1)-\eta +Q^2+\beta  (\eta -1) S^2}\nonumber\\
& =\frac{2-\varepsilon (2 \beta +\eta )}{\varepsilon  (2 \beta  (\eta -1)-3 \eta
   )+2 (\eta -1)}.
\end{align*}
The quotient, $\frac{p_c}{\mu_c}$ is a gravitational strength parameter. In GR where the parameter $\beta=1$, the maximal value of the gravitational strength, $\frac{1}{\eta+1}$, is obtained when $\varepsilon_1=0$, which corresponds to the subset $Q=1$. However, in the Einstein-aether theory the parameter $\beta$ is a freely specifiable parameter, and for $\eta >1, \beta >\frac{3 \eta }{2 \eta -2}, \frac{\varepsilon_1}{\varepsilon_2} >\frac{2 \eta
   -2}{2 \beta  \eta -2 \beta -3 \eta }$, the maximal strength is not $\frac{p_c}{\mu_c}=\frac{1}{\eta+1}$ anymore as it is in GR.   

We see that there exists solutions with a regular center but negative pressure, so that we have to impose the condition
\begin{equation*}
\frac{2-\varepsilon (2 \beta +\eta )}{\varepsilon  (2 \beta  (\eta -1)-3 \eta
   )+2 (\eta -1)}>0,
\end{equation*}
that is:
\begin{enumerate}
\item $\eta >1, \beta \leq -\frac{\eta }{2}, \varepsilon >0$, or 
\item $\eta >1,  -\frac{\eta }{2}<\beta \leq \frac{3 \eta }{2 \eta -2}, 
   0<\varepsilon <\frac{2}{2 \beta +\eta }$, or
\item $\eta >1, \beta >\frac{3 \eta }{2 \eta -2}, 0<\varepsilon <\frac{2}{2
   \beta +\eta }$, or 
\item $\eta >1, \beta >\frac{3 \eta }{2 \eta -2}, \varepsilon >\frac{2 \eta
   -2}{2 \beta  \eta -2 \beta -3 \eta }$.
\end{enumerate}
This condition is reduced in GR, to $\eta >1, 0<\varepsilon <\frac{2}{\eta +2}$, when $\beta=1$.

For  $C-(Q-S)^2>0$, the first and second Buchdahl conditions are satisfied at the solution as $\lambda\rightarrow -\infty$, if
\begin{align*}
&\frac{(\beta -1) (\eta +2) \varepsilon }{\varepsilon  (6 \beta +\eta
   -4)+2 (\eta -1)}\geq 0,\\
  &   \frac{\eta  (\eta +2) \varepsilon  (\mu_{0}+(\eta -1) p_c)}{\mu_{0} (\varepsilon  (6 \beta +\eta
   -4)+2 (\eta -1))}\leq 1.
\end{align*} 
Additionally, taking the limit $\lambda\rightarrow -\infty$ we have
\begin{equation*}
\frac{1}{9} \left(7 C-3 Q^2+3 \beta  S^2\right)+2 \sqrt{C \left(C+3 Q^2-3
   \beta  S^2\right)}-C+(Q-S)^2\rightarrow \frac{40}{9}>0,
\end{equation*}
such that the third Buchdahl condition is also satisfied.
Thus, combining these conditions we have the conditions for the existence of regular solution at the center associated to $P_3^{+}$.

\item The equilibrium point $P_4^{+}$ generalizes the so called Tolman point (which corresponds to $\beta=1$), which now is promoted to a 1-parameter solution. This solution exists for $0\leq \beta \leq \frac{1}{8} (\eta +2)^2$. 
Following the same method as for the analysis of $P_3^{+}$ we have explored approximated solutions related to $P_4^{+}$ by constructing the unstable manifold of this equilibrium point. 

{\bf{Case 1:}}\\
When $\lambda_{2}$, $\lambda_{3}$ are both reals and negative, that is whenever $\frac{63}{64}<\beta \leq \frac{9}{8}, 1<\eta \leq -2+\frac{8 \sqrt{\beta }}{\sqrt{7}}$, or 
 $\beta >\frac{9}{8}, 2 \sqrt{2} \sqrt{\beta }-2<-2+\eta \leq \frac{8 \sqrt{\beta }}{\sqrt{7}}$, we obtain that any solution near the unstable manifold of $P_4^{+}$, satisfies 
\begin{small}
\begin{align*}
&Q=1-\frac{(\eta +2)^2 (-4 \beta +\eta  (2 \eta +3)+2)}{2 \left((4 \beta +\eta -2) \left((\eta +2)^2-8 \beta \right)\right)} \varepsilon e^{\frac{2 \eta  \lambda }{\eta +2}} ,\\
&S=\frac{2}{\eta +2}-\frac{ (\eta +2) \left(-8 \beta ^2-2 (\eta -2) \beta +\eta  (\eta +2)^2\right)}{2 \left(\beta  (4 \beta +\eta -2) \left((\eta +2)^2-8
   \beta \right)\right)} \varepsilon e^{\frac{2 \eta  \lambda }{\eta +2}} + \mathcal{O}(\varepsilon ^2) e^{\frac{4 \eta  \lambda }{\eta +2}},\\
   & C=1-\frac{8 \beta }{(\eta +2)^2}+\varepsilon e^{\frac{2 \eta}{\eta +2}\lambda}+ \mathcal{O}(\varepsilon ^2) e^{\frac{4 \eta  \lambda }{\eta +2}}.
\end{align*} 
\end{small}
This expansion is accurate as long as $\lambda\rightarrow -\infty$. 

Using this solution, we find 
\begin{small}
\begin{align*}
& \frac{p_c}{\mu_c}=\lim_{\lambda \rightarrow -\infty }\frac{C-Q^2+\beta  S^2}{C (\eta -1)-\eta +Q^2+\beta  (\eta -1) S^2}\nonumber\\
   & = \lim_{\lambda \rightarrow -\infty } \frac{1}{\eta-1}-\frac{\varepsilon  \left(\eta  (\eta +2)^4 (-4 \beta +\eta  (2 \eta +3)+2) e^{\frac{2 \eta  \lambda }{\eta +2}}\right)}{4 \left(\beta  (\eta -1)^2 (4 \beta +\eta -2) \left((\eta +2)^2-8 \beta  \right)\right)}+O\left(\varepsilon ^2\right)\nonumber\\
   &    = \frac{1}{\eta-1} >0.
\end{align*}
\end{small}
Furthermore, the Buchdahl conditions that can be expressed as 
\begin{small}
\begin{align*}
& 1\geq \frac{-(\eta -1) \left(C+\beta  S^2\right)+\eta -Q^2}{3 \left(C-(Q-S)^2\right)},\\
& 1\leq \frac{\eta  \left(1-Q^2\right) (\mu_{0}+(\eta -1) p_c)}{3 \mu_{0} \left(C-(Q-S)^2\right)},\\
& \frac{1}{9} \left(7 C-3 Q^2+3 \beta  S^2\right)+2 \sqrt{C \left(C+3 Q^2-3
   \beta  S^2\right)}-C+(Q-S)^2\geq 0.
\end{align*}
\end{small}
And as $\lambda\rightarrow -\infty$, applying the above conditions we have that the second one is satisfied; and the first and third imply
\begin{small}
\begin{align*}
& \frac{\beta -\beta  \eta }{6 \beta -3 (\eta +1)}\geq 1,\\
& \frac{4 (7 \beta +9)}{9 (\eta +2)^2}+2 \sqrt{\left(4-\frac{20 \beta }{(\eta +2)^2}\right) \left(1-\frac{8 \beta }{(\eta +2)^2}\right)}-\frac{4}{\eta
   +2}+\frac{4}{9}\geq 0.
\end{align*}
\end{small}
These conditions are not satisfied for $\beta=1$ (that is, for GR). But in AE-theory $\beta$ is a free  parameter, such that the above inequalities can be satisfied for $\eta >1, \frac{3 \eta +3}{\eta +5}\leq \beta <\frac{\eta +1}{2}, 64 \beta -7 (\eta +2)^2\geq 0$.

{\bf{Case 2:}}\\
For the choice  $0<\beta \leq \frac{63}{64}, \eta >1$, or 
 $\beta >\frac{63}{64}, \eta >-2 +\frac{8 \sqrt{\beta }}{\sqrt{7}}$, the eigenvalues $\lambda_2, \lambda_3$ are complex conjugates with negative real part.  We obtain that any solution near the unstable manifold of $P_4^{+}$ satisfies 
\begin{small}
\begin{align*}
& Q= 1-\frac{(\eta +2)^2  (-4 \beta +\eta  (2 \eta +3)+2) }{2 (4 \beta +\eta -2) \left(8 \beta -(\eta +2)^2\right)}\varepsilon e^{\frac{2 \eta  \lambda }{\eta +2}},\\
& S=\frac{2}{\eta +2}+\frac{(\eta +2) \varepsilon  \left(-8 \beta ^2-2 \beta  (\eta
   -2)+\eta  (\eta +2)^2\right) e^{\frac{2 \eta  \lambda }{\eta +2}}}{2 \beta  (4 \beta +\eta -2) \left(8 \beta -(\eta +2)^2\right)}+ \mathcal{O}(\varepsilon ^2) e^{\frac{4 \eta  \lambda }{\eta +2}},\\
& C=1-\frac{8 \beta }{(\eta +2)^2}+\varepsilon  e^{\frac{2 \eta  \lambda }{\eta +2}} + \mathcal{O}(\varepsilon ^2) e^{\frac{4 \eta  \lambda
   }{\eta +2}},
	\end{align*}
	\end{small}
where we have substituted the approximated solution $u_1= \varepsilon e^{\frac{2 \eta}{\eta +2}\lambda}$, that is obtained by integrating the linearized equation along the unstable direction. This expansion is accurate as long as $\lambda\rightarrow -\infty$. At the stable manifold the orbits spiral in and tend asymptotically to the origin with modes $\cos(\frac{\sqrt{7 (\eta +2)^2-64 \beta }}{2 (\eta +2)} \lambda ) e^{-\frac{\lambda}{2}}$, $\sin(\frac{\sqrt{7 (\eta +2)^2-64 \beta }}{2 (\eta +2)} \lambda ) e^{-\frac{\lambda}{2}}$. 

We have the estimates
\begin{small}
\begin{align*}
& \frac{p_c}{\mu_c}=\lim_{\lambda \rightarrow -\infty }\frac{C-Q^2+\beta  S^2}{C (\eta -1)-\eta +Q^2+\beta  (\eta -1) S^2}\nonumber\\
   & = \lim_{\lambda \rightarrow -\infty } \frac{1}{\eta -1}+\frac{\eta  (\eta +2)^4 \varepsilon  \left(-4 \beta +2 \eta ^2+3 \eta +2\right) e^{\frac{2 \eta  \lambda }{\eta +2}}}{4 \beta  (\eta -1)^2 (4 \beta +\eta -2) \left(-8 \beta +\eta ^2+4 \eta
   +4\right)}+O\left(\varepsilon ^2\right)
\nonumber\\
   &    = \frac{1}{\eta - 1} >0.
\end{align*}
\end{small}
 Furthermore, for  $C-(Q-S)^2>0$, the Buchdahl conditions reduce to 
\begin{small}
\begin{equation*}
\frac{2 (\beta -1) \beta }{-2 \beta +\eta +1}+\beta \leq 3, 2 \beta \mu_0\geq (\eta +1) \mu_0,
\end{equation*} 
\begin{equation*}
\frac{9 \sqrt{40 \beta ^2-13 \beta  (\eta +2)^2+(\eta +2)^4}+7 \beta +\eta ^2-5 \eta -5}{2 \beta -\eta
   -1}\leq 0,
	\end{equation*}
	\end{small}
		as $\lambda\rightarrow -\infty$, respectively. 
	
	That is, when $1<\eta \leq 1.04725, \beta <\frac{1}{64} \left(7 \eta ^2+28 \eta +28\right), \mu_0\leq 0$ or $\eta >1.04725, \beta \leq \frac{3 \eta +3}{\eta +5}, \mu_0\leq 0$. We are assuming $\mu_0\geq 0$, therefore the conditions are fulfilled if $\mu_0=0$.

\end{enumerate}

\subsection{Summary of sources and sinks}

\subsubsection{Perfect fluid}

For this analysis we have used the formulation $\{Q, S, C\}$ given by the model \eqref{inv:qt_evol}, which represents the evolution of a perfect fluid has the EoS $\mu =\mu _{0}+\left( \eta -1\right) p,~\eta >1$ in the static Eistein-aether theory. We have found the following summary of sources/ sinks:
\begin{enumerate}
\item $P_1^{+}$ is a source for $\beta=1, 1\leq\eta<2$. Since the conditions \eqref{Condition36} are fulfilled this solution has  a regular center as $\lambda\rightarrow -\infty$. Because of $C=0$ it belongs to the plane-symmetric boundary set. Furthermore, it belongs to the scale invariant boundary $Q=+1$.

\item $P_1^{-}$ is a sink for $\beta=1, 1\leq\eta<2$. Because of $C=0$ it belongs to the plane-symmetric boundary set. Furthermore, it belongs to the scale invariant boundary $Q=-1$. Since the conditions \eqref{asymp_flatness} are fulfilled this solution is asymptotically flat as $\lambda \rightarrow +\infty$. 

\item $P_2^{+}$ is a source for $\beta=1, \eta\geq 1$.  Since the first inequality of \eqref{Condition36} is not fulfilled, this solution does not have a regular center. Because of $C=0$ it belongs to the plane-symmetric boundary set. Furthermore, it belongs to the scale invariant boundary $Q=+1$. 

\item $P_2^{-}$ is a sink for $\beta=1, \eta\geq 1$. Because of $C=0$ it belongs to the plane-symmetric boundary set. Furthermore, it belongs to the scale invariant boundary $Q=-1$. This solution is not asymptotically flat  since the conditions \eqref{asymp_flatness} are not fulfilled as $\lambda \rightarrow +\infty$. 

\item $P_2^{+}$ is a sink for $\beta=-\frac{\eta+2}{4}\leq 1, \eta\geq 1$. Because of $C=0$ it belongs to the plane-symmetric boundary set. Furthermore, it belongs to the scale invariant boundary $Q=+1$. This solution is not asymptotically flat  since the conditions \eqref{asymp_flatness} are not fulfilled as $\lambda \rightarrow +\infty$. 

\item $P_2^{-}$ is a source for $\beta=-\frac{\eta+2}{4}\leq 1, \eta\geq 1$. Since the first inequality of \eqref{Condition36} is not fulfilled, this solution does not have a regular center as $\lambda\rightarrow -\infty$. Because of $C=0$ it belongs to the plane-symmetric boundary set. Furthermore, it belongs to the scale invariant boundary $Q=-1$. 

\item $P_5^{+}$ is a sink  for $\eta \geq 1, \beta<0$. Because of $C=0$ it belongs to the plane-symmetric boundary set. Furthermore, it belongs to the scale invariant boundary $Q=-1$.  Furthermore, it belongs to the scale invariant boundary $Q=+1$.  This solution is not asymptotically flat  since the conditions \eqref{asymp_flatness} are not fulfilled as $\lambda \rightarrow +\infty$. 

\item $P_5^{-}$ is a source for $\eta \geq 1, \beta<0$. Since the conditions \eqref{Condition36} are not fulfilled, this solution does not have a regular center as $\lambda\rightarrow -\infty$. Because of $C=0$ it belongs to the plane-symmetric boundary set. Furthermore, it belongs to the scale invariant boundary $Q=-1$. 

\item $P_6^{+}$ is a source for $\eta \geq 1, 16 \beta \geq (\eta +2)^2$. It has a regular center as $\lambda\rightarrow -\infty$ only when $\beta=1$ (i.e., when this point coincides with $P_1^{+}$). Otherwise the conditions \eqref{Condition36} are not fulfilled, and the solution does not have a regular center as $\lambda\rightarrow -\infty$. Because of $C=0$ it belongs to the plane-symmetric boundary set. Furthermore, it belongs to the scale invariant boundary $Q=+1$.
\item $P_6^{-}$ is a sink for $\eta \geq 1, 16 \beta \geq (\eta +2)^2$.  Because of $C=0$ it belongs to the plane-symmetric boundary set. Furthermore, it belongs to the scale invariant boundary $Q=-1$. It is not asymptotically flat as $\lambda \rightarrow +\infty$ unless $\beta=1$ (i.e., when $P_6^{-}$ merge with $P_1^{-}$ ).

\item $P_7^{+}$ is a source for $\eta \geq 1, \beta>0$. The conditions \eqref{Condition36} are not fulfilled, and the solution does not have a regular center  as $\lambda \rightarrow -\infty$. Because of $C=0$ it belongs to the plane-symmetric boundary set. Furthermore, it belongs to the scale invariant boundary $Q=+1$.

\item $P_7^{-}$ is a sink for  $\eta \geq 1, \beta>0$. Because of $C=0$ it belongs to the plane-symmetric boundary set. Furthermore, it belongs to the scale invariant boundary $Q=-1$.

\item $P_8^{+}$ is a sink for $\beta<0$. It is not asymptotically flat as $\lambda\rightarrow +\infty$. 

\item $P_8^{-}$ is a source for $\beta<0$. It has a regular center as $\lambda \rightarrow -\infty$ if $\eta >1, \beta <0, \mu_{0}>0,  p_c\geq  \frac{\mu_0\left(6 \beta-\beta  \eta  -3\right)}{\beta \eta(\eta-1)}.$
   
\end{enumerate}

\subsubsection{Perfect fluid plus a scalar field with exponential potential}

On the other hand, we have taken a natural extension of the previous analysis, as in the General Relativistic  case \cite{Nilsson:2000zg}, by studying the model \eqref{inv:qt_evol2}, which corresponds to a stationary comoving aether with perfect fluid and scalar field with exponential potential  in 
a static metric. And we have presented the following summary of sources/ sinks:
\begin{enumerate}
\item $P_5^{+}$ is a sink for $\beta<0, \eta \geq 1$. 
\item $P_5^{-}$ is source for $\beta<0, \eta \geq 1$. 
\item $P_8^{+}$ is non-hyperbolic with a 4D stable manifold. 
\item $P_8^{-}$ is non-hyperbolic with a 4D unstable manifold.

\item The following subsets (arcs, or specific equilibrium points) of the line $P_{10}^{+}(S_c)$ are unstable for the given conditions: 
 \begin{enumerate}
   \item $S_c<0, \beta <\frac{1}{{S_c}^2}, \eta >1, k<\sqrt{\frac{-2 {S_c}^2+8 S_c-8}{\beta {S_c}^2-1}}$, or 
  \item $S_c<0, \beta =\frac{1}{{S_c}^2}, \eta >1, k\in \mathbb{R}$, or         
  \item $S_c=0, \beta \in \mathbb{R}, \eta>1, k<2 \sqrt{2}$, or 
  \item $0<S_c<\frac{4}{3}, \beta <\frac{1}{{S_c}^2}, 1< \eta <\frac{4-2 S_c}{S_c},  k<\sqrt{\frac{-2 {S_c}^2+8 S_c-8}{\beta  {S_c}^2-1}}$, or  
  \item $0<S_c<\frac{4}{3}, \beta =\frac{1}{{S_c}^2}, 1< \eta <\frac{4-2
  S_c}{S_c}, k\in \mathbb{R}$.
 \end{enumerate}

\item The following subsets (arcs, or specific equilibrium points) of the line $P_{10}^{-}(S_c)$ are stable for the given conditions: 
 \begin{enumerate}
   \item $S_c<0, \beta <\frac{1}{{S_c}^2}, \eta >1, k<\sqrt{\frac{-2 {S_c}^2+8 S_c-8}{\beta {S_c}^2-1}}$, or 
  \item $S_c<0, \beta =\frac{1}{{S_c}^2}, \eta >1, k\in \mathbb{R}$, or         
  \item $S_c=0, \beta \in \mathbb{R}, \eta>1, k<2 \sqrt{2}$, or 
  \item $0<S_c<\frac{4}{3}, \beta <\frac{1}{{S_c}^2}, 1< \eta <\frac{4-2 S_c}{S_c},  k<\sqrt{\frac{-2 {S_c}^2+8 S_c-8}{\beta  {S_c}^2-1}}$, or  
  \item $0<S_c<\frac{4}{3}, \beta =\frac{1}{{S_c}^2}, 1< \eta <\frac{4-2
  S_c}{S_c}, k\in \mathbb{R}$.
 \end{enumerate}

\item The following subsets (arcs, or specific equilibrium points) of the line $P_{11}^{+}(S_c)$ are unstable for the given conditions: 
\begin{enumerate}
   \item ${S_c}<0, \beta <\frac{1}{{S_c}^2}, \eta
   \geq 1, k>-\sqrt{2} \sqrt{-\frac{{S_c}^2-4 {S_c}+4}{\beta  {S_c}^2-1}}$, or 
   \item ${S_c}<0, \beta =\frac{1}{{S_c}^2}, \eta \geq 1, k\in \mathbb{R}$, or 
   \item ${S_c}=0, \beta \in \mathbb{R}, \eta \geq 1, k>-2 \sqrt{2}$, or 
  \item $0<{S_c}<\frac{4}{3}, \beta <\frac{1}{{S_c}^2}, 1\leq \eta <\frac{4-2{S_c}}{{S_c}}, k>-\sqrt{2} \sqrt{-\frac{{S_c}^2-4 {S_c}+4}{\beta  {S_c}^2-1}}$, or
  \item $0<{S_c}<\frac{4}{3}, \beta =\frac{1}{{S_c}^2}, 1\leq \eta
   <\frac{4-2 {S_c}}{{S_c}}, k\in \mathbb{R}$.
\end{enumerate}

\item The following subsets (arcs, or specific equilibrium points) of the line  $P_{11}^{-}(S_c)$ are stable for the given conditions: 
\begin{enumerate}
   \item ${S_c}<0, \beta <\frac{1}{{S_c}^2}, \eta
   \geq 1, k>-\sqrt{2} \sqrt{-\frac{{S_c}^2-4 {S_c}+4}{\beta  {S_c}^2-1}}$, or 
   \item ${S_c}<0, \beta =\frac{1}{{S_c}^2}, \eta \geq 1, k\in \mathbb{R}$, or 
   \item ${S_c}=0, \beta \in \mathbb{R}, \eta \geq 1, k>-2 \sqrt{2}$, or 
  \item $0<{S_c}<\frac{4}{3}, \beta <\frac{1}{{S_c}^2}, 1\leq \eta <\frac{4-2{S_c}}{{S_c}}, k>-\sqrt{2} \sqrt{-\frac{{S_c}^2-4 {S_c}+4}{\beta  {S_c}^2-1}}$, or
  \item $0<{S_c}<\frac{4}{3}, \beta =\frac{1}{{S_c}^2}, 1\leq \eta
   <\frac{4-2 {S_c}}{{S_c}}, k\in \mathbb{R}$.
\end{enumerate}

\item $P_{12}^{+}$ is non-hyperbolic with a 4D unstable for 
   \begin{enumerate}
   \item $\beta <0, \eta \geq 1, k>-\sqrt{2} \sqrt{\frac{(8 \beta -\eta -2)^2}{\beta  \left(16 \beta -\eta ^2-4 \eta -4\right)}}$, or 
    \item $\beta >\frac{9}{16}, 1\leq \eta <2 \left(2
   \sqrt{\beta }-1\right), k<\sqrt{2} \sqrt{\frac{(8 \beta -\eta -2)^2}{\beta  \left(16 \beta -\eta ^2-4 \eta -4\right)}}$. 
    \end{enumerate}

\item $P_{12}^{-}$ is non-hyperbolic with a 4D stable manifold for 
   \begin{enumerate}
   \item $\beta <0, \eta \geq 1, k>-\sqrt{2} \sqrt{\frac{(8 \beta -\eta -2)^2}{\beta  \left(16 \beta -\eta ^2-4 \eta -4\right)}}$, or 
    \item $\beta >\frac{9}{16}, 1\leq \eta <2 \left(2
   \sqrt{\beta }-1\right), k<\sqrt{2} \sqrt{\frac{(8 \beta -\eta -2)^2}{\beta  \left(16 \beta -\eta ^2-4 \eta -4\right)}}$. 
    \end{enumerate}
		
		\item  $P_{13}^{+}$ is non-hyperbolic with a 4D unstable manifold for
  \begin{enumerate}
  \item $\beta <0, \eta \geq 1, k<\sqrt{2} \sqrt{\frac{(8 \beta -\eta -2)^2}{\beta  \left(16 \beta -\eta ^2-4 \eta -4\right)}}$, or 
  \item $\beta >\frac{9}{16}, 1\leq \eta <2 \left(2
   \sqrt{\beta }-1\right), k>-\sqrt{2} \sqrt{\frac{(8 \beta -\eta -2)^2}{\beta  \left(16 \beta -\eta ^2-4 \eta -4\right)}}$.
  \end{enumerate}

\item $P_{13}^{-}$ is non-hyperbolic with a  4D  stable manifold for 
  \begin{enumerate}
  \item $\beta <0, \eta \geq 1, k<\sqrt{2} \sqrt{\frac{(8 \beta -\eta -2)^2}{\beta  \left(16 \beta -\eta ^2-4 \eta -4\right)}}$, or 
  \item $\beta >\frac{9}{16}, 1\leq \eta <2 \left(2
   \sqrt{\beta }-1\right), k>-\sqrt{2} \sqrt{\frac{(8 \beta -\eta -2)^2}{\beta  \left(16 \beta -\eta ^2-4 \eta -4\right)}}$.
  \end{enumerate}

\item $P_{16}^{+}$ is a source for 
\begin{enumerate}
\item $\eta \geq 1, 0<\beta <\frac{\eta +2}{8}, -\sqrt{\frac{\eta }{\beta }}<k<-2 \sqrt{2} \sqrt{\frac{\eta ^2}{-16 \beta +\eta ^2+4 \eta +4}}$, or
\item $\eta \geq 1, 0<\beta <\frac{\eta +2}{8}, 2 \sqrt{2} \sqrt{\frac{\eta ^2}{-16 \beta +\eta ^2+4 \eta +4}}<k<\sqrt{\frac{\eta }{\beta }}$.
\end{enumerate}

\item $P_{16}^{-}$ is a sink for 
\begin{enumerate}
\item $\eta \geq 1, 0<\beta <\frac{\eta +2}{8}, -\sqrt{\frac{\eta }{\beta }}<k<-2 \sqrt{2} \sqrt{\frac{\eta ^2}{-16 \beta +\eta ^2+4 \eta +4}}$, or
\item $\eta \geq 1, 0<\beta <\frac{\eta +2}{8}, 2 \sqrt{2} \sqrt{\frac{\eta ^2}{-16 \beta +\eta ^2+4 \eta +4}}<k<\sqrt{\frac{\eta }{\beta }}$.
\end{enumerate}

\item $P_{18}^{+}$ is a source  for 
\begin{enumerate}
\item $\eta \geq 1, \beta <0, k<-\sqrt{2} \sqrt{\frac{4 \beta -1}{\beta }}$, or
\item $\eta \geq 1, \beta <0, k>\sqrt{2} \sqrt{\frac{4 \beta -1}{\beta }}$, or
\item $\eta \geq 1, 0<\beta \leq \frac{\eta +2}{8}, k<-\sqrt{\frac{\eta }{\beta }}$, or
\item $\eta \geq 1, 0<\beta \leq \frac{\eta +2}{8}, k>\sqrt{\frac{\eta }{\beta }}$, or
\item $\eta \geq 1, \beta >\frac{\eta +2}{8}, k<-\sqrt{2} \sqrt{\frac{4 \beta -1}{\beta }}$, or 
\item $\eta \geq 1, \beta >\frac{\eta +2}{8}, k>\sqrt{2} \sqrt{\frac{4 \beta -1}{\beta }}$. 
\end{enumerate}

\item $P_{18}^{-}$ is a sink for 
\begin{enumerate}
\item $\eta \geq 1, \beta <0, k<-\sqrt{2} \sqrt{\frac{4 \beta -1}{\beta }}$, or
\item $\eta \geq 1, \beta <0, k>\sqrt{2} \sqrt{\frac{4 \beta -1}{\beta }}$, or
\item $\eta \geq 1, 0<\beta \leq \frac{\eta +2}{8}, k<-\sqrt{\frac{\eta }{\beta }}$, or
\item $\eta \geq 1, 0<\beta \leq \frac{\eta +2}{8}, k>\sqrt{\frac{\eta }{\beta }}$, or
\item $\eta \geq 1, \beta >\frac{\eta +2}{8}, k<-\sqrt{2} \sqrt{\frac{4 \beta -1}{\beta }}$, or 
\item $\eta \geq 1, \beta >\frac{\eta +2}{8}, k>\sqrt{2} \sqrt{\frac{4 \beta -1}{\beta }}$. 
\end{enumerate}

\end{enumerate}

\subsection{Universal horizons}

In the Einstein-aether theory there are spherical black hole solutions formed
by gravitational collapse for all viable parameter
values of the theory. However, due to the  Lorentz-violating nature of the theory, these solutions are quite different from the standard black holes in GR, since the broken Lorentz invariance completely modifies the causal structure of gravity, and the Killing horizon does not capture the notion of the causal boundary. Indeed, Lorentz-violating theories now 
admit superluminal excitations, which can cross the Killing horizon and escape to spatial infinity.  In some particular Lorentz-violating theories, like the Einstein-aether theory, the static, spherically-symmetric, black hole solutions contain a special hypersurface called the ``universal horizon'' that acts as a genuine causal boundary because it traps all excitations, even those which could be traveling at arbitrarily high velocities \cite{Barausse:2011pu,Newref1}. Consequently, still there is a causally disconnected region in black hole solutions but now being bounded by a universal horizon not far inside the metric horizon, so that a notion of black hole persists \cite{Barausse:2011pu,Newref5}.

For studying the causal structure of spacetimes with a causally preferred 
foliation, a framework was developed that allows 
for rigorously defined concepts such as black/white holes and also formalizes the notion of a universal horizon introduced previously in the simpler setting of static and spherically symmetric geometries \cite{Newref5}.
The question of what happens to the universal horizon in the extremal limit, where no such region exists 
any longer, has also been investigated \cite{Newref6}. 
In addition, Hawking radiation has been found to be associated with the universal horizon. 
These  absolute causal boundaries  are not Killing horizons but still obey a first law of black hole mechanics \cite{[12]}  and must consequently have an entropy if they do not violate a generalized second law. 
At these horizons, the Hawking radiation is thermal with 
a temperature proportional to its surface gravity. 
The viability of the first law (and hence a thermodynamical interpretation) has been studied
for several known exact universal horizon solutions  \cite{Newref4}
and calculations do, indeed, appear to predict the emission of a thermal flux  \cite{[14]}.

Therefore, there are absolute causal boundaries in gravitational theories with broken Lorentz invariance, in which there exists a surface located at a finite $r=r_{uh}$ called a universal horizon (and which always lies inside the Killing horizon) which acts like a one-way membrane, so that particles even with infinitely large speed cannot escape from it once they are inside it. 
In stationary spacetimes it has been shown that the universal horizon can be characterized by the local coordinate and gauge invariant condition 
\begin{equation}
u_a \xi^a=0 \; \text{at}\; r=r_{uh},
\end{equation}    
where $\xi^a$ denotes the asymptotically time-like Killing vector associated with stationarity and $u_a$ is the four-velocity of the aether \cite{Newref5,Lin:2017jvc}. Since $u_a$ is time-like by definition, the condition $u_a \xi^a=0$ can only be satisfied in the region of the spacetime where $\xi^a$ is spacelike. 

Unfortunately, the gauge and coordinates used in the qualitative analysis in this paper are not well suited for studying the possible existence of a universal horizon unless, due to a topological pathology, it is located at $r=0$ ($\lambda \rightarrow - \infty$) or $r\rightarrow \infty$ ($\lambda \rightarrow - \infty$) and characterized by one of the equilibrium points studied earlier. 
Here we are interested in $N(r_{uh})=0$ at finite  $r=r_{uh}$. Assuming that $N$ is analytic at $r=r_{uh}$, we can write $N=(r-r_{uh})^{n^2} f(r)$, where $f(r)=a^2+b (r-r_{uh})+ \ldots $ close to $r=r_{uh}$, so that  
\begin{equation}
y=r \frac{N'}{N}= \frac{n^2 r_{uh}}{(r-r_{uh})}+ \ldots \rightarrow \infty
\end{equation}
 as $r\rightarrow r_{uh}$. We can study the behavior of $N$ in terms of the variables $\sigma=y$ and $\theta=y-x$ subject to the constraint \eqref{static_sf} or the normalized variables  defined by eqn. \eqref{inv:qt}, e.g., 
\begin{equation}
S=\frac{\sigma}{\sqrt{\frac{\mu_0}{\eta}+\theta^2}}=\frac{1}{\sqrt{\frac{\mu_0}{\eta y^2}+\left(1-\frac{x}{y}\right)^2}},  
\end{equation}
and subject to the constraint
\begin{equation}
\label{C}
1-C-\beta S^2\geq 0,
\end{equation}
where $C\geq 0$, so that $Q, S, C$ are all bounded when $\beta>0$.
Combining all of the restrictions we find as $r\rightarrow r_{uh}$: 

\begin{enumerate}
\item  If $x\rightarrow c y$: then $(c-1)^2 -\beta >0$. If $c=1$, then $\beta$ is necessarily negative and $S\rightarrow \infty$. If $c\neq 1$, then $S\rightarrow S_0$, where $S_0$ is defined in terms of the other parameters, and $p+K$ diverges (and assuming that $p$ does not diverges at an horizontal horizon, this implies that $K\rightarrow \infty$). 
\item If $x\gg y$: $x^2\simeq p+K$ diverges and $S\rightarrow 0$ (and $C$ bounded with $C<1$); note that equilibrum point $P_3$ has $S=0, C=1$. 
\item If $y\gg x$: $S\rightarrow 1$, and $\beta\leq 1$; for the GR value $\beta=1$, $C\rightarrow 0$.
\end{enumerate}

\section*{Acknowledgments}

G. L. was funded by Comisi\'on Nacional de Investigaci\'on Cient\'{\i}fica y Tecnol\'ogica (CONICYT) through FONDECYT Iniciaci\'on grant no.
11180126 and by
Vicerrector\'ia de Investigaci\'on y Desarrollo Tecnol\'ogico at Universidad
Cat\'olica del Norte.  A. C. was supported by NSERC of Canada and partially supported by FONDECYT Iniciaci\'on grant no.
11180126. 
Andronikos Paliathanasis is acknowledged for useful comments.   
\begin{appendix}

\section{Regularity conditions}

In this appendix are summarized  some regularity conditions that must satisfy the relevant physical solutions, especially if they are expected to be used as star models. 

\subsection{Perfect fluid with linear equation of state}\label{regular_pf}

\subsubsection{Conditions for regularity at the origin and asymptotic flatness}

Using the coordinate change $(t, r)\rightarrow (t, \rho)$, where $\rho$ is a new radial coordinate, such that the line element 
\begin{equation}
ds^{2}=-N^{2}(r)dt^{2}+r^{-2}dr^{2}+K^{-1}(r)(d\vartheta ^{2}+\sin
^{2}\vartheta d\varphi ^{2}),  \label{met2A}
\end{equation}%
becomes
\begin{equation}
ds^{2}=-N(r(\rho))^2 dt^{2}+\frac{d{\rho}^{2}}{1-\frac{2 m(\rho)}{\rho}}+\rho^2 (d\vartheta ^{2}+\sin
^{2}\vartheta d\varphi ^{2}),  \label{met2B}
\end{equation}%
we have the identifications
\begin{align}
&K=\rho^{-2}, \\
&\frac{d r}{r}=\frac{d \rho}{\sqrt{1-\frac{2m(\rho)}{\rho}}}, \label{EQ_54}
\end{align}
where $m(\rho)$ denotes the mass up to the radius $\rho$.

Then, we define the  Misner-Sharp mass \cite{Misner:1964je}
\begin{align}
\mathcal{M}(\rho):=\frac{m(\rho)}{\rho}=\frac{C-(Q-S)^2}{2 C}.
\end{align}

As a first approach, we impose regularity at the center, that is, as $\lambda\rightarrow -\infty$, by extrapolating the conditions for relativistic stars as given by the Buchdahl inequalities \cite{Buchdahl:1959zz,hartle1978}, which in units where $8\pi G=1$ are expressed as
\begin{equation}
\label{Buchdahl_cond}
\mathcal{M}\geq \frac{1}{6}\rho^2 \mu, \quad 
 \mathcal{M}\leq \frac{1}{6}\rho^2 \mu_c, \quad 
 \mathcal{M}\leq \frac{2}{9}\left(1-\frac{3}{4} \rho^2 p+\sqrt{1+\frac{3}{4} \rho^2 p}\right),
\end{equation}
where $\mu_c$ is the energy density at the center of the star and $\rho$ is a radial variable. \\
To find the generalized regularity conditions we have to integrate the
full equations which determine the star's structure and the geometry in the static spherically symmetric Einstein-aether theory for a perfect fluid starting from the center $\rho=0$ with central density $\mu_c$, out to the surface $\rho=\rho*$ where the pressure vanishes. That is, we have to consider the boundary conditions 

\begin{equation}
 p(0)=p(\rho_c)=p_c, \quad 
 m(0)=0, \quad e^{2 \phi(\rho_*)}=1-2 \frac{m(\rho_*)}{\rho_*}
\end{equation}
and follow the same strategy to find estimates for the mass as in \cite{Buchdahl:1959zz,hartle1978}. 
Because we have assumed the linear equation of state $\mu=\mu_{0} +(\eta-1)p,$ the energy density at the surface of zero pressure is $\mu_0$.
The central energy density and central pressure are related through $\mu_c=\mu_{0} +(\eta-1)p_c$.

Notice the additional relations for the radial coordinate;
\begin{equation}
\rho^2=\frac{\eta(1-Q^2)}{\mu_0 C},
\end{equation}
and for the matter energy density:
\begin{equation}
\mu:=\mu_0+(\eta-1)p=
\frac{\mu_0 \left(\eta -Q^2- (\eta -1) (C+\beta S^2)\right)}{\eta  \left(1-Q^2\right)}.
\end{equation}

\begin{enumerate}
\item
Thus,  the Buchdahl conditions \eqref{Buchdahl_cond} can be expressed  in terms of the variables $Q, S, C$, as 
{\small{
\begin{align}
& \frac{\eta -Q^2- (\eta -1) (C+\beta S^2)}{3}\leq  C-(Q-S)^2 \nonumber\\
\leq 
&\min\left\{\frac{\eta(1-Q^2\left(\mu_0+(\eta-1)p_{c}\right))}{3 \mu_0} ,  \frac{1}{9}\left(\left(7 C-3 Q^2+3 \beta  S^2\right)+2\sqrt{ C \left(C+3 Q^2-3 \beta  S^2\right)}\right)\right\}, \label{Condition36}
\end{align}
}}
as $\lambda\rightarrow -\infty$,
where $p_c$ is the central pressure of the star.

\item Asymptotic flatness as $\rho \rightarrow +\infty$ or, equivalently, as $\lambda \rightarrow +\infty$: 
\begin{equation}
\label{asymp_flatness}
\lim_{\lambda \rightarrow +\infty}  [C-(Q-S)^2]=0, \quad 
\lim_{\lambda \rightarrow +\infty}  [Q^2-C-\beta S^2]=0. 
\end{equation}
 The first condition corresponds to $\lim_{\lambda \rightarrow + \infty} {\cal M}=0$. 
The second condition implies from \eqref{EQ-40}  that $$\lim_{\lambda\rightarrow +\infty} e^{\phi}=\alpha,$$ and that the surface of zero pressure is reached. The constant $\alpha$ is absorbed by a time redefinition.  This means that asymptotically we obtain the Minkowski metric.

\item From the relations \eqref{rel-30}, vacuum ($\mu_0=0, \mu=0, p=0$) corresponds to 
\begin{equation}
Q^2=1, \quad 
C+\beta S^2=1. 
\end{equation}

\end{enumerate}

\subsubsection{Stars}

 To obtain physically reasonable spherically symmetric models with non-negative pressure one matches each interior solution with the exterior Schwarzschild vacuum solution \begin{equation}\label{eq_57A}
ds^{2}=-\left(1-\frac{2M}{\rho}\right) dt^{2}+\frac{d{\rho}^{2}}{\left(1-\frac{2M}{\rho}\right)}+\rho^2 (d\vartheta ^{2}+\sin
^{2}\vartheta d\varphi ^{2}).
\end{equation}
when the radius, $\rho_*=\sqrt{\frac{\eta(1-{Q_*}^2)}{\mu_0 ({Q_*}^2-\beta {S_*}^2)}}$, where the pressure becomes zero \footnote{
In our set up, the solutions in their way from $Q=+1$ to $Q=-1$, all intersect the surface of vanishing pressure $C+\beta S^2-Q^2=0$ at an interior point $(Q_*, S_*,C_*)$.}.
When the radio $\rho_*$ is reached, $\frac{1-{Q_*}^2}{1-{C_*}-\beta {S_*}^2}=1$,  $e^\phi=\alpha$. This fixes $\alpha=\sqrt{1-2 M/R}=\sqrt{\frac{(Q_*-S_*)^2}{Q_*^2-\beta  S_*^2}}$, where $M$ is the total mass of the star as given by 
\begin{equation}
M=\frac{S_{*} (2 Q_{*}-(\beta+1)  S_{*})}{2 \left(Q_{*}^2-\beta  S_{*}^2\right)} \sqrt{\frac{\eta(1-{Q_*}^2)}{\mu_0 ({Q_*}^2-\beta {S_*}^2)}}.
\end{equation}
The interior solution (evaluated at the surface of zero pressure)
\begin{equation}
ds^{2}=-\frac{(Q_*-S_*)^2}{{Q_*}^2-\beta {S_*}^2}  dt^{2}+\frac{{Q_*}^2-\beta {S_*}^2}{(Q_*-S_*)^2}d{\rho}^{2}+\rho^2 (d\vartheta ^{2}+\sin
^{2}\vartheta d\varphi ^{2}).  \label{met2B2B}
\end{equation}  
is matched at $\rho=\rho_*$ with the static vacuum spacetime described by \eqref{eq_57A}.

\section{Equilibrium points in the finite region of the phase space for a perfect fluid with linear equation of state}\label{phys-intepretation}

We now try to find some
asymptotic expansions for all the equilibrium points of \eqref{inv:qt_evol}. By convenience, we introduce the radial rescaling 
$r=e^{\ell}$, such that $\ell\rightarrow -\infty$ as $r\rightarrow 0$ and $\ell\rightarrow \infty$ as $r\rightarrow \infty$. Hence,  Eq.  \eqref{met2A} becomes
\begin{equation}
ds^{2}=-N^{2}({\ell})dt^{2}+d{\ell}^{2}+K^{-1}(\ell)(d\vartheta ^{2}+\sin
^{2}\vartheta d\varphi ^{2}),  \label{met2A2}
\end{equation}%

The equilibrium points of the system \eqref{inv:qt_evol} are 
\begin{enumerate}
\item $P_1^{\pm}: (Q,S,C)=(\pm 1, \pm 1, 0)$ exist for $\beta=1$ or $\beta=\frac{\eta +2}{4}\leq 0$. For $\beta=1, 1\leq\eta<2$, $P_1^{+}$ (respectively, $P_1^{-}$) is a source (respectively, a sink); for $\beta=1, \eta>2$, $P_1^{\pm}$ are saddles and for $\beta=\frac{\eta +2}{4}\leq 0$, $P_1^{\pm}$ are saddles.   On substitution of the values of $Q$ and $S$ into \eqref{eqN}, \eqref{eqK} and integration we obtain $N=\bar{N}_0e^{\pm\lambda}$ and $K=\bar{K}_0 = \text{constant}$.  After we evaluate the values of $Q,S,C$ in \eqref{eqr} and \eqref{eqy}, it follows that $\frac{d{\ell}}{d\lambda}=\pm \frac{1}{y}$ and $\frac{d y}{d\lambda}=\left\{\begin{array}{cc}
\mp 1, & \beta=1\\
\mp \frac{\eta}{2}, & \beta=\frac{\eta +2}{4}
\end{array}.\right.$ Then, $y(\lambda )= \left\{\begin{array}{cc}
c_1 e^{-\lambda  \epsilon }, & \beta=1\\
c_1 e^{-\frac{1}{2} \eta  \lambda  \epsilon }, & \beta=\frac{\eta +2}{4}
\end{array},\right.  {\ell}(\lambda )=  \left\{\begin{array}{cc}
\frac{e^{\lambda  \epsilon }}{c_1}+c_2, & \beta=1\\
\frac{2 e^{\frac{\eta  \lambda  \epsilon }{2}}}{c_1 \eta }+c_2, & \beta=\frac{\eta +2}{4}
\end{array},\right.$ where $\epsilon=\pm 1$ and $c_1, c_2$ are constants of integration integration. For $\beta=1$ the metric becomes $ds^2= -\bar{N}_0^2 e^{\pm 2\lambda}
dt^2+ \frac{e^{\pm 2\lambda}}{c_1^2}d\lambda^2 +{\bar{K}_0}^{-1}(d\vartheta ^{2}+\sin ^{2}\vartheta d\varphi ^{2})$ or $ds^2= -\bar{N}_0^2 c_1^2 \rho_\pm^2
dt^2+ d\rho_\pm^2 +{\bar{K}_0}^{-1}(d\vartheta ^{2}+\sin ^{2}\vartheta d\varphi ^{2})$ under the coordinate transformation $\rho_\pm=\frac{e^{\pm \lambda}}{c_1}$.
They correspond to $P_1^{\pm}$ in  \cite{Nilsson:2000zf}, which are the Kasner's plane-symmetric vacuum solutions \cite{kasner}. For  $\beta=\frac{\eta +2}{4}\leq 0$ the metric becomes $ds^2= -\bar{N}_0^2 e^{\pm 2\lambda}
dt^2+ \frac{e^{\pm \eta \lambda}}{c_1^2}d\lambda^2 +{\bar{K}_0}^{-1}(d\vartheta ^{2}+\sin ^{2}\vartheta d\varphi ^{2})$ or $ds^2= -\bar{N}_0^2 c_1^{\frac{4}{\eta}}\rho_\pm^{\frac{4}{\eta}}
dt^2+ \frac{4}{\eta^2}d\rho_\pm^2 +{\bar{K}_0}^{-1}(d\vartheta ^{2}+\sin ^{2}\vartheta d\varphi ^{2})$ under the coordinate transformation $\rho_\pm=\frac{e^{\pm\frac{\eta}{2} \lambda}}{c_1}$. In this case the Ricci scalar is $R=\frac{(\eta-2)}{\rho_\pm ^2}+2 \bar{K}_0$. Thus for $\eta\neq 2$ there is a naked singularity at $\rho_\pm=0^+$.
\item  $P_2^{\pm}: (Q,S,C)=(\pm 1, \mp 1, 0)$ exist for $\beta=1$ or $4 \beta +\eta +2=0,  \beta \leq 1$.  For $\beta=1, \eta\geq 1$, $P_2^{+}$ (respectively, $P_2^{-}$) is a source (respectively, a sink). For $4 \beta +\eta +2=0,  \beta \leq 1, \eta\geq 1$, $P_2^{+}$ (respectively, $P_2^{-}$) is a sink (respectively, a source). On substitution of the values of $Q$ and $S$ into \eqref{eqN}, \eqref{eqK} and integration we obtain $N=\bar{N}_0e^{\mp\lambda}$ and $K=\bar{K}_0 e^{\mp 4\lambda}$. After we evaluate the values of $Q,S,C$ in \eqref{eqr} and \eqref{eqy}, it follows that $\frac{d{\ell}}{d\lambda}=\mp \frac{1}{y}$ and $\frac{d y}{d\lambda}=\left\{\begin{array}{cc}
\mp 3, & \beta=1\\
\pm \frac{\eta}{2}, & 4 \beta +\eta +2=0
\end{array}.\right.$ Then, $y(\lambda )= \left\{\begin{array}{cc}
c_1 e^{-3 \lambda  \epsilon }, & \beta=1\\
c_1 e^{\frac{\eta  \lambda  \epsilon }{2}}, & 4 \beta +\eta +2=0
\end{array},\right.$\\  
${\ell}(\lambda )=  \left\{\begin{array}{cc}
c_2-\frac{e^{3 \lambda  \epsilon }}{3 c_1}, & \beta=1\\
\frac{2 e^{-\frac{1}{2} \eta  \lambda  \epsilon }}{c_1 \eta }+c_2, & 4 \beta +\eta +2=0
\end{array},\right.$ where $\epsilon=\pm 1$ and $c_1, c_2$ are constants of integration. For $\beta=1$ the metric becomes\\$ds^2= -\bar{N}_0^2 e^{\mp 2\lambda}
dt^2+ \frac{e^{\pm 6\lambda}}{c_1^2}d\lambda^2 +{\bar{K}_0}^{-1} e^{\pm 4\lambda}(d\vartheta ^{2}+\sin ^{2}\vartheta d\varphi ^{2})$ or $ds^2= -\frac{\bar{N}_0^2}{\rho_\pm}
dt^2+ \frac{\rho_\pm}{4 c_1^2}  d\rho_\pm^2 +\frac{\rho_\pm^2}{\bar{K}_0}(d\vartheta ^{2}+\sin ^{2}\vartheta d\varphi ^{2})$ under the coordinate transformation $\rho_\pm=e^{\pm 2 \lambda}$. The Ricci Scalar becomes $R=\frac{2 \bar{K}_0}{\rho_\pm^2}$.  Thus at $\rho_\pm=0^+$ we have a singularity.  The equilibrium points  $P_2^{\pm}$ correspond to $P_2^\pm$ in  \cite{Nilsson:2000zf}, which are the Kasner's plane-symmetric vacuum solutions \cite{kasner}. For $4 \beta +\eta +2=0$ the metric becomes $ds^2= -\bar{N}_0^2 e^{\mp 2\lambda}
dt^2+ \frac{e^{\mp \eta \lambda}}{c_1^2}d\lambda^2 +{\bar{K}_0}^{-1} e^{\pm 4\lambda}(d\vartheta ^{2}+\sin ^{2}\vartheta d\varphi ^{2})$ or $ds^2= -\frac{\bar{N}_0^2}{\rho_\pm}
dt^2+ \frac{\rho_\pm^{-\frac{(4+\eta)}{2}}}{4 c_1^2}  d\rho_\pm^2 +\frac{\rho_\pm^2}{\bar{K}_0}(d\vartheta ^{2}+\sin ^{2}\vartheta d\varphi ^{2})$ under the coordinate transformation $\rho_\pm=e^{\pm 2 \lambda}$. The Ricci Scalar becomes $R=\frac{2 \bar{K}_0}{\rho_\pm^2}-\frac{3}{4}
   c_1^2 (\eta+6) \rho_\pm ^{\eta/2}$.  Thus at $\rho_\pm=0^+$ and at $\rho_\pm=+\infty$ we have singularities.
\item $P_3^{\pm}: (Q,S,C)= 	(\pm 1, 0, 1)$, always exist and are saddles.	 When we substitute the values of $Q$ and $S$ into \eqref{eqN}, \eqref{eqK} and integrate, we obtain $N=\bar{N}_0=\text{constant}$ and $K=\bar{K}_0 e^{\mp 2\lambda}.$ From  \eqref{eqr} and the definition of $C$ it follows that $d{\ell}^2=K^{-1}d\lambda^2$. Thus the line element \eqref{met2} becomes $ds^{2}=-\bar{N}_0^{2}dt^{2}+\bar{K}_0^{-1} e^{\pm 2\lambda}d\lambda^2 +\bar{K}_0^{-1} e^{\pm 2\lambda}\left(d\vartheta ^{2}+\sin ^{2}\vartheta d\varphi ^{2}\right)$. Defining $\rho_\pm=\frac{e^{\pm \lambda}}{\sqrt{\bar{K}_0}}$, we get $ds^{2}=-\bar{N}_0^{2}dt^{2}+d\rho_\pm^2 +\rho_\pm^2\left(d\vartheta ^{2}+\sin^{2}\vartheta d\varphi ^{2}\right)$, which corresponds to Minkowski spacetime on explicitly spherically symmetric form  \cite{Nilsson:2000zf}. These points are the analogues of $P_3^{\pm}$ investigated in \cite{Nilsson:2000zf}.
\item $P_4^{\pm}: (Q,S,C)= 	\left(\pm 1, \pm\frac{2}{\eta +2},1-\frac{8 \beta}{(\eta +2)^2}\right)$ exist for	$\eta \geq 1, 0\leq \beta \leq \frac{1}{8} (\eta +2)^2$ and are saddles. 	When we substitute the values of $Q$ and $S$ into \eqref{eqN}, \eqref{eqK} and integrate, we obtain $N=\bar{N}_0e^{\pm \frac{2\lambda}{2+\eta}}$ and $K=\bar{K}_0 e^{\mp \frac{2\eta\lambda}{2+\eta}}$. From  \eqref{eqr} and the definition of $C$ it follows that $d{\ell}^2=\frac{1}{\bar{K}_0}\left(1-\frac{8 \beta}{(\eta +2)^2}\right)e^{\pm \frac{2\eta\lambda}{2+\eta}}d\lambda^2$. Thus the line element \eqref{met2} becomes\\$ds^{2}=-\bar{N}_0^{2}e^{\pm \frac{4\lambda}{2+\eta}}dt^{2}+\frac{1}{\bar{K}_0}\left(1-\frac{8\beta}{(\eta +2)^2}\right)e^{\pm \frac{2\eta\lambda}{2+\eta}}d\lambda^2 +\frac{1}{\bar{K}_0} e^{\pm \frac{2\eta\lambda}{2+\eta}}\left(d\vartheta ^{2}+\sin ^{2}\vartheta d\varphi ^{2}\right)$. Defining \\$\rho_\pm=\frac{1}{\sqrt{\bar{K}_0}} e^{\pm \frac{\eta\lambda}{2+\eta}}, T=\bar{N}_0\bar{K}_0^{\frac{1}{\eta}} t$, we obtain\\$ds^{2}=-\rho_\pm^{\frac{4}{\eta}}dT^{2}+\left(\frac{(\eta +2)^2-8 \beta }{\eta ^2}\right)d\rho_\pm^2 +\rho_\pm^2 \left(d\vartheta ^{2}+\sin ^{2}\vartheta d\varphi ^{2}\right)$. For $\beta=1$ these points are the analogues of $P_4^{\pm}$ investigated in \cite{Nilsson:2000zf}, which corresponds to a nonregular self-similar perfect fluid solution discussed in \cite{Tolman:1939jz,Oppenheimer:1939ne,Misner:1964zz}. One interesting feature is that following the ARS algorithm, the dominant terms found in the previous part of this Section correspond to the points $P_4^{\pm}$.
\item $P_5^{\pm}: (Q,S,C)= \left(\pm 1,\pm\frac{\eta +2}{4 \beta},0\right)$ exist for $\eta\geq 1, \beta<0$ or $\eta >1, 16 \beta \geq (\eta +2)^2$. $P_5^{+}$ (respectively, $P_5^{-}$) is a sink (respectively, a source) for  $\eta \geq 1, \beta<0$.   Otherwise they are saddles. When we substitute the values of $Q$ and $S$ into \eqref{eqN}, \eqref{eqK} and integrate, we obtain $N=\bar{N}_0e^{\pm \frac{(\eta+2)\lambda}{4\beta}}$ and $K=\bar{K}_0 e^{\mp 2 \left(1-\frac{\eta+2}{4\beta}\right)\lambda}$. From  \eqref{eqr} and the definition of $S$ it follows that $\frac{d{\ell}}{d\lambda}=\pm\frac{\eta +2}{4 \beta y}$. On the other hand, when we evaluate the values of $Q,S,C$ in \eqref{eqy}, it follows that $\frac{d y}{d\lambda}=\mp \frac{\eta(\eta+2)}{8\beta}y$. Then we have $y (\lambda )= c_ 1 e^{-\frac{\eta  (\eta +2) \lambda  \epsilon }{8\beta}}, {\ell} (\lambda )= \frac{2 e^{\frac{\eta  (\eta +2) \
\lambda  \epsilon }{8 \beta}}}{c_ 1 \eta }+c_ 2$, where $\epsilon=\pm 1$ and $c_1, c_2$ are constants of integration.
The metric \eqref{met2} becomes\\$ds^{2}=-\bar{N}_0^2 e^{\pm\frac{(\eta+2)\lambda}{2\beta}}dt^{2}+\frac{(\eta +2)^2 e^{\pm\frac{\eta  (\eta +2) \lambda}{4 \beta}}}{16 c_1^2 \beta^2}d\lambda^2+{\bar{K}_0}^{-1} e^{\pm 2 \left(1-\frac{\eta+2}{4\beta}\right)\lambda}(d\vartheta ^{2}+\sin
^{2}\vartheta d\varphi ^{2})$. On the introduction of $\rho_\pm=e^{\pm\left(1-\frac{\eta+2}{4\beta}\right)\lambda}$ the line element becomes\\$-\bar{N}_0^2\rho_\pm ^{\frac{2 (\eta +2)}{4 \beta -\eta -2}}dt^2+ \frac{(\eta+2)^2}{2 c_1^2 (4\beta-\eta-2)^2}\rho_\pm ^{-\frac{16\beta-\eta ^2-6 \eta +24}{2 (4\beta-\eta-2)}}d\rho_\pm^2+\frac{\rho_\pm^2}{\bar{K}_0}(d\vartheta ^{2}+\sin^{2}\vartheta d\varphi ^{2})$. For $\beta=1$ these solutions are the analogues of $P_5^{\pm}$ investigated in \cite{Nilsson:2000zf}, which correspond to self-similar plane-symmetric perfect fluid models. For $\eta \geq 1, 16 \beta \geq (\eta +2) (\eta +4)$ the exponent of the  $tt$   component is positive and the exponent of the $\rho_\pm\rho_\pm$ component is negative.  Thus the singularity has an horizon at  $\rho_\pm= 0^+$ .
\item	$P_6^{\pm}: (Q,S,C)= \left(\pm 1, \pm \frac{1}{\sqrt{\beta}}, 0\right)$. They exist for $\eta\geq 1, \beta>0$. $P_6^{+}$ (respectively, $P_6^{-}$) is a source (respectively, a sink) for $\eta \geq 1, 
16 \beta \geq (\eta +2)^2$. They are saddles for  $\eta \geq 1, 0<\beta< \frac{1}{4}$ or $\eta \geq 1, \frac{1}{4}<\beta<\frac{1}{16} (\eta +2)^2$.  Otherwise they are non-hyperbolic.  When we substitute the values of $Q$ and $S$ into \eqref{eqN}, \eqref{eqK} and integrate, we obtain $N=\bar{N}_0e^{\pm \frac{\lambda}{\sqrt{\beta}}}$ and $K=\bar{K}_0 e^{\mp \frac{2 \left(\sqrt{\beta}-1\right)\lambda}{\sqrt{\beta}}}$. From  \eqref{eqr} and the definition of $S$ it follows that $\frac{d{\ell}}{d\lambda}=\pm \frac{1}{y \sqrt{\beta}}$. On the other hand, when we evaluate the values of $Q,S,C$ in \eqref{eqy}, it follows that $\frac{d y}{d\lambda}=\pm \frac{1-2 \sqrt{\beta}}{\sqrt{\beta}}y$. Then $y(\lambda )= c_1 e^{\left(-\frac{1}{\sqrt{\beta}}-2\right) \lambda  \epsilon }, {\ell}(\lambda )= c_2-\frac{e^{\left(\frac{1}{\sqrt{\beta}}+2\right) \lambda  \epsilon }}{c_1 \left(2 \sqrt{\beta}+1\right)}$, where $\epsilon=\pm 1$ and $c_1, c_2$ are constants of integration. The line element \eqref{met2} becomes
$ds^2= -\bar{N}_0^2 e^{\pm \frac{2\lambda}{\sqrt{\beta}}}
dt^2+ \frac{ e^{\pm\frac{2 \left(2 \sqrt{\beta}+1\right) \lambda}{\sqrt{\beta}}}}{c_1^2 \beta}{d\lambda}^2 +{\bar{K}_0}^{-1} e^{\pm \frac{2 \left(\sqrt{\beta}-1\right)\lambda}{\sqrt{\beta}}}(d\vartheta ^{2}+\sin
^{2}\vartheta d\varphi ^{2}).$ Under the transformation $\rho_\pm=e^{\pm \frac{\left(\sqrt{\beta}-1\right)\lambda}{\sqrt{\beta}}}$, the metric becomes
$ds^2= -\bar{N}_0^2 \rho_\pm^{\frac{2(1+\sqrt{\beta})}{\beta-1}}
dt^2+ \frac{ \rho_\pm^{\frac{2(2+\beta+3\sqrt{\beta})}{\beta-1}}}{c_1^2 (1+\beta-2\sqrt{\beta})}{d\rho_\pm}^2 +\frac{\rho_\pm^2}{\bar{K}_0}(d\vartheta ^{2}+\sin
^{2}\vartheta d\varphi ^{2}).$ Because the exponents of the $tt$ and $\rho_\pm\rho_\pm$ components are of the same sign for $\beta>0$, $\rho_\pm= 0^+$ is a naked singularity. For $\beta=1$ these points correspond to $P_1^{\pm}$ in  \cite{Nilsson:2000zf}, which are Kasner's plane-symmetric vacuum solutions \cite{kasner}.
\item $P_7^{\pm}: (Q,S,C)= \left(\pm 1, \mp \frac{1}{\sqrt{\beta}}, 0\right)$ exist for $\eta\geq 1, \beta>0$. $P_7^{+}$ (respectively, $P_7^{-}$) is a source (respectively, a sink) for $\eta \geq 1, \beta>0$.  When we substitute the values of $Q$ and $S$ in \eqref{eqN}, \eqref{eqK} and integrate, we obtain $N=\bar{N}_0e^{\mp \frac{\lambda}{\sqrt{\beta}}}$ and $K=\bar{K}_0 e^{\mp \frac{2 \left(1+\sqrt{\beta}\right)\lambda}{\sqrt{\beta}}}$. From  \eqref{eqr} and the definition of $S$ it follows that $\frac{d{\ell}}{d\lambda}=\mp \frac{1}{y \sqrt{\beta}}$. On the other hand, when we evaluate the values of $Q,S,C$ in \eqref{eqy}, it follows that $\frac{d y}{d\lambda}=\mp \frac{1+2 \sqrt{\beta}}{\sqrt{\beta}}y$. Then $y(\lambda )= c_1 e^{\left(-\frac{1}{\sqrt{\beta}}-2\right) \lambda  \epsilon }, {\ell}(\lambda )= c_2-\frac{e^{\left(\frac{1}{\sqrt{\beta}}+2\right) \lambda  \epsilon }}{c_1 \left(2\sqrt{\beta}+1\right)}$, where $\epsilon=\pm 1$. The line element \eqref{met2} becomes $ds^{2}=-\bar{N}_0^2e^{\mp \frac{2\lambda}{\sqrt{\beta}}}dt^{2}+\frac{e^{\pm \frac{2 \left(1+2\sqrt{\beta}\right)\lambda}{\sqrt{\beta}}}}{c_1^2 \beta}d\lambda^2 +{\bar{K}_0}^{-1}  e^{\pm \frac{2 \left(1+\sqrt{\beta}\right)\lambda}{\sqrt{\beta}}}(d\vartheta ^{2}+\sin^{2}\vartheta d\varphi ^{2})$. Under the change of variables $\rho_\pm=e^{\pm \frac{\left(1+\sqrt{\beta}\right)\lambda}{\sqrt{\beta}}}$, the metric becomes
$-\bar{N}_0^2\rho_\pm^{\frac{2(1-\sqrt{\beta})}{\beta-1}}dt^2+\frac{\rho_\pm^{\frac{2(\beta-\sqrt{\beta})}{\beta-1}}}{c_1^2(1+\beta+2 \sqrt{\beta})}d\rho_\pm^2+\frac{\rho_\pm^2}{\bar{K}_0}(d\vartheta ^{2}+\sin^{2}\vartheta d\varphi ^{2})$.  As the exponents of the $tt$ and $\rho_\pm\rho_\pm$ components are both negative for $\beta>0$, $\rho_\pm= 0^+$ is a naked singularity. For $\beta=1$ these points correspond to $P_2^\pm$ in  \cite{Nilsson:2000zf} and these are the Kasner's plane-symmetric vacuum solutions \cite{kasner}.
\item $P_8^{\pm}: (Q,S,C)=\left(\pm\frac{1}{\sqrt{1-\beta}}, \pm\frac{1}{\sqrt{1-\beta}},\frac{2 \beta -1}{\beta -1}\right)$. They exist for $\beta\leq 0$. 
$P_8^{+}$ (respectively, $P_8^{-}$) is a sink (respectively, a source) for $\beta<0$. When we substitute the values of $Q$ and $S$ into \eqref{eqN}, \eqref{eqK} and integrate, we obtain $N=\bar{N}_0e^{\pm \frac{\lambda}{\sqrt{1-\beta}}}$, and $K=\bar{K}_0 = \text{constant}$. From  \eqref{eqr} and the definition of $Q$ it follows that $d{\ell}^2=\frac{\eta}{\mu_0}\left(\frac{\beta }{\beta -1}\right)d\lambda^2$. The metric becomes $ds^2= -\bar{N}_0^2 e^{\pm \frac{2\lambda}{\sqrt{1-\beta}}}
dt^2+ \frac{\eta}{\mu_0}\left(\frac{\beta }{\beta -1}\right)d\lambda^2 +{\bar{K}_0}^{-1}(d\vartheta ^{2}+\sin
^{2}\vartheta d\varphi ^{2}).$ These are regular solutions with constant curvature $R=2 \left(\frac{\mu_0}{\beta \eta}+\bar{K}_0\right)$ for $\beta<0$. 
\item 	$P_9^{\pm}: (Q,S,C)= \left(\pm 2 \sqrt{\beta}, \pm \frac{1}{\sqrt{\beta}}, 0\right)$ exist for $\eta \geq 1, 0<\beta \leq \frac{1}{4}$. They are saddles for $\eta \geq 1, 0<\beta \leq \frac{1}{4}$ and non-hyperbolic for $\beta=\frac{1}{4}$ (numerically it is the saddle in Fig. \ref{fig:Fig1f}).  When we substitute the values of $Q$ and $S$ into \eqref{eqN}, \eqref{eqK} and integrate, we obtain $N=\bar{N}_0e^{\pm \frac{\lambda}{\sqrt{\beta}}}$ and $K=\bar{K}_0 e^{\mp \frac{2 \left(2\beta-1\right)\lambda}{\sqrt{\beta}}}$. For $\beta<\frac{1}{4}$ it follows from \eqref{eqr} and the definition of $Q$ that $d{\ell}^2=\frac{\eta}{\mu_0}(1-4\beta)d\lambda^2$.
The line element \eqref{met2} becomes $ds^{2}=-\bar{N}_0^2e^{\pm \frac{2\lambda}{\sqrt{\beta}}}dt^{2}+\frac{\eta}{\mu_0}(1-4\beta)d\lambda^2+{\bar{K}_0}^{-1} e^{\pm \frac{2 \left(2\beta-1\right)\lambda}{\sqrt{\beta}}}(d\vartheta ^{2}+\sin^{2}\vartheta d\varphi ^{2})$. Under the change of variables $\rho_\pm=\frac{1}{\sqrt{\bar{K}_0}} e^{\pm \frac{\left(2\beta-1\right)\lambda}{\sqrt{\beta}}}, T=\bar{N}_0{\bar{K}_0}^{\frac{1}{2(2\beta-1)}}t$, the metric becomes $ds^{2}=-\rho_\pm^{\frac{2}{2\beta-1}}dT^{2}+\frac{\eta(1-4\beta)\beta}{\mu_0(2\beta-1)^2}\frac{d\rho_\pm^2}{\rho_\pm^2}+\rho_\pm^2(d\vartheta ^{2}+\sin^{2}\vartheta d\varphi ^{2})$. Because the exponents of the $TT$ and $\rho_\pm\rho_\pm$ components are both negative for $0<\beta< \frac{1}{4}$, $\rho_\pm= 0^+$ is a naked singularity.
\end{enumerate}

\section{Equilibrium points in the finite region of the phase space for the exponential potential}\label{phys-intepretation2}

The system admits  the equilibrium points $P_1^{\pm}$-$P_9^{\pm}$  discussed before in the invariant set $A_\varphi=A_{W}=0$. 
For further details about the  physical interpretation of the equilibrium points $P_1^{\pm}$-$P_9^{\pm}$ we submit the reader to Appendix
\ref{phys-intepretation}, where we have represented the line elements of their corresponding cosmological solutions.
The stability conditions change slightly due to the new axis $A_\varphi$, $A_{W}$. 
 \begin{enumerate}

\item $P_1^{\pm}$ as discussed in the previous section.

\item  $P_2^{\pm}$ as discussed in the previous section.

\item $P_3^{\pm}:=\left(\pm 1, 0, 1, 0, 0\right)$. They always exist and are saddle since the eigenvalues are:\\ $\mp 1, \mp 1, \pm 1, \pm 2, \pm 2$.

\item  $P_4^{\pm}:=\left(\pm 1, \pm \frac{2}{2+\eta }, 1-\frac{8 \beta}{(2+\eta )^2}, 0, 0\right)$. Exists for $0\leq  \beta \leq \frac{1}{8} (\eta +2)^2, \eta \geq 1$. 
\\ The eigenvalues are $\pm \frac{\eta }{\eta +2}, \pm \frac{2 \eta }{\eta +2}, \mp 1, \mp\frac{\eta +2+\sqrt{64 \beta -7 (\eta +2)^2}}{2 (\eta +2)}, \mp \frac{\eta +2-\sqrt{64 \beta -7 (\eta +2)^2}}{2 (\eta +2)}$. 
This point is a saddle (at least two eigenvalues have different signs).

\item  $P_5^{\pm}:=\left(\pm 1, \pm \frac{2+\eta }{4\beta}, 0, 0, 0\right)$. Exist for $\eta\geq 1,\beta<0$ or $\eta \geq 1,  \beta \geq \frac{1}{16}(\eta +2)^2$.
\\ The eigenvalues are  $\pm\frac{\eta  (\eta +2)}{8 \beta }, \pm \frac{(\eta +2)^2-8 \beta }{4 \beta }, \pm \frac{(\eta +2)^2-16 \beta }{8 \beta }, \pm \frac{(\eta +2)^2-16 \beta }{8 \beta }, \pm \frac{\eta  (\eta +2)}{4 \beta }$.
$P_5^{+}$ (respectively, $P_5^{-}$) is a sink (respectively, a source), for $\beta<0, \eta \geq 1$. 
Otherwise, they are saddles. 

\item  $P_6^{\pm}$ as discussed in the previous section are not isolated anymore, and belong to the lines of equilibrium points $P_{10}^{\pm}, P_{11}^{\pm}$ as we will see below. This is different to the results in Appendix  \ref{phys-intepretation}.

\item  $P_7^{\pm}$ as discussed in the previous section are not isolated anymore, and belong to the lines of equilibrium points $P_{10}^{\pm}, P_{11}^{\pm}$ as we will see below. This a different to the results in Appendix \ref{phys-intepretation}.

\item $P_8^{\pm}:=\left(\pm\frac{1}{\sqrt{1- \beta}}, \pm \frac{1}{\sqrt{1- \beta}}, \frac{2 \beta -1}{\beta -1}, 0, 0\right)$. They exist for $\eta \geq 1, \beta \leq 0$. 
The eigenvalues are\\
$0,\mp \frac{1}{\sqrt{1-\beta }}, \mp \frac{1-\sqrt{16 \beta -7}}{2 \sqrt{1-\beta }},\mp \frac{1+\sqrt{16 \beta -7}}{2 \sqrt{1-\beta }}, \mp \frac{\eta }{\sqrt{1-\beta }}$. 
These points are non-hyperbolic. For $P_8^{+}$ (respectively, $P_8^{-}$) and given $\beta<0$ (we have assumed $\eta\geq 1$), there are two negative (respectively, positive) eigenvalues, and two complex eigenvalues with negative (respectively, positive) real parts. So $P_8^{+}$ (respectively, $P_8^{-}$) has a 4D stable (respectively, unstable) manifold.

\item $P_9^{\pm}:=\left(\pm 2 \sqrt{\beta}, \pm \frac{1}{\sqrt{\beta}}, 0, 0, 0\right)$. Exist for $\eta \geq 1, 0<\beta \leq \frac{1}{4}$.\\
The eigenvalues are $0,\pm \frac{2(1-2 \beta) }{\sqrt{\beta }}, \pm \frac{1-4 \beta }{\sqrt{\beta }}, \pm \frac{1-4 \beta }{\sqrt{\beta }},\mp \frac{\eta }{\sqrt{\beta }}$. 
The equilibrium point is saddle for $\eta \geq 1, 0<\beta<\frac{1}{4}$, and non-hyperbolic when $\beta=\frac{1}{4}$. 

\end{enumerate}

Now, let's discuss the new equilibrium points, due to the extra coordinates $A_\varphi$ and $A_{W}$ related to the scalar field. They are:
\begin{enumerate}
  \setcounter{enumi}{9}

\item Line of equilibrium points $P_{10}^{\pm}(S_c):=\left(\pm 1, \pm S_c, 0, \pm \sqrt{1-\beta S_c^2}, 0\right)$, where we have explicitly shown the dependence on the parameter $S_c$ of the lines. They exist for $\beta\leq 0, S_c\in\mathbb{R}$ or $\beta>0, -\frac{1}{\sqrt{\beta}}\leq S_c\leq \frac{1}{\sqrt{\beta}}$. \\
Eigenvalues: $0, \pm 2, \pm 2(2- S_c), \pm\left(2-S_c-\frac{k \sqrt{1-\beta S_c^2}}{\sqrt{2}}\right), \pm \left(4-S_c (2+\eta )\right)$.
These lines cover the points $P_7^{\pm}$ and $P_6^{\pm}$ in the previous section for the choices $S_c=-\frac{1}{\sqrt{\beta}}$ and $S_c=\frac{1}{\sqrt{\beta}}$ respectively. Since $Q^2=1$ ($\mu_0=0$), it follows from the definition of $\lambda$ that the metric can be written as\\ 
$ds^2=-N^2dt^2+\frac{S_c^2}{y^2}d\lambda^2+K^{-1} \left(d\vartheta ^{2}+\sin ^{2}\vartheta d\varphi ^{2}\right)$.  When we substitute the values of $Q,S, C, A_\varphi$ and $A_{W}$ into \eqref{eqN-exp}, \eqref{eqK-exp} , and \eqref{eqy-exp} and integrating, we get  $N=\bar{N}_0e^{\pm S_c \lambda}, K=\bar{K}_0 e^{\mp 2(1-S_c)\lambda}, y=\bar{y}_0 e^{\mp (2-S_c)\lambda}$. 
Thus, the metric can be written as \\$ds^2=-\bar{N}_0e^{\pm 2 S_c \lambda}dt^2+\frac{S_c^2}{\bar{y}_0^2} e^{\pm 2(2-S_c)\lambda} d\lambda^2+{\bar{K}_0}^{-1} e^{\pm 2(1-S_c)\lambda} \left(d\vartheta ^{2}+\sin ^{2}\vartheta d\varphi ^{2}\right)$.

The following subsets (arcs, or specific equilibrium points) of the line $P_{10}^{+}(S_c)$ (respectively, $P_{10}^{-}(S_c)$) are unstable  (respectively, stable) for the given conditions: 
 \begin{enumerate}
   \item $S_c<0, \beta <\frac{1}{{S_c}^2}, \eta >1, k<\sqrt{\frac{-2 {S_c}^2+8 S_c-8}{\beta {S_c}^2-1}}$, or 
  \item $S_c<0, \beta =\frac{1}{{S_c}^2}, \eta >1, k\in \mathbb{R}$, or         
  \item $S_c=0, \beta \in \mathbb{R}, \eta>1, k<2 \sqrt{2}$, or 
  \item $0<S_c<\frac{4}{3}, \beta <\frac{1}{{S_c}^2}, 1< \eta <\frac{4-2 S_c}{S_c},  k<\sqrt{\frac{-2 {S_c}^2+8 S_c-8}{\beta  {S_c}^2-1}}$, or  
  \item $0<S_c<\frac{4}{3}, \beta =\frac{1}{{S_c}^2}, 1< \eta <\frac{4-2
  S_c}{S_c}, k\in \mathbb{R}$.
 \end{enumerate}

\item Line of equilibrium points $P_{11}^{\pm}(S_c):=\left(\pm 1, \pm S_c, 0,\mp \sqrt{1-\beta S_c^2}, 0\right)$, where we have explicitly shown the dependence on the parameter $S_c$ of the lines. They exist for $\beta\leq 0, S_c\in\mathbb{R}$ or $\beta>0, -\frac{1}{\sqrt{\beta}}\leq S_c\leq \frac{1}{\sqrt{\beta}}$. \\ 
Eigenvalues: $0, \pm 2, \pm 2(2- S_c), \pm\left(2-S_c+\frac{k \sqrt{1-\beta S_c^2}}{\sqrt{2}}\right), \pm\left(4-S_c (2+\eta )\right)$. 
These lines cover the points $P_7^{\pm}$ and $P_6^{\pm}$ in the previous section for the choices $S_c=-\frac{1}{\sqrt{\beta}}$ and $S_c=\frac{1}{\sqrt{\beta}}$ respectively. When we substitute the values of $Q,S, C, A_\varphi$ and $A_{W}$ into \eqref{eqN-exp}, \eqref{eqK-exp} , and \eqref{eqy-exp} and integrating, we get  $N=\bar{N}_0e^{\pm S_c \lambda}, K=\bar{K}_0 e^{\mp 2(1-S_c)\lambda}, y=\bar{y}_0 e^{\mp (2-S_c)\lambda}$. 
Thus, the metric can be written as $ds^2=-\bar{N}_0e^{\pm 2 S_c \lambda}dt^2+\frac{S_c^2}{\bar{y}_0^2} e^{\pm 2(2-S_c)\lambda} d\lambda^2+{\bar{K}_0}^{-1} e^{\pm 2(1-S_c)\lambda} \left(d\vartheta ^{2}+\sin ^{2}\vartheta d\varphi ^{2}\right)$.

The following subsets (arcs, or specific equilibrium points) of the line $P_{11}^{+}(S_c)$ (respectively, $P_{11}^{-}(S_c)$) are unstable  (respectively, stable) for the given conditions: 
\begin{enumerate}
   \item ${S_c}<0, \beta <\frac{1}{{S_c}^2}, \eta
   \geq 1, k>-\sqrt{2} \sqrt{-\frac{{S_c}^2-4 {S_c}+4}{\beta  {S_c}^2-1}}$, or 
   \item ${S_c}<0, \beta =\frac{1}{{S_c}^2}, \eta \geq 1, k\in \mathbb{R}$, or 
   \item ${S_c}=0, \beta \in \mathbb{R}, \eta \geq 1, k>-2 \sqrt{2}$, or 
  \item $0<{S_c}<\frac{4}{3}, \beta <\frac{1}{{S_c}^2}, 1\leq \eta <\frac{4-2{S_c}}{{S_c}}, k>-\sqrt{2} \sqrt{-\frac{{S_c}^2-4 {S_c}+4}{\beta  {S_c}^2-1}}$, or
  \item $0<{S_c}<\frac{4}{3}, \beta =\frac{1}{{S_c}^2}, 1\leq \eta
   <\frac{4-2 {S_c}}{{S_c}}, k\in \mathbb{R}$.
\end{enumerate}

\item $P_{12}^{\pm}:=\left(\pm 1, \pm \frac{2+\eta }{4\beta}, 0, \pm \frac{\sqrt{16 \beta -(\eta +2)^2}}{4 \sqrt{\beta }}, 0\right)$. 
These exist for $\eta \geq 1, \beta<0$; or
   $\eta \geq 1, \beta\geq \frac{1}{16} (\eta+2)^2$. \\
	The eigenvalues are: $0, \pm 2, \mp \frac{-16 \beta +2 (\eta +2)+\sqrt{2} \sqrt{\beta } k \sqrt{16 \beta -(\eta +2)^2}}{8 \beta }, \mp\frac{-8 \beta +\eta +2}{2 \beta },\mp \frac{(\eta +2)^2-16 \beta }{4 \beta }$. 
Substituting the values of $Q,S, C, A_\varphi$ and $A_{W}$ into \eqref{eqN-exp}, \eqref{eqK-exp} , and \eqref{eqy-exp} and integrating, we get  $N=\bar{N}_0e^{\pm \frac{(\eta +2) \lambda}{4 \beta}}, K=\bar{K}_0 e^{\mp \frac{\lambda   (4 \beta-\eta -2)}{2 \beta}}, y=\bar{y}_0 e^{\mp \frac{\eta  \lambda  (-8 \beta+\eta  (\eta +4)+4)}{4 \beta (\eta +2)}}$. It follows from the definition of $\lambda$ that the metric can be written as\\
$ds^2=-{\bar{N}_0}^2 e^{\pm \frac{(\eta +2) \lambda}{2 \beta}}dt^2+\frac{(2+\eta)^2 }{16 \bar{y}_0^2 \beta^2} e^{\pm \frac{\eta  \lambda  (-8 \beta+\eta  (\eta +4)+4)}{2 \beta (\eta +2)}}d\lambda^2+{\bar{K}_0}^{-1} e^{\pm \frac{\lambda   (4 \beta-\eta -2)}{2 \beta}} \left(d\vartheta ^{2}+\sin ^{2}\vartheta d\varphi ^{2}\right)$.
The equilibrium point $P_{12}^{+}$ (respectively, $P_{12}^{-}$) has a 4D unstable (respectively, stable) manifold for 
   \begin{enumerate}
   \item $\beta <0, \eta \geq 1, k>-\sqrt{2} \sqrt{\frac{(8 \beta -\eta -2)^2}{\beta  \left(16 \beta -\eta ^2-4 \eta -4\right)}}$, or 
    \item $\beta >\frac{9}{16}, 1\leq \eta <2 \left(2
   \sqrt{\beta }-1\right), k<\sqrt{2} \sqrt{\frac{(8 \beta -\eta -2)^2}{\beta  \left(16 \beta -\eta ^2-4 \eta -4\right)}}$. 
    \end{enumerate}

\item $P_{13}^{\pm}:=\left(\pm 1, \pm \frac{2+\eta }{4\beta}, 0, \mp \frac{\sqrt{16 \beta -(\eta +2)^2}}{4 \sqrt{\beta }}, 0\right)$. 
These exist for $\eta \geq 1, \beta<0$; or
   $\eta \geq 1, \beta\geq \frac{1}{16} (\eta+2)^2$. \\
   The eigenvalues are: $0, \pm 2, \pm \frac{16 \beta -2 (\eta +2)+\sqrt{2} \sqrt{\beta } k \sqrt{16 \beta -(\eta +2)^2}}{8 \beta }, \mp\frac{-8 \beta +\eta +2}{2 \beta }, \mp\frac{(\eta +2)^2-16 \beta }{4 \beta }$. 
Substituting the values of $Q,S, C, A_\varphi$ and $A_{W}$ into \eqref{eqN-exp}, \eqref{eqK-exp} , and \eqref{eqy-exp} and integrating, we get  $N=\bar{N}_0e^{\pm \frac{(\eta +2) \lambda}{4 \beta}}, K=\bar{K}_0 e^{\mp \frac{\lambda   (4 \beta-\eta -2)}{2 \beta}}, y=\bar{y}_0 e^{\mp \frac{\eta  \lambda  (-8 \beta+\eta  (\eta +4)+4)}{4 \beta (\eta +2)}}$. It follows from the definition of $\lambda$ that the metric can be written as 
\\$ds^2=-{\bar{N}_0}^2 e^{\pm \frac{(\eta +2) \lambda}{2 \beta}}dt^2+\frac{(2+\eta)^2 }{16 \bar{y}_0^2 \beta^2} e^{\pm \frac{\eta  \lambda  (-8 \beta+\eta  (\eta +4)+4)}{2 \beta (\eta +2)}}d\lambda^2+{\bar{K}_0}^{-1} e^{\pm \frac{\lambda   (4 \beta-\eta -2)}{2 \beta}} \left(d\vartheta ^{2}+\sin ^{2}\vartheta d\varphi ^{2}\right)$.
These points are non-hyperbolic (one zero eigenvalue). The equilibrium point $P_{13}^{+}$ (respectively, $P_{13}^{-}$) has a 4D unstable (respectively, stable) manifold for 
  \begin{enumerate}
  \item $\beta <0, \eta \geq 1, k<\sqrt{2} \sqrt{\frac{(8 \beta -\eta -2)^2}{\beta  \left(16 \beta -\eta ^2-4 \eta -4\right)}}$, or 
  \item $\beta >\frac{9}{16}, 1\leq \eta <2 \left(2
   \sqrt{\beta }-1\right), k>-\sqrt{2} \sqrt{\frac{(8 \beta -\eta -2)^2}{\beta  \left(16 \beta -\eta ^2-4 \eta -4\right)}}$.
  \end{enumerate}

\item $P_{14}^{\pm}:=\left(\pm 1, \pm \frac{k \sqrt{\beta  \left(k^2-8\right)+2}+4}{\beta  k^2+2}, 0,
 \mp\frac{\sqrt{2} \left(\sqrt{\beta  \left(k^2-8\right)+2}-2 \beta  k\right)}{\beta  k^2+2}, 0\right)$.
These exist for 
  \begin{enumerate}
  \item $\beta <0, -\sqrt{2} \sqrt{\frac{4 \beta -1}{\beta }}\leq k<-\sqrt{2} \sqrt{-\frac{1}{\beta }}, \eta \geq 1$, or
  \item $\beta <0,  -\sqrt{2} \sqrt{-\frac{1}{\beta }}<k<\sqrt{2} \sqrt{-\frac{1}{\beta }}, \eta \geq 1$, or
  \item $\beta <0, \sqrt{2} \sqrt{-\frac{1}{\beta }}<k\leq \sqrt{2} \sqrt{\frac{4 \beta -1}{\beta }}, \eta \geq 1$, or
  \item $0\leq \beta \leq \frac{1}{4}, k\in\mathbb{R}, \eta \geq 1$ or
  \item $\beta >\frac{1}{4}, k\leq -\sqrt{2} \sqrt{\frac{4 \beta -1}{\beta }}, \eta \geq 1$, or 
  \item $\beta >\frac{1}{4}, k\geq \sqrt{2} \sqrt{\frac{4 \beta -1}{\beta }}, \eta \geq 1$.
  \end{enumerate}
The eigenvalues are: $0,0,\pm 2,\mp\frac{2 k \left(\sqrt{\beta  \left(k^2-8\right)+2}-2 \beta  k\right)}{\beta  k^2+2}, \pm \frac{-4 \eta -(\eta +2) k \sqrt{\beta  \left(k^2-8\right)+2}+4 \beta  k^2}{\beta  k^2+2}$.
Substituting the values of $Q,S, C, A_\varphi$ and $A_{W}$ into \eqref{eqN-exp}, \eqref{eqK-exp} , and \eqref{eqy-exp} and integrating, we get \\$N=\bar{N}_0 \exp \left(\pm\frac{\lambda    \left(k \sqrt{\beta\left(k^2-8\right)+2}+4\right)}{\beta k^2+2}\right)$, 
$K=\bar{K}_0 \exp \left(\mp\frac{2 k \lambda   \left(\beta k - \sqrt{\beta\left(k^2-8\right)+2}-2\right)}{\beta k^2+2}\right)$,\\
$y=\bar{y}_0 \exp \left(\mp\frac{k \lambda  \left(2 \beta k-
\sqrt{\beta\left(k^2-8\right)+2}\right)}{\beta k^2+2}\right)$. It follows from the definition of $\lambda$ that the metric can be written as \\
$ds^2=-{\bar{N}_0}^2 e^{\pm\frac{2\lambda    \left(k \sqrt{\beta
   \left(k^2-8\right)+2}+4\right)}{\beta k^2+2}}dt^2+\frac{1}{\bar{y}_0^2}\left(\frac{4+k \sqrt{\beta
   \left(k^2-8\right)+2}}{2+\beta k^2}\right)^2 e^{\pm\frac{2 k \lambda  \left(2 \beta k-\sqrt{\beta \left(k^2-8\right)+2}\right)}{\beta k^2+2}}d\lambda^2+$\\${\bar{K}_0}^{-1} e^{\pm\frac{2 k \lambda   \left(\beta k - \sqrt{\beta\left(k^2-8\right)+2}-2\right)}{\beta k^2+2}} \left(d\vartheta ^{2}+\sin ^{2}\vartheta d\varphi ^{2}\right)$.
The equilibrium points are non-hyperbolic (two zero eigenvalues). The equilibrium point $P_{14}^{+}$ (respectively, $P_{14}^{-}$) has a 3D unstable (respectively, stable) manifold in the following cases: 
\begin{enumerate}
	\item $\beta >\frac{9}{16}, \eta \geq 2 (4 \beta -1), k<-2 \sqrt{2} \sqrt{\frac{\eta ^2}{-16 \beta +\eta ^2+4 \eta +4}}$, or 
    \item $\frac{3}{8}<\beta <\frac{9}{16}, \eta \geq 2 (4 \beta
   -1), k<-2 \sqrt{2} \sqrt{\frac{\eta ^2}{-16 \beta +\eta ^2+4 \eta +4}}$, or 
   \item $\beta <-\frac{1}{4}, \eta \geq \frac{2 (4 \beta -1)}{4 \beta +1}, -\sqrt{2} \sqrt{\frac{4 \beta -1}{\beta }}\leq k<-2 \sqrt{2} \sqrt{\frac{\eta ^2}{-16 \beta +\eta ^2+4 \eta +4}}$, or 
   \item $-\frac{1}{4}\leq \beta <0, \eta \geq 1, -\sqrt{2} \sqrt{-\frac{1}{\beta }}<k<-2 \sqrt{2}
   \sqrt{\frac{\eta ^2}{-16 \beta +\eta ^2+4 \eta +4}}$, or 
   \item $0\leq \beta \leq \frac{3}{8}, \eta \geq 1, k<-2 \sqrt{2} \sqrt{\frac{\eta ^2}{-16 \beta +\eta ^2+4 \eta +4}}$, or
   \item $\beta <-\frac{1}{4}, 1\leq \eta <\frac{2 (4 \beta -1)}{4 \beta +1}, -\sqrt{2} \sqrt{-\frac{1}{\beta }}<k<-2 \sqrt{2} \sqrt{\frac{\eta ^2}{-16 \beta +\eta ^2+4 \eta +4}}$, or
   \item $\beta =\frac{9}{16}, \eta =1, k\geq \frac{2 \sqrt{10}}{3}$, or 
   \item $\beta =\frac{9}{16}, 1<\eta <\frac{5}{2}, \frac{2 \sqrt{10}}{3}\leq k<2 \sqrt{2} \sqrt{\frac{\eta ^2}{\eta ^2+4 \eta -5}}$, or 
  \item $\beta =\frac{9}{16}, 1\leq \eta <\frac{5}{2}, k\leq -\frac{2 \sqrt{10}}{3}$, or 
   \item $\beta =\frac{9}{16}, \eta \geq \frac{5}{2}, k<-2
   \sqrt{2} \sqrt{\frac{\eta ^2}{\eta ^2+4 \eta -5}}$, or 
   \item $\beta >\frac{9}{16}, 1\leq \eta \leq 2 \left(2 \sqrt{\beta }-1\right), k\geq \sqrt{2} \sqrt{\frac{4 \beta -1}{\beta}}$, or 
   \item $\beta >\frac{9}{16}, 2 \left(2 \sqrt{\beta }-1\right)<\eta <2 (4 \beta -1), \sqrt{2} \sqrt{\frac{4 \beta -1}{\beta }}\leq k<2 \sqrt{2} \sqrt{\frac{\eta ^2}{-16 \beta
   +\eta ^2+4 \eta +4}}$, or 
   \item $\beta >\frac{9}{16}, 2 \left(2 \sqrt{\beta }-1\right)<\eta <2 (4 \beta -1), k\leq -\sqrt{2} \sqrt{\frac{4 \beta -1}{\beta }}$, or 
   \item $\beta >\frac{9}{16}, 1\leq \eta \leq 2 \left(2 \sqrt{\beta }-1\right), k\leq -\sqrt{2} \sqrt{\frac{4 \beta -1}{\beta }}$, or 
   \item $\beta <-\frac{1}{4}, \eta \geq \frac{2 (4 \beta -1)}{4 \beta +1}, \sqrt{2} \sqrt{-\frac{1}{\beta }}<k\leq \sqrt{2} \sqrt{\frac{4 \beta -1}{\beta }}$, or 
   \item $\beta <0, \eta \geq 1, \sqrt{2} \sqrt{-\frac{1}{\beta }}<k\leq \sqrt{2}
   \sqrt{\frac{4 \beta -1}{\beta }}$, or 
   \item $-\frac{1}{4}\leq \beta <0, \eta \geq 1, -\sqrt{2} \sqrt{\frac{4 \beta -1}{\beta }}\leq k<-\sqrt{2} \sqrt{-\frac{1}{\beta }}$, or
   \item $\beta <-\frac{1}{4}, 1\leq \eta <\frac{2 (4 \beta -1)}{4 \beta +1}, -\sqrt{2} \sqrt{\frac{4 \beta -1}{\beta }}\leq k<-\sqrt{2} \sqrt{-\frac{1}{\beta }}$, or
   \item $\frac{3}{8}<\beta <\frac{9}{16}, 1\leq \eta <2 (4 \beta -1), \sqrt{2} \sqrt{\frac{4 \beta -1}{\beta }}\leq k<2 \sqrt{2} \sqrt{\frac{\eta ^2}{-16 \beta +\eta ^2+4 \eta +4}}$, or
   \item $\frac{3}{8}<\beta <\frac{9}{16}, 1\leq \eta <2 (4 \beta -1), k\leq -\sqrt{2} \sqrt{\frac{4 \beta -1}{\beta }}$.	
\end{enumerate}

\item $P_{15}^{\pm}:=\left(\pm 1, \pm \frac{4-k \sqrt{\beta  \left(k^2-8\right)+2}}{2+\beta k^2},
 0, \pm \frac{\sqrt{2} \left(2 \beta k+\sqrt{\beta  \left(k^2-8\right)+2}\right)}{2+\beta k^2}, 0\right)$. These  exist  for
 \begin{enumerate}
 \item $\beta <0, -\sqrt{2} \sqrt{\frac{4 \beta -1}{\beta }}\leq k<-\sqrt{2} \sqrt{-\frac{1}{\beta }}, \eta \geq 1$, or
 \item $\beta <0, -\sqrt{2} \sqrt{-\frac{1}{\beta }}<k<\sqrt{2} \sqrt{-\frac{1}{\beta }}, \eta \geq 1$, or 
 \item $\beta <0, \sqrt{2} \sqrt{-\frac{1}{\beta }}<k\leq \sqrt{2} \sqrt{\frac{4 \beta -1}{\beta }}, \eta \geq 1$, or
 \item $0\leq \beta \leq \frac{1}{4}, \eta \geq 1, k\in \mathbb{R}$, or
 \item $\beta >\frac{1}{4}, k\leq -\sqrt{2} \sqrt{\frac{4 \beta -1}{\beta }}, \eta \geq 1$, or
 \item $\beta >\frac{1}{4}, k\geq \sqrt{2} \sqrt{\frac{4 \beta -1}{\beta }}, \eta \geq 1$.
 \end{enumerate}
The eigenvalues are: $0, 0, \pm 2, \pm \frac{2 k \left(\sqrt{\beta  \left(k^2-8\right)+2}+2 \beta  k\right)}{\beta  k^2+2}, \pm \frac{-4 \eta +(\eta +2) k \sqrt{\beta  \left(k^2-8\right)+2}+4 \beta  k^2}{\beta  k^2+2}$.\\
Substituting the values of $Q,S, C, A_\varphi$ and $A_{W}$ into \eqref{eqN-exp}, \eqref{eqK-exp} , and \eqref{eqy-exp} and integrating, 
we get  $N=\bar{N}_0 \exp \left(\pm\frac{\lambda   \left(4-k \sqrt{\beta\left(k^2-8\right)+2}\right)}{\beta k^2+2}\right)$,\\
$K=\bar{K}_0 \exp \left(\mp\frac{2 k \lambda   \left(\beta k + \sqrt{\beta
   \left(k^2-8\right)+2}-2\right)}{\beta k^2+2}\right)$,\\ $y=\bar{y}_0 \exp \left(\mp\frac{k \lambda  \left(2 \beta k+\sqrt{\beta\left(k^2-8\right)+2}\right)}{\beta k^2+2}\right)$.  
	It follows from the definition of $\lambda$ that the metric can be written as \\
$ds^2=-{\bar{N}_0}^2 e^{\pm\frac{2\lambda    \left(4-k \sqrt{\beta
   \left(k^2-8\right)+2}\right)}{\beta k^2+2}}dt^2+\frac{1}{\bar{y}_0^2}\left(\frac{4-k \sqrt{\beta
   \left(k^2-8\right)+2}}{2+\beta}\right)^2 e^{\pm\frac{2 k \lambda  \left(2 \beta k+\sqrt{\beta
   \left(k^2-8\right)+2}\right)}{\beta k^2+2}}d\lambda^2+$\\
   ${\bar{K}_0}^{-1} e^{\pm\frac{2 k \lambda   \left(\beta k + \sqrt{\beta \left(k^2-8\right)+2}-2\right)}{\beta k^2+2}} \left(d\vartheta ^{2}+\sin ^{2}\vartheta d\varphi ^{2}\right)$.
The equilibrium points are non-hyperbolic (two zero eigenvalues). The equilibrium point $P_{15}^{+}$ (respectively, $P_{15}^{-}$) has a 3D unstable (respectively, stable) manifold in the following cases: 
\begin{enumerate}
	\item $\beta <-\frac{1}{4}, 1\leq \eta <\frac{2 (4 \beta -1)}{4 \beta +1}, 2 \sqrt{2} \sqrt{\frac{\eta ^2}{-16 \beta +\eta ^2+4 \eta +4}}<k<\sqrt{2} \sqrt{-\frac{1}{\beta }}$, or
    \item $\beta <-\frac{1}{4}, 1\leq \eta <\frac{2 (4 \beta -1)}{4 \beta +1}, -\sqrt{2} \sqrt{\frac{4 \beta -1}{\beta }}\leq k<-\sqrt{2} \sqrt{-\frac{1}{\beta }}$, or 
    \item $\beta<-\frac{1}{4}, \eta \geq \frac{2 (4 \beta -1)}{4 \beta +1}, -\sqrt{2} \sqrt{\frac{4 \beta -1}{\beta }}\leq k<-\sqrt{2} \sqrt{-\frac{1}{\beta }}$, or 
    \item $\beta <-\frac{1}{4}, 1\leq \eta <\frac{2 (4 \beta -1)}{4 \beta +1}, \sqrt{2} \sqrt{-\frac{1}{\beta }}<k\leq \sqrt{2} \sqrt{\frac{4 \beta -1}{\beta }}$, or 
    \item $\beta <-\frac{1}{4}, \eta \geq \frac{2 (4
   \beta -1)}{4 \beta +1}, 2 \sqrt{2} \sqrt{\frac{\eta ^2}{-16 \beta +\eta ^2+4 \eta +4}}<k\leq \sqrt{2} \sqrt{\frac{4 \beta -1}{\beta }}$, or 
   \item $-\frac{1}{4}\leq \beta <0, \eta \geq 1, 2 \sqrt{2} \sqrt{\frac{\eta ^2}{-16 \beta +\eta ^2+4 \eta +4}}<k<\sqrt{2} \sqrt{-\frac{1}{\beta }}$, or 
   \item $-\frac{1}{4}\leq \beta <0, \eta \geq 1, -\sqrt{2} \sqrt{\frac{4
   \beta -1}{\beta }}\leq k<-\sqrt{2} \sqrt{-\frac{1}{\beta }}$, or 
   \item $-\frac{1}{4}\leq \beta <0, \eta \geq 1, \sqrt{2} \sqrt{-\frac{1}{\beta }}<k\leq \sqrt{2} \sqrt{\frac{4 \beta-1}{\beta }}$, or 
   \item $0\leq \beta \leq \frac{3}{8}, \eta \geq 1, k>2 \sqrt{2} \sqrt{\frac{\eta ^2}{-16 \beta +\eta ^2+4 \eta +4}}$, or 
   \item $\frac{3}{8}<\beta
   <\frac{9}{16}, 1\leq \eta <2 (4 \beta -1), k\geq \sqrt{2} \sqrt{\frac{4 \beta -1}{\beta }}$, or 
   \item $\frac{3}{8}<\beta <\frac{9}{16}, 1\leq \eta <2 (4 \beta -1), -2 \sqrt{2}
   \sqrt{\frac{\eta ^2}{-16 \beta +\eta ^2+4 \eta +4}}<k\leq -\sqrt{2} \sqrt{\frac{4 \beta -1}{\beta }}$, or 
   \item $\frac{3}{8}<\beta <\frac{9}{16}, \eta \geq 2 (4 \beta -1), k>2 \sqrt{2}
   \sqrt{\frac{\eta ^2}{-16 \beta +\eta ^2+4 \eta +4}}$, or 
   \item $\beta =\frac{9}{16}, \eta =1, k\leq -\frac{2 \sqrt{10}}{3}$, or
   \item $\beta =\frac{9}{16}, 1\leq \eta <\frac{5}{2}, k\geq \frac{2 \sqrt{10}}{3}$, or 
   \item $\beta =\frac{9}{16}, 1<\eta <\frac{5}{2}, -2 \sqrt{2} \sqrt{\frac{\eta ^2}{\eta ^2+4 \eta -5}}<k\leq -\frac{2\sqrt{10}}{3}$, or 
   \item $\beta =\frac{9}{16}, \eta \geq \frac{5}{2}, k>2 \sqrt{2} \sqrt{\frac{\eta ^2}{\eta ^2+4 \eta -5}}$, or 
   \item $\beta >\frac{9}{16}, 2 \left(2
   \sqrt{\beta }-1\right)<\eta <2 (4 \beta -1), k\geq \sqrt{2} \sqrt{\frac{4 \beta -1}{\beta }}$, or 
   \item $\beta >\frac{9}{16}, 1\leq \eta \leq 2 \left(2 \sqrt{\beta }-1\right), k\geq
   \sqrt{2} \sqrt{\frac{4 \beta -1}{\beta }}$, or 
   \item $\beta >\frac{9}{16}, 2 \left(2 \sqrt{\beta }-1\right)<\eta <2 (4 \beta -1), -2 \sqrt{2} \sqrt{\frac{\eta ^2}{-16 \beta +\eta ^2+4
   \eta +4}}<k\leq -\sqrt{2} \sqrt{\frac{4 \beta -1}{\beta }}$, or
   \item $\beta >\frac{9}{16}, 1\leq \eta \leq 2 \left(2 \sqrt{\beta }-1\right), k\leq -\sqrt{2} \sqrt{\frac{4 \beta-1}{\beta }}$, or
   \item $\beta >\frac{9}{16}, \eta \geq 2 (4 \beta -1), k>2 \sqrt{2} \sqrt{\frac{\eta ^2}{-16 \beta +\eta ^2+4 \eta +4}}$.	
\end{enumerate}

\item $P_{16}^{\pm}:=\left(\pm 1 , \pm \frac{(\eta +2) k^2}{2 \left(\eta ^2+2 \beta  k^2\right)}, 0, \pm \frac{\eta  (\eta +2) k}{2 \sqrt{2} \left(\eta ^2+2 \beta  k^2\right)}, 
   \frac{\sqrt{\eta } \sqrt{\eta +2} \sqrt{k^2 \left((\eta +2)^2-16 \beta \right)-8 \eta ^2}}{2 \sqrt{2} \left(\eta ^2+2 \beta  k^2\right)}\right)$. \\ These exist for 
   \begin{enumerate}
   \item $\eta \geq 1, \beta =0, k\geq \frac{2 \sqrt{2} \eta }{\eta +2} $, or
   \item $\eta \geq 1, \beta =0, k\leq -\frac{2 \sqrt{2} \eta }{\eta +2}$, or 
   \item $\eta \geq 1, \beta <0, -\frac{\sqrt{-\frac{\eta ^2}{\beta }}}{\sqrt{2}}<k\leq -2 \sqrt{2} \sqrt{\frac{\eta ^2}{-16 \beta +\eta ^2+4 \eta +4}}$, or 
   \item $\eta \geq 1, \frac{\eta +2}{8}\leq \beta
   <\frac{1}{16} \left(\eta+2\right)^2, k=\frac{2 \eta }{\sqrt{\frac{1}{2} (\eta +2)^2-8 \beta }}$, or 
   \item $\eta \geq 1, \frac{\eta +2}{8}\leq \beta <\frac{1}{16}\left(\eta+2\right)^2, k=-\frac{2 \eta }{\sqrt{\frac{1}{2} (\eta +2)^2-8 \beta }}$, or 
   \item $\eta \geq 1, \beta <0, 2 \sqrt{2} \sqrt{\frac{\eta ^2}{-16 \beta +\eta ^2+4 \eta +4}}\leq
   k<\frac{\sqrt{-\frac{\eta ^2}{\beta }}}{\sqrt{2}}$, or 
   \item $\eta \geq 1, 0<\beta <\frac{\eta +2}{8}, -\sqrt{\frac{\eta }{\beta }}\leq k\leq -2 \sqrt{2} \sqrt{\frac{\eta ^2}{-16 \beta +\eta ^2+4 \eta +4}}$, or 
   \item $\eta \geq 1, 0<\beta <\frac{\eta +2}{8}, 2 \sqrt{2} \sqrt{\frac{\eta ^2}{-16 \beta +\eta ^2+4 \eta +4}}\leq k\leq \sqrt{\frac{\eta }{\beta }}$.
   \end{enumerate}

The eigenvalues are: $\mp \frac{8 \eta ^2+k^2 \left(16 \beta -(\eta +2)^2\right)}{4 \left(\eta ^2+2 \beta  k^2\right)}, \mp \frac{4 \eta ^2+k^2 \left(8 \beta -(\eta +2)^2\right)}{2 \left(\eta ^2+2 \beta 
   k^2\right)}, \pm \frac{\eta  (\eta +2) k^2}{2 \left(\eta
   ^2+2 \beta  k^2\right)}$,\\
   $\mp \frac{\sqrt{\beta } \left(8 \eta ^2+k^2 \left(16 \beta -(\eta +2)^2\right)\right)-\sqrt{8 \eta ^2+k^2 \left(16 \beta -(\eta +2)^2\right)} \sqrt{8 \eta ^2 (\beta +\eta +2)+\beta  k^2 (16
   \beta -(\eta +2) (9 \eta +2))}}{8 \sqrt{\beta } \left(\eta ^2+2 \beta  k^2\right)}$, \\
   $\mp \frac{\sqrt{\beta } \left(8 \eta ^2+k^2 \left(16 \beta -(\eta +2)^2\right)\right)+\sqrt{8 \eta ^2 (\beta +\eta
   +2)+\beta  k^2 (16 \beta -(\eta +2) (9 \eta +2))} \sqrt{8 \eta ^2+k^2 \left(16 \beta -(\eta +2)^2\right)}}{8 \sqrt{\beta } \left(\eta ^2+2 \beta  k^2\right)}$.
\\
Substituting the values of $Q,S, C, A_\varphi$ and $A_{W}$ into \eqref{eqN-exp}, \eqref{eqK-exp} , and \eqref{eqy-exp} and integrating, we get
$N=\bar{N}_0 e^{\pm\frac{(\eta +2) k^2 \lambda }{2(2 \beta k^2+\eta ^2)}}, K=\bar{K}_0 e^{\mp \frac{\lambda\left(k^2 (4 \beta-\eta -2)+2 \eta ^2\right)}{2 \beta k^2+\eta ^2}}, y=\bar{y}_0 e^{\mp \frac{\eta  (\eta +2) k^2 \lambda }{4 \left(2\beta k^2+\eta^2\right)}}$. From the definition of $\lambda$ the metric can be written as 
\\$ds^2=-{\bar{N}_0}^2 e^{\pm\frac{(\eta +2) k^2 \lambda }{2 \beta k^2+\eta ^2}}dt^2+\frac{1}{\bar{y}_0^2}\left(\frac{k^2 (2+\eta )}{4\beta k^2+2 \eta ^2}\right)^2 e^{\pm \frac{\eta  (\eta +2) k^2 \lambda }{2 \left(2 \beta k^2+\eta^2\right)}}d\lambda^2+{\bar{K}_0}^{-1} e^{\pm \frac{\lambda\left(k^2 (4 \beta-\eta -2)+2 \eta ^2\right)}{2 \beta k^2+\eta ^2}} \left(d\vartheta ^{2}+\sin ^{2}\vartheta d\varphi ^{2}\right)$. \\
$P_{16}^{+}$ (respectively, $P_{16}^{-}$) is a source (respectively, a sink) for 
\begin{enumerate}
\item $\eta \geq 1, 0<\beta <\frac{\eta +2}{8}, -\sqrt{\frac{\eta }{\beta }}<k<-2 \sqrt{2} \sqrt{\frac{\eta ^2}{-16 \beta +\eta ^2+4 \eta +4}}$, or
\item $\eta \geq 1, 0<\beta <\frac{\eta +2}{8}, 2 \sqrt{2} \sqrt{\frac{\eta ^2}{-16 \beta +\eta ^2+4 \eta +4}}<k<\sqrt{\frac{\eta }{\beta }}$.
\end{enumerate}
 It is a saddle for 
\begin{enumerate}
\item $\eta \geq 1, \beta <0, -\frac{\sqrt{-\frac{\eta ^2}{\beta }}}{\sqrt{2}}<k<-2 \sqrt{2} \sqrt{\frac{\eta ^2}{-16 \beta +\eta ^2+4 \eta +4}}$ or 
\item $\eta \geq 1, \beta <0, 2 \sqrt{2} \sqrt{\frac{\eta ^2}{-16 \beta +\eta ^2+4 \eta +4}}<k<\frac{\sqrt{-\frac{\eta ^2}{\beta }}}{\sqrt{2}}$.
\end{enumerate} 
non-hyperbolic otherwise.

\item  $P_{17}^{\pm}:=\left(\pm 1,\pm\frac{(\eta +2) k^2}{2 \left(\eta ^2+2 \beta  k^2\right)}, 0, \pm \frac{\eta  (\eta +2) k}{2 \sqrt{2} \left(\eta ^2+2 \beta  k^2\right)}, - \frac{\sqrt{\eta } \sqrt{\eta +2} \sqrt{k^2 \left((\eta +2)^2-16 \beta \right)-8 \eta ^2}}{2 \sqrt{2} \left(\eta ^2+2 \beta  k^2\right)}\right)$. \\ These exist for 
\begin{enumerate}
\item $\eta \geq 1, \beta <0, k<-\frac{\sqrt{-\frac{\eta ^2}{\beta }}}{\sqrt{2}}$, or 
\item $\eta \geq 1, \beta <0, k=-2 \sqrt{2} \sqrt{\frac{\eta ^2}{-16 \beta +\eta ^2+4 \eta +4}}$, or
\item $\eta \geq 1, \beta <0, k=2 \sqrt{2} \sqrt{\frac{\eta ^2}{-16 \beta +\eta ^2+4 \eta +4}}$, or 
\item $\eta \geq 1, \beta <0, k>\frac{\sqrt{-\frac{\eta ^2}{\beta }}}{\sqrt{2}}$, or 
\item $\eta \geq 1, 0\leq \beta <\frac{1}{16} (\eta +2)^2, k=-2 \sqrt{2} \sqrt{\frac{\eta ^2}{-16 \beta +\eta ^2+4 \eta +4}}$, or 
\item $\eta \geq 1, 0\leq \beta <\frac{1}{16} (\eta +2)^2, k=2 \sqrt{2} \sqrt{\frac{\eta ^2}{-16 \beta +\eta ^2+4 \eta +4}}$. 
\end{enumerate}

Eigenvalues: $\mp \frac{8 \eta ^2+k^2 \left(16 \beta -(\eta +2)^2\right)}{4 \left(\eta ^2+2 \beta  k^2\right)}, \mp \frac{4 \eta ^2+k^2 \left(8 \beta -(\eta +2)^2\right)}{2 \left(\eta ^2+2 \beta 
   k^2\right)}, \pm \frac{\eta  (\eta +2) k^2}{2 \left(\eta
   ^2+2 \beta  k^2\right)}$, \\
   $\mp\frac{\sqrt{\beta } \left(8 \eta ^2+k^2 \left(16 \beta -(\eta +2)^2\right)\right)-\sqrt{8 \eta ^2+k^2 \left(16 \beta -(\eta +2)^2\right)} \sqrt{8 \eta ^2 (\beta +\eta +2)+\beta  k^2 (16
   \beta -(\eta +2) (9 \eta +2))}}{8 \sqrt{\beta } \left(\eta ^2+2 \beta  k^2\right)}$, \\
   $\mp\frac{\sqrt{\beta } \left(8 \eta ^2+k^2 \left(16 \beta -(\eta +2)^2\right)\right)+\sqrt{8 \eta ^2 (\beta +\eta
   +2)+\beta  k^2 (16 \beta -(\eta +2) (9 \eta +2))} \sqrt{8 \eta ^2+k^2 \left(16 \beta -(\eta +2)^2\right)}}{8 \sqrt{\beta } \left(\eta ^2+2 \beta  k^2\right)}$. 
   
Substituting the values of $Q,S, C, A_\varphi$ and $A_{W}$ into \eqref{eqN-exp}, \eqref{eqK-exp} , and \eqref{eqy-exp} and integrating, we get
$N=\bar{N}_0 e^{\pm\frac{(\eta +2) k^2 \lambda }{4 \beta k^2+2 \eta ^2}}, K=\bar{K}_0 e^{\mp \frac{\lambda\left(k^2 (4 \beta-\eta -2)+2 \eta ^2\right)}{2 \beta k^2+\eta ^2}}, y=\bar{y}_0 e^{\mp \frac{\eta  (\eta +2) k^2 \lambda }{4 \left(2 \beta k^2+\eta^2\right)}}$. It follows from the definition of $\lambda$ that the metric can be written as 
\\$ds^2=-{\bar{N}_0}^2 e^{\pm\frac{(\eta +2) k^2 \lambda }{2\beta k^2+\eta ^2}}dt^2+\frac{1}{\bar{y}_0^2}\left(\frac{k^2 (2+\eta )}{4 \beta k^2+2 \eta ^2}\right)^2 e^{\pm \frac{\eta  (\eta +2) k^2 \lambda }{2 \left(2 \beta k^2+\eta^2\right)}}d\lambda^2+{\bar{K}_0}^{-1} e^{\pm \frac{\lambda\left(k^2 (4 \beta-\eta -2)+2 \eta ^2\right)}{2 \beta k^2+\eta ^2}} \left(d\vartheta ^{2}+\sin ^{2}\vartheta d\varphi ^{2}\right)$.

They are a saddle for 
\begin{enumerate}
\item $\eta \geq 1, \beta <0, k<-\frac{\sqrt{-\frac{\eta ^2}{\beta }}}{\sqrt{2}}$, or
\item $\eta \geq 1, \beta <0, k>\frac{\sqrt{-\frac{\eta ^2}{\beta }}}{\sqrt{2}}$.
\end{enumerate}
 non-hyperbolic, otherwise. 

\item $P_{18}^{\pm}:=\left(\pm 1, \pm \frac{1}{2 \beta }, 0, \pm\frac{k}{2 \sqrt{2}}, \frac{\sqrt{\beta  \left(k^2-8\right)+2}}{2 \sqrt{2} \sqrt{\beta }}\right)$. \\   These exist for 
\begin{enumerate}
\item $\eta \geq 1, 0<\beta \leq \frac{1}{4}, k\in\mathbb{R}$, or 
\item $\eta \geq 1, \beta >\frac{1}{4}, k\leq -\sqrt{2} \sqrt{\frac{4 \beta -1}{\beta }}$, or
\item $\eta \geq 1, \beta >\frac{1}{4}, k\geq \sqrt{2} \sqrt{\frac{4 \beta -1}{\beta }}$.
\end{enumerate}
The eigenvalues are: $\pm\frac{k^2}{2}, \pm\frac{-4 \beta +\beta  k^2+2}{2 \beta }, \pm \frac{-8 \beta +\beta  k^2+2}{4 \beta }, \pm \frac{-8 \beta +\beta  k^2+2}{4 \beta }, \pm \frac{\beta  k^2-\eta }{2 \beta }$.\\
Substituting the values of $Q,S, C,A_\varphi$ and $A_{W}$ into \eqref{eqN-exp}, \eqref{eqK-exp} , and \eqref{eqy-exp} and integrating, we get
$N=\bar{N}_0 e^{\pm\frac{\lambda }{2 \beta}}, K=\bar{K}_0 e^{\mp\frac{(2 \beta -1) \lambda }{\beta}}, y=\bar{y}_0 e^{\mp \frac{1}{4} k^2 \lambda}$.  It follows from the definition of $\lambda$ that the metric can be written as 
$ds^2=-{\bar{N}_0}^2 e^{\pm\frac{\lambda }{\beta}}dt^2+\frac{1}{4\bar{y}_0^2 \beta^2} e^{\pm \frac{k^2 \lambda }{2}}d\lambda^2+{\bar{K}_0}^{-1} e^{\pm \frac{(2\beta -1) \lambda }{\beta}} \left(d\vartheta ^{2}+\sin ^{2}\vartheta d\varphi ^{2}\right)$.

$P_{18}^{+}$ (respectively, $P_{18}^{-}$) is a source (respectively, a sink) for 
\begin{enumerate}
\item $\eta \geq 1, \beta <0, k<-\sqrt{2} \sqrt{\frac{4 \beta -1}{\beta }}$, or
\item $\eta \geq 1, \beta <0, k>\sqrt{2} \sqrt{\frac{4 \beta -1}{\beta }}$, or
\item $\eta \geq 1, 0<\beta \leq \frac{\eta +2}{8}, k<-\sqrt{\frac{\eta }{\beta }}$, or
\item $\eta \geq 1, 0<\beta \leq \frac{\eta +2}{8}, k>\sqrt{\frac{\eta }{\beta }}$, or
\item $\eta \geq 1, \beta >\frac{\eta +2}{8}, k<-\sqrt{2} \sqrt{\frac{4 \beta -1}{\beta }}$, or 
\item $\eta \geq 1, \beta >\frac{\eta +2}{8}, k>\sqrt{2} \sqrt{\frac{4 \beta -1}{\beta }}$. 
\end{enumerate}

It is non-hyperbolic for 

\begin{enumerate}
\item $\eta \geq 1,  0<\beta \leq \frac{1}{4},  k\in\left\{-\sqrt{\frac{\eta }{\beta }}, 0, \sqrt{\frac{\eta }{\beta }}\right\}$, or
\item $\eta \geq 1, \frac{1}{4}<\beta <\frac{\eta +2}{8}, k\in\left\{-\sqrt{\frac{\eta }{\beta }}, -\sqrt{2} \sqrt{\frac{4 \beta -1}{\beta }},\sqrt{2} \sqrt{\frac{4 \beta -1}{\beta }}, \sqrt{\frac{\eta }{\beta }}\right\}$, or 
\item $\eta \geq 1, \beta \geq \frac{\eta +2}{8}, k\in\left\{-\sqrt{2} \sqrt{\frac{4 \beta -1}{\beta }}, \sqrt{2} \sqrt{\frac{4 \beta -1}{\beta }}\right\}$
\end{enumerate}

Otherwise, they are saddle. 

\item $P_{19}^{\pm}:=\left( \pm 1, \pm \frac{2}{\beta  k^2+2}, \frac{\beta  \left(k^2-4\right)+2}{\beta  k^2+2}, \pm \frac{\sqrt{2} \beta  k}{\beta  k^2+2}, \sqrt{-\frac{2\beta
   }{\beta  k^2+2}}\right)$. \\
	There exist for 
	\begin{enumerate}
	\item $\eta\geq 1, \beta=0, k\in\mathbb{R}$, or 
	\item $\eta\geq 1, \beta<0, -\sqrt{2} \sqrt{-\frac{1}{\beta}}<k<\sqrt{2} \sqrt{-\frac{1}{\beta}}$.
	\end{enumerate}
	
	For $\beta=0$ these points reduces to $P_8^{\pm}$. 
Substituting the values of $Q,S, C,A_\varphi$ and $A_{W}$ into \eqref{eqN-exp}, \eqref{eqK-exp} , and \eqref{eqy-exp} and integrating, we get
$N=\bar{N}_0 e^{\pm \frac{2 \lambda}{\beta k^2+2}}, K=\bar{K}_0 e^{\mp \frac{2 \beta k^2 \lambda}{\beta k^2+2}}, y=\bar{y}_0 e^{\mp\frac{\beta k^2 \lambda }{\epsilon  \left(\beta k^2+2\right)}}$. Since $Q^2=1, C\neq 0$, it follows from the definition of $\lambda$ that the metric can be written as\\
$ds^2=-{\bar{N}_0}^2 e^{\pm \frac{4 \lambda}{\beta k^2+2}}dt^2+{\bar{K}_0}^{-1} e^{\pm \frac{2 \beta k^2 \lambda}{\beta k^2+2}} \left[ \left(\frac{\beta  \left(k^2-4\right)+2}{\beta  k^2+2}\right)d\lambda^2+ \left(d\vartheta ^{2}+\sin ^{2}\vartheta d\varphi ^{2}\right)\right]$.
For $\beta\neq 0$, the eigenvalues are \\$\mp 1, \pm \frac{2 \beta  k^2}{\beta  k^2+2},  \pm \frac{2 \left(\beta  k^2-\eta \right)}{\beta  k^2+2}, \mp\frac{\beta  k^2+2+\sqrt{64 \beta -7 \beta ^2 k^4+32 \beta ^2 k^2-28 \beta  k^2-28}}{2
   \left(\beta  k^2+2\right)}$,\\ 
	$\mp\frac{\beta  k^2+2-\sqrt{64 \beta -7 \beta ^2 k^4+32 \beta ^2 k^2-28 \beta  k^2-28}}{2
   \left(\beta  k^2+2\right)}$. 
 Hence, the point $P_{19}^{+}$ (respectively, $P_{19}^{-}$) behaves as a saddle.
\end{enumerate}

\end{appendix}


\begin{thebibliography}{120}
\bibitem{Jacobson:2000xp} T.~Jacobson and D.~Mattingly,
Phys.\ Rev.\ D \textbf{64}, 024028 (2001). 

\bibitem{Eling:2004dk} C.~Eling, T.~Jacobson and D.~Mattingly,
gr-qc/0410001. 

\bibitem{DJ} W.~Donnelly and T.~Jacobson,
Phys.\ Rev.\ D \textbf{82}, 064032 (2010). 

\bibitem{kann} S. Kanno and J. Soda, Phys. Rev. D \textbf{74}, 063505 (2006).


\bibitem{Zlosnik:2006zu} T.~G.~Zlosnik, P.~G.~Ferreira and G.~D.~Starkman,
Phys.\ Rev.\ D \textbf{75}, 044017 (2007). 


\bibitem{CarrJ} I. Carruthers and T. Jacobson, Phys Rev D {\bf 83} 024034
	(2011).


	\bibitem{Jacobson}
	T.~Jacobson,
	PoS QG {\bf -PH}, 020 (2007).
	%


\bibitem{Carroll:2004ai} S.~M.~Carroll and E.~A.~Lim,
Phys.\ Rev.\ D \textbf{70}, 123525 (2004). 


\bibitem{Garfinkle:2011iw} D.~Garfinkle and T.~Jacobson,
Phys.\ Rev.\ Lett.\ \textbf{107} (2011) 191102. 

\bibitem{Horava} P. Horava,
	Phys. Rev. D79 (2009) 084008; T. Jacobson,  Phys. Rev. D81 101502 (2010).
	
\bibitem{TJab13} T. Jacobson, Phys.\ Rev.\ D {\bf 89}, 081501 (2014). 
	



\bibitem{Wang:2017brl}
  A.~Wang,
  Int.\ J.\ Mod.\ Phys.\ D {\bf 26} (2017) no.07,  1730014.
	
			
\bibitem{Sarbach:2019yso} 
  O.~Sarbach, E.~Barausse and J.~A.~Preciado-López,
  arXiv:1902.05130 [gr-qc].
	
\bibitem{Oost:2018oww} 
  J.~Oost, M.~Bhattacharjee and A.~Wang,
  Gen.\ Rel.\ Grav.\  {\bf 50}, no. 10, 124 (2018).
	
	
	
\bibitem{Nilsson:2018knn} 
  N.~A.~Nilsson and E.~Czuchry,
  Phys.\ Dark Univ.\  {\bf 23}, 100253 (2019).

	

	
\bibitem{Trinh:2018pcb} 
  D.~Trinh, F.~Pace, R.~A.~Battye and B.~Bolliet,
  Phys.\ Rev.\ D {\bf 99}, no. 4, 043515 (2019).
	

	
	\bibitem{Barrow:2012qy} J.~D.~Barrow,
Phys.\ Rev.\ D \textbf{85}, 047503 (2012). 


\bibitem{Sandin:2012gq} P.~Sandin, B.~Alhulaimi and A.~Coley,
Phys.\ Rev.\ D \textbf{87}, no. 4, 044031 (2013).




\bibitem{Alhulaimi:2013sha} B.~Alhulaimi, A.~Coley and P.~Sandin,
J.\ Math.\ Phys.\ \textbf{54}, 042503 (2013).


\bibitem{Coley:2015qqa} A.~A.~Coley, G.~Leon, P.~Sandin and J.~Latta,
JCAP \textbf{1512}  010 (2015). 


\bibitem{Latta:2016jix} 
  J.~Latta, G.~Leon and A.~Paliathanasis,
  JCAP {\bf 1611}, no. 11, 051 (2016).
	
	

\bibitem{Campista:2018gfi} 
  M.~Campista, R.~Chan, M.~F.~A.~da Silva, O.~Goldoni, V.~H.~Satheeshkumar and J.~F.~V.~da Rocha,
  arXiv:1807.07553 [gr-qc].
	


	

	
\bibitem{VanDenHoogen:2018anx} 
  R.~J.~Van Den Hoogen, A.~A.~Coley, B.~Alhulaimi, S.~Mohandas, E.~Knighton and S.~O'Neil,
  JCAP {\bf 1811}, no. 11, 017 (2018).
	
\bibitem{Alhulaimi:2017ocb} 
  B.~Alhulaimi, R.~J.~Van Den Hoogen and A.~A.~Coley,
  JCAP {\bf 1712}, no. 12, 045 (2017).
	

		
	\bibitem{WE}
	J. Wainwright and G. F. R. Ellis (editors), {\em Dynamical Systems
		in Cosmology} (Cambridge University Press, 1997).
		
		



\bibitem{Coley:2003mj} A.~A.~Coley, ``Dynamical systems and cosmology,''
(Astrophysics and Space Science Library. 291. ISBN 1-4020-1403-1).


\bibitem{Copeland:1997et} E.~J.~Copeland, A.~R.~Liddle and D.~Wands,
Phys.\ Rev.\ D \textbf{57}, 4686 (1998);
P.~G.~Ferreira and M.~Joyce,
Phys.\ Rev.\ Lett.\ \textbf{79}, 4740 (1997);
X.~m.~Chen, Y.~g.~Gong and E.~N.~Saridakis,
JCAP \textbf{0904}, 001 (2009); 
C.~Xu, E.~N.~Saridakis and G.~Leon,
JCAP \textbf{1207}, 005 (2012).




\bibitem{Nilsson:2000zf} U.~S.~Nilsson and C.~Uggla,
Annals Phys.\ \textbf{286}, 278 (2001). 




\bibitem{Tolman:1939jz} R.~C.~Tolman,
Phys.\ Rev.\ \textbf{55}, 364 (1939). 


\bibitem{Oppenheimer:1939ne} J.~R.~Oppenheimer and G.~M.~Volkoff,
Phys.\ Rev.\ \textbf{55} (1939) 374. 

\bibitem{Misner:1964zz} C.~W.~Misner and H.~S.~Zapolsky,
Phys.\ Rev.\ Lett.\ \textbf{12} (1964) 635. 
[Erratum, Phys.\ Rev.\ Lett.\ \textbf{13} (1964) 122.].


	%
	
	
\bibitem{kasner} Edward Kasner.
Trans. Amer. Math. Soc. \textbf{27} (1925), 101-105.



\bibitem{Ganguly:2014qia}
  A.~Ganguly, R.~Gannouji, R.~Goswami and S.~Ray,
  Class.\ Quant.\ Grav.\  {\bf 32}, no. 10, 105006 (2015).

\bibitem{Clarkson:2002jz}
  C.~A.~Clarkson and R.~K.~Barrett,
  Class.\ Quant.\ Grav.\  {\bf 20}, 3855 (2003).
\bibitem{Clarkson:2007yp}
  C.~Clarkson,
  Phys.\ Rev.\ D {\bf 76}, 104034 (2007).



	\bibitem{Barausse:2011pu}
	E.~Barausse, T.~Jacobson and T.~P.~Sotiriou,
	Phys.\ Rev.\ D {\bf 83}, 124043 (2011).

	\bibitem{Eling:2006ec}
	C.~Eling and T.~Jacobson,
	Class.\ Quant.\ Grav.\  {\bf 23}, 5643 (2006).


		\bibitem{tilt} A.A. Coley and S. Hervik,  {Class. Quant. Grav.}
	\textbf{22} 579 (2005);
	A.A. Coley, S. Hervik and W.C. Lim,  {Class. Quant. Grav.}
	\textbf{23} 3573 (2006).
	
	\bibitem{Eling:2006df}
	C.~Eling and T.~Jacobson,
	Class.\ Quant.\ Grav.\  {\bf 23}, 5625 (2006).
    

  

	
	
	\bibitem{Seifert:2007fr}
	M.~D.~Seifert,
	Phys.\ Rev.\  D {\bf 76}, 064002 (2007).
	
		
	
		\bibitem{Gao:2013im}
	C.~Gao and Y.~-G.~Shen,
	arXiv:1301.7122 [gr-qc].
	
	

	\bibitem{Tamaki:2007kz}
	T.~Tamaki and U.~Miyamoto,
	arXiv:0709.1011 [gr-qc].
	
	
	\bibitem{Eling:2007xh}
	C.~Eling, T.~Jacobson and M.~Coleman Miller,
	Phys.\ Rev.\  D {\bf 76}, 042003 (2007).
	
\bibitem{Garfinkle:2007bk} D.~Garfinkle, C.~Eling and T.~Jacobson,
Phys.\ Rev.\ D \textbf{76} (2007) 024003. 
	
	\bibitem{Konoplya:2006rv}
	R.~A.~Konoplya and A.~Zhidenko,
	Phys.\ Lett.\  B {\bf 644}, 186 (2007);
	Phys.\ Lett.\  B {\bf 648}, 236 (2007).
	
\bibitem{Nilsson:2000zg} 
  U.~S.~Nilsson and C.~Uggla,
  Annals Phys.\  {\bf 286}, 292 (2001).
	
\bibitem{Carr:1999rv} 
  B.~J.~Carr, A.~A.~Coley, M.~Goliath, U.~S.~Nilsson and C.~Uggla,
  Class.\ Quant.\ Grav.\  {\bf 18}, 303 (2001).
	
	

\bibitem{Leon:2019jnu} 
  G.~Leon, A.~Coley and A.~Paliathanasis,
  ``Static Spherically Symmetric Einstein-aether models II: Integrability and the Modified Tolman-Oppenheimer-Volkoff approach,''
  arXiv:1906.05749 [gr-qc].
	
	
	\bibitem{Abl} M.J. Ablowitz, A. Ramani and H. Segur, \ Lettere al Nuovo
Cimento \textbf{23} 333 (1978);  M.J. Ablowitz, A. Ramani and H. Segur, J. Math. Phys. \textbf{%
21} 715 (1980); M.J. Ablowitz, A. Ramani and H. Segur, J. Math. Phys. \textbf{%
21} 1006 (1980).

\bibitem{paperIII} Genly Leon, Alan Coley, Joey Latta, and Alfredo Millano,  ``Static Spherically Symmetric Einstein-Aether models III: perfect fluids  plus an scalar field with an harmonic potential'' (under preparation).

\bibitem{Alho:2015cza}  A.~Alho, J.~Hell and C.~Uggla, 
Class.\ Quant.\ Grav.\ \textbf{32}, no. 14, 145005 (2015). 


	

	
	
\bibitem{Jacobson:2004ts} 
  T.~Jacobson and D.~Mattingly,
  Phys.\ Rev.\ D {\bf 70}, 024003 (2004).



\bibitem{Elliott:2005va} 
  J.~W.~Elliott, G.~D.~Moore and H.~Stoica,
  JHEP {\bf 0508}, 066 (2005).



\bibitem{Foster:2005dk} 
  B.~Z.~Foster and T.~Jacobson,
  Phys.\ Rev.\ D {\bf 73}, 064015 (2006).

	
\bibitem{Lefschetz} S. Lefschetz, Differential Equations: Geometric Theory
(Dover, New York, 1977).

\bibitem{Lynch} S. Lynch, Dynamical Systems with Applications using Maple:
(Birkhauser, Boston, 2010).


\bibitem{Perko} L. Perko, \textit{Differential equations and dynamical
systems, third edition} (Springer-Verlag, New York, 2001).


\bibitem{Newref1}
B Cropp, S Liberati, A Mohd, M Visser, Phys.  Rev.  D 89, 064061 (2014). 



\bibitem{Newref5}
N Franchini, M Saravani, T P. Sotiriou, Phys. Rev. D 96, 104044 (2017). 


\bibitem{Newref6}
J Bhattacharyya, M Colombo, T P. Sotiriou,
Class. Quantum Grav. 33, 235003 (2016).  


	
\bibitem{[12]} P.  Berglund, J.  Bhattacharyya and D.  Mattingly, 
Phys.  Rev.  D 85 (2012) 124019. 
	

	
\bibitem{Newref4}
S Liberati, C Pacilio, Phys. Rev. D 96, 104060 (2017).

 \bibitem{[14]} P.  Berglund, J.  Bhattacharyya and D.  Mattingly, Phys. Rev.  Lett.  110, no.  7, 071301 (2013). 


\bibitem{Lin:2017jvc} 
  K.~Lin, S.~Mukohyama, A.~Wang and T.~Zhu,
  Phys.\ Rev.\ D {\bf 95}, no. 12, 124053 (2017).
	
	
\bibitem{Misner:1964je} 
  C.~W.~Misner and D.~H.~Sharp,
  Phys.\ Rev.\  {\bf 136}, B571 (1964).
	
\bibitem{Buchdahl:1959zz} 
  H.~A.~Buchdahl,
  Phys.\ Rev.\  {\bf 116}, 1027 (1959).
	
\bibitem{hartle1978}	J. B. Hartle. 
Phys. Rep.,
{\bf 46} :201-247, 1978.

	


\end{thebibliography}

\end{document}